\documentclass[fleqn,usenatbib]{mnras}

\usepackage{newtxtext,newtxmath}
\usepackage{graphicx}
\usepackage{physics}
\usepackage{epstopdf}
\usepackage{amsmath}
\usepackage{comment,xcolor}
\usepackage[toc,page]{appendix}
\usepackage{setspace}
\usepackage[export]{adjustbox}
\usepackage{natbib}
\usepackage{multirow, multicol}
\usepackage{float}
\usepackage[T1]{fontenc}

\DeclareRobustCommand{\VAN}[3]{#2}
\let\VANthebibliography\thebibliography
\def\thebibliography{\DeclareRobustCommand{\VAN}[3]{##3}\VANthebibliography}

\usepackage[margin=10pt,font=small,labelfont={sf,bf},format=plain,indention=0pt]{caption}

\usepackage[caption=false]{subfig}
\DeclareCaptionListOfFormat{myparens}{#2}
\usepackage{capt-of}

\usepackage[inline,shortlabels]{enumitem}
\makeatletter

\newcommand{\myItemCol}[2]{\item[\textcolor{#2}{#1}]\protected@edef\@currentlabel{#1}}

\newcommand{\myItem}[1]{\item[#1]\protected@edef\@currentlabel{#1}}

\makeatother

\usepackage{dashrule}
\usepackage{multirow, ltxtable, tabularx, longtable, makecell}
\newcolumntype{S}{>{\hsize=0.8\hsize\centering\arraybackslash}X}
\newcolumntype{L}{>{\hsize=1\hsize\centering\arraybackslash}X}

\usepackage{boldline}
\usepackage{tabularx}
\usepackage{hyperref}

\makeatletter
\let\saved@includegraphics\includegraphics
\AtBeginDocument{\let\includegraphics\saved@includegraphics}
\renewenvironment*{figure}{\@float{figure}}{\end@float}
\makeatother

\usepackage{xspace}

\newcommand{\eg}{e.g.\/,\xspace}

\newcommand{\Sim}{\sim\kern-0.2em\xspace}

\newcommand{\eq}{Eq.\/\xspace}

\newcommand{\Sec}{Section}
\newcommand{\Secs}{Sections}

\newcommand{\equ}[1]{\eq~(\ref{equ:#1})}
\newcommand{\figu}[1]{Fig.~\ref{fig:#1}}

\newcommand{\bi}{\begin{itemize}}
\newcommand{\ei}{\end{itemize}}

\newcommand{\software}[1]{\textsc{#1}}

\title[ll-GRBs as VHE gamma-ray and UHECR sources]{Multi-wavelength radiation models for low-luminosity GRBs, and the implications for UHECRs}
\author[A. Rudolph et al]{
A. Rudolph,$^{1}$\thanks{E-mail: annika.rudolph@desy.de}
\v Z. Bo\v snjak,$^{2}$\thanks{E-mail: Zeljka.Bosnjak@fer.hr}
A. Palladino,$^{1}$
I. Sadeh,$^{1}$ 
and W. Winter$^{1}$
\\
$^{1}$ Deutsches Elektronen-Synchrotron DESY, Platanenallee 6, 15738 Zeuthen, Germany\\
$^{2}$ Faculty of Electrical Engineering and Computing, University of Zagreb, Unska ul. 3, 10000 Zagreb, Croatia\\
}

\date{Accepted 2022 February 11. Received 2022 January 21; in original form 2021 July 30}

\pubyear{2022}

\begin{document}
\label{firstpage}
\pagerange{\pageref{firstpage}--\pageref{lastpage}}
\maketitle
\begin{abstract}

We study the prompt phase of low-luminosity Gamma-Ray Bursts (ll-GRBs) as potential source of very-high-energy (VHE) gamma rays and Ultra-High-Energy Cosmic Rays (UHECRs). Within the internal shock model we choose parameters for the relativistic outflow such that our representative events have observed properties similar to GRBs 980425, 100316D and 120714B and self-consistently calculate the full spectral and temporal properties in a leptonic synchrotron self-Compton scenario.
To investigate the conditions under which inverse Compton radiation may lead to a peak in the GeV--TeV range, we vary the fraction of internal energy supplying the magnetic field.  Further, we determine the maximal energies achievable for UHECR nuclei and derive constraints on the baryonic loading/typical duration by comparing to the extragalactic gamma-ray background. We find that ll-GRBs are potential targets for multiwavelength studies and in reach for Imaging Atmospheric Cherenkov Telescopes (IACTs) and optical/UV instruments. For comparable sub-MeV emission and similar dynamical evolution of the outflow, weak (strong) magnetic fields induce high (low) fluxes in the VHE regime and low (high) fluxes in the optical. VHE emission may be suppressed by $\gamma \gamma$-absorption close to the engine or interactions with the extragalactic background light for redshifts $z > 0.1$. For UHECRs, the maximal energies of iron nuclei (protons) can be as high as $\simeq 10^{11}$~GeV ($10^{10}$~GeV) if the magnetic energy density is large (and the VHE component is correspondingly weak). These high energies are possible by decoupling the production regions of UHECR and gamma-rays in our multizone model. Finally, we find basic consistency with the energy budget needed to accommodate the UHECR origin from ll-GRBs.
\end{abstract} 
\section{Introduction}

During the past two decades the interest in low-luminosity gamma-ray bursts (ll-GRBs) 
has been rising due to the new observations indicating that these events 
constitute a sub-population having significantly different characteristics with  respect to commonly 
observed long  gamma-ray bursts. In addition, their high  local rate (\citealt{Liang:2006ci}) makes them the key targets for 
the multimessenger astronomy (e.g. \citealt{Murase:2006mm}; \citealt{Boncioli:2018lrv}; \citealt{liu2012}; \citealt{siellez2018}).

Long GRBs, most probably associated with core-collapse supernovae, 
typically last 10s of seconds, and are characterized by
high isotropic equivalent luminosities,   
$ 10^{50} \sim 10^{52} \mathrm{erg\,s^{-1}}$ (for a recent review  see  e.g. \citealt{Kumar:2014upa}). 
ll-GRBs
exhibit substantially lower
luminosities 
$L_{\text{iso}} \lesssim 10^{49} \mathrm{erg\,s^{-1}}$, and in some cases substantially longer durations, of
up to several $10^{3}$~seconds~\cite{Liang:2006ci, Virgili:2008gp, Sun:2015bda}. 
Additionally, these events are characterized
by relatively low Lorentz factors during the prompt phase (${\Gamma \lesssim 50}$)
and 
lower peak energies (E$_{\textit{p}} \lesssim$ 100~keV)~(\cite{Ghirlanda:2017opl, Cano:2016ccp}).

Due to their low luminosity, ll-GRBs are mostly detected at low
redshifts, ${z \lesssim 0.1}$.  
Only a small number of ll-GRBs have been observed to date, due to
the limited sensitivity of currently running X-ray instruments to
low-E$_{\textit{p}}$ events~(\cite{Liang:2006ci, Sun:2015bda}). Observations by instruments with the ability to detect very faint GRBs (\cite{foley08, Virgili:2008gp})
showed that the population of ll-GRBs is distinct from high-luminosity events, and dominates the local GRB population. It is estimated
that the local rate of occurrence of these events is 
$\rho_0 \sim 200 \ \rm Gpc^{-3} yr^{-1}$,
roughly two orders of magnitude
higher than that of high-luminosity GRBs (GRB-HL).  

Several interpretations of ll-GRBs have been proposed. One possibility is that 
they could be accommodated within the same framework as commonly observed long GRBs (having high luminosities), if they were observed off-axis 
(\citealt{Pescalli:2014qja}; \citealt{Aloy:2018czj}). This however implies the existence of a very bright afterglow, which is inconsistent with observations (\cite{Daigne:2007qz}).
Additionally, an off-axis interpretation is not straightforward in various
particular cases, given e.g. the 
chromatic afterglow characteristics ~(\cite{2010ApJ...713L..55M}).
It has been alternatively proposed that some ll-GRBs are indeed
intrinsically less energetic than commonly observed GRBs~(\cite{Amati:2006ru}).
In this case, various scenarios may be invoked to explain
the observed emission: one possibility is that of  relativistic
\textit{shock breakouts}~(\cite{Waxman:2007rr, 2012ApJ...747...88N, Bromberg:2011, nakar2015}). In such cases, jets are choked; they
do not penetrate  the environment of their
progenitors entirely, and instead dissipate their energy into the
surrounding medium. \cite{nakar2015} proposed a unified view, where the differences between common and the low-luminosity events arise from the existence of an extended low-mass envelope in the case of ll-GRBs, in which the jet deposits all of its energy. 
The observed thermal X-ray emission accompanying the prompt non-thermal emission in several cases can then be interpreted as the breakout of a supernova shock from the effective photosphere of the progenitor star, e.g. \cite{Waxman:2007rr}; \cite{ sparre2012}.

Alternatively, collimated emission  has also
been observed in some ll-GRBs, indicating event topologies
similar to those of HL-GRBs, albeit having mildly-relativistic
jets~(\cite{Daigne:2007qz, Ghisellini:2007ya, Zhang:2012jc, Irwin:2015rbf}).
An interesting example is that of GRB~190829A,
which was observed at late times during
the afterglow phase by H.E.S.S. at very-high
(${> 100}$~GeV) energies.
For this event, a combination of a shock breakout and a collimated
emission episode have been proposed
for the prompt phase~(\cite{Chand:2020wqt}). Collimated emission for the prompt phase has also been explored in \cite{Fraija:2020vsa, Zhang:2020qbt}.
This indicates that both the jetted and the breakout
physical pictures may be realized in nature, depending
on the properties of a particular progenitor.

In this work we explore the possibility that the emission in low-luminosity events can be attributed to intrinsically weak GRB jets seen on-axis. The feasibility of a similar scenario has been discussed for GRB~980425 in \cite{Daigne:2007qz} and for GRB 060218 in \cite{Irwin:2015rbf}, where the prompt X-rays are attributed to a  long-lived jet. 
Unlike these studies, which do not perform time-dependent multiwavelength modelling, we make specific predictions of light curves and spectra. 

We model three examples for ll-GRBs with characteristic light curves and emission time-scales, and describe them within the internal shock framework (\cite{Daigne:1998xc, reesmeszaros94,Kobayashi:1997jk}), a multizone model which accounts for different emission sites along the jet. 
By varying the fraction of energy supplying the magnetic field (thus effectively imposing different magnetic field strengths) and accounting for the corresponding impact on the shape of the predicted spectrum, we study the efficiency of the inverse Compton scatterings at GeV--TeV energies. 
Our predictions indicate that ll-GRBs are potential targets for the Cherenkov Telescope Array (CTA).
We also model the time-dependent evolution of the light curves in different energy bands. 
We pay special attention to the fluxes in the optical and high-energy regime, which complement the X-ray signatures more commonly observed. 

ll-GRBs have also been discussed as a possible population which may power the Ultra-High Energy Cosmic Ray (UHECR) and neutrino fluxes~(\cite{Murase:2006mm,Murase:2008mr,Liu:2011cua,Murase:2013ffa, Fraija:2013cha, Senno:2015tsn, Zhang:2017moz,Boncioli:2018lrv}). 
However, \cite{Samuelsson:2018fan, Samuelsson:2020upt} have argued that cosmic-ray nuclei might not be able to reach 
utra-high energies required for UHECR fits. These findings were based on the comparison of predicted synchrotron to the observed optical and X-ray flux of a specific GRB. For the theoretical predictions, equipartition parameters (as 
the fraction of energy transferred to non-thermal electrons) usually found in the literature at face value and an energy budget high enough to power the UHECR flux were assumed (\cite{Samuelsson:2020upt}).
Additionally, in this (simplified) one-zone model it is assumed that all messengers come from the same region. Investigating the limitations of this approach, we use the multizone results of our radiation modelling to calculate the maximally achievable energies for UHECR nuclei, as well as the implications from the contribution to the Extragalactic diffuse Gamma-ray Background (EGB), such as constraints on the duration versus baryonic loading. 

The paper is organized as follows: In Section~\ref{sec:reference_grbs}, we introduce our reference events that we select from the list of observed ll-GRBs with confirmed supernova association. Section~\ref{sec:model} describes our modelling scheme, and the internal shock model for prompt GRB emission. In Section~\ref{sec:parameters} we summarize our (input) assumptions for the modelling used to reproduce the reference events, in Section~\ref{sec:theory_he_emission} we derive analytical estimates on the expected luminosity of a high-energy component.
The results of our simulations are introduced in Section~\ref{sec:results}, while we derive the maximal energies attainable for UHECR nuclei in Section~\ref{sec:uhecr}. The concluding summary can be found in  Section~\ref{sec:summary}.

\section{Reference GRBs}
\label{sec:reference_grbs}
\begin{table*}
    \begin{minipage}[c]{1\textwidth}
      \begin{center}
        \begin{tabularx}{0.95\textwidth}{LLLLLLLL}
          GRB & $E_{\gamma, \text{iso}}$~[erg] & $L_{\gamma, \text{iso}}$~[erg/s] & $E_{\text{peak}}$~[keV] & $T_{90}$~[s] & $z$ & SN & Prototype \vspace{2pt}\\ 
          \hline \hline 
          \textbf{980425} & $\mathbf{1.6 \cdot 10^{48}}$ & $\mathbf{4.6 \cdot 10^{46}}$ & $\mathbf{122}$ 
          & \textbf{34.9} 
          & \textbf{0.0085} & \textbf{1998bw} & \textbf{sp-GRB} \\
          031203 & ${1.2 \cdot 10^{49}}$ & ${3.6 \cdot 10^{47}}$ & $291$ & 37
          & 0.105  & 2003lw \\
          060218 & ${4.3 \cdot 10^{49}}$ & ${2.1 \cdot 10^{46}}$ & $4.7  
          $ & 2100
          & 0.0335 & 2006aj \\
          \textbf{100316D} & $\mathbf{3.9 \cdot 10^{49}}$ & $\mathbf{3.2 \cdot 10^{46}}$ & $\mathbf{30}$ & \textbf{1300} & \textbf{0.0591} & \textbf{2010bh} & \textbf{ul-GRB}  \\
          120121B & ${1.4 \cdot 10^{48}}$ & ${7.7 \cdot 10^{46}}$ & $92$ & 18.4 & 0.017 & 2012ba \\
          120422A & ${4.5 \cdot 10^{49}}$ & ${1.1 \cdot 10^{49}}$ & $53$ & 5.4 & 0.283 & 2012bz \\
          \textbf{120714B} & $\mathbf{{5.9 \cdot 10^{50}}}$ & $\mathbf{{5.2 \cdot 10^{48}}}$ & $\mathbf{101}$ &  \textbf{159} & \textbf{0.3984} & \textbf{2012eb} & \textbf{hl-GRB} \\
          130702A & ${6.6 \cdot 10^{50}}$ & ${1.3 \cdot 10^{49}}$ & $15$ & 59 & 0.145 & 2013dx \\
          161219B & ${8.5 \cdot 10^{49}}$ & ${1.4 \cdot 10^{49}}$ & $106$ & 7 & 0.1475 & 2016jca \\
          171205A & ${2.2 \cdot 10^{49}}$ & ${1.2 \cdot 10^{47}}$ & $125$ & 190 & 0.0368 & 2017iuk \\
          190829A & ${1.9 \cdot 10^{50}}$ & ${1.7 \cdot 10^{49}}$ & $11$ & 11 & 0.0785 &  2019oyw \\
            201015A & ${1.1 \cdot 10^{50}}$ & ${1.6 \cdot 49^{50}}$ & $10$ & 10 & 0.426 & AT2020wyy \\
          \hline
        \end{tabularx}
        \vspace{5pt}
        \captionsetup{type=table}

      \end{center}
    \end{minipage}\hfill
    \begin{minipage}[c]{1\textwidth}
      \begin{center}
        \caption{ \label{TBLll-GRBs}  List of ll-GRBs with associated supernovae. 
        $E_{\gamma, \text{iso}}$ is the isotropic equivalent emitted energy, $T_{90}$ the observed duration, $z$ the redshift, $E_\text{peak}$ denotes the observed peak energy and we derive ${L_{\gamma, \text{iso}} \equiv E_{\gamma, \text{iso}}(1+z)/T_{90}}$.
        The GRBs which will serve as references for our models are marked bold, and the prototype names are listed in the last column.
         References:
         980425~\citep{Ghisellini:2006zh, Kaneko:2006mt};
         031203~\citep{Ghisellini:2006zh, Kaneko:2006mt};
         060218~\citep{Campana:2006qe, Kaneko:2006mt};
         100316D~\citep{Starling:2010ed, Cano:2016ccp};
         120121B~\citep{Kovacevic:2014isa};
         120422A~\citep{Schulze:2014fia};
         120714B~\citep{Klose:2018ftc};
         130702A~\citep{Volnova:2016lal, Singer:2013xha};
         161219B~\citep{Cano:2017sab};
         171205A~\citep{DElia:2018xrz};
         190829A~\citep{Chand:2020wqt, Collaboration:2021fro};
         201015A~\citep{Suda:2021e7}. 
        }
      \end{center}
    \end{minipage}\hfill
\end{table*}

The classification of a specific event as a ll-GRB is not always straightforward, primarily due to the difficulty in performing effective followup observations. We list the entire sample of events
that, to our knowledge, have been confidently identified as ll-GRBs in Table~\ref{TBLll-GRBs}.
The sample is restricted to those bursts which have strongly been associated 
with a known supernova, which supports their interpretation as core-collapse events.
We select three events as benchmark scenarios for study, GRB~980425; GRB~100316D; and GRB~120714B. These cases reflect different properties of the known sample of ll-GRBs, such as duration, luminosity, and redshift. The properties of these three events are additionally summarized in Table~\ref{tab:params_modelinput}.
Our intent 
is not  to reproduce the light curves and spectra of these specific bursts exactly; rather, we use them as characteristic examples of the observed properties of ll-GRBs. The following paragraphs summarize the observations of the selected GRBs. 
Note that the papers that we will cite in the following do not report the same level of accuracy e.g., for the fitted spectral parameters.

For our first prototype, \textit{single-peaked GRB (sp-GRB)} we use the observed properties (spectral peak energy, luminosity, redshift, duration, and light curve) of GRB~980425, which has been studied within the internal shock scenario in past studies (see for example \cite{Daigne:2007qz} who propose its origin to be a mildly-relativistic low-energy jet instead of a normal GRB being observed off-axis). Given its relatively high peak energy and low isotropic energy, it is an outlier to known GRB correlations (\cite{Amati:2006ky}). Its smooth single-peaked light curve makes it a good prototype for studying single light-curve peaks. The time-integrated spectrum has a peak energy of $122 \pm 17$~keV \cite{Kaneko:2006mt}. The spectral index below the peak is found to be $\alpha = - 0.78 \pm 0.27 $ (\cite{Frontera:1999ew}) with a peak energy of $E_\mathrm{peak} = 68 \pm 40$~keV for GRBM (onboard BeppoSAX) observation during the maximum of the pulse, if the spectrum is fit by a Band-function (\cite{Band:1993eg}). For an exponential cutoff power-law fit it is determined as $\alpha \approx - 1.16 \pm  0.09 $ with $E_\mathrm{peak} = 133 \pm 8$~keV, during the maximum of the pulse as observed by BATSE LAD oboard the Compton Gamma-Ray Observatory (CGRO, \cite{Kaneko:2006mt}). 
We additionally summarize the fluences for the various instruments: BeppoSAX reports a fluence of $(2.8 \pm 0.5) \cdot 10^{-6} \mathrm{erg/cm^2}$ in the 40-700~keV band and $(1.8 \pm 0.3) \cdot 10^{-6} \, \mathrm{erg/cm^2}$ in the 2-26~keV range (\cite{Pian:1999ec}). \cite{Kaneko:2006mt} performed an analysis of the BATSE LAD observations and fluences of $ 1.99 \cdot 10^{-6} \, \mathrm{erg/cm^2}$ in the X-ray regime (2-30~keV) and a fluence of $3.40 \cdot 10^{-6} \, \mathrm{erg/cm^2}$ in the $\gamma$-ray regime (30-400~keV).

Our second prototype, \textit{ultra-long GRB (ul-GRB)}  will have observed properties similar to GRB~100316D, an ultra-long ll-GRB.
In contrast to the well-studied ultra-long GRB~060218 (where the black-body component compromises 13 per cent of the prompt spectrum) it has a sub-dominant black-body component contributing only 3 per cent to the the X-Ray flux (0.3 - 10~keV) (\citealt{Starling:2010ed}). This makes it a 
suitable candidate for the internal shock model. Also, while GRB~060218 has a very low peak energy of only $\approx$ 5~keV, the peak energy of GRB~100316D is $\approx$ 30~keV. The light curve comprises multiple peaks with maximal photon fluxes decreasing with time.  The spectral index below the peak (for a cutoff power-law fit) is found to be $\alpha \approx - 1.4$ (\cite{Starling:2010ed}), which is comparable to the one for GRB~060218 (\cite{Kaneko:2006mt} find $\alpha = - 1.44 \pm 0.006$). 
The reported fluence in the \textit{Swift} BAT range (15-350~keV) is $ (5.1 \pm 0.39) \cdot 10^{-6} \mathrm{erg/cm^2}$ (\cite{Starling:2010ed}). 
The UltraViolet Optical Telescope (UVOT) onboard the \textit{Swift} satellite reported non-detection in the $u$-band for three different time-intervals (of exposure times 35~s, 194~s and 36~s with mid-times 324~s, 440~s and 634~s after the BAT trigger) \cite{Starling:2010ed}. In \cite{Fan:2010br}, these are translated into time-averaged limits between $1.9 \cdot 10^{-13}$~erg/$\mathrm{cm^2}$ and $6.3 \cdot 10^{-13}$~erg/$\mathrm{cm^2}$, where absorption in our own and the host galaxy are accounted for.

Finally, for our third prototype \textit{high-luminosity GRB (hl-GRB)} we use the observed properties of
GRB~120714B. This GRB has a higher luminosity when comparing it to GRBs~980425 and 100316D, making it a very plausible candidate for an engine-driven scenario (\cite{Zhang:2012jc}). The BAT analysis (\cite{GCN13481}) reports a relatively high peak energy of $101$~keV and a best fit with a power-law of index $\alpha = - 1.52 \pm 0.17$. The light curve is simple, single-peaked with $T_\mathrm{90}$=159~s. Being the most distant ll-GRB in our table ($z=0.3964$), we expect a larger impact of absorption by the Extragalactic Background Light (EBL) on the observed very-high-energy (VHE) spectrum.
This burst was observed by Swift BAT, who report a fluence of $ (1.2 \pm 0.1) \cdot 10^{-6} \mathrm{erg/cm^2}$ in the 15-150~keV band \cite{GCN13481}. 

\section{Multiwavelength time-dependent radiation model}
\label{sec:model}

\begin{figure*}
\centering
\makebox[\textwidth][c]{
\includegraphics[width=.8 \textwidth]{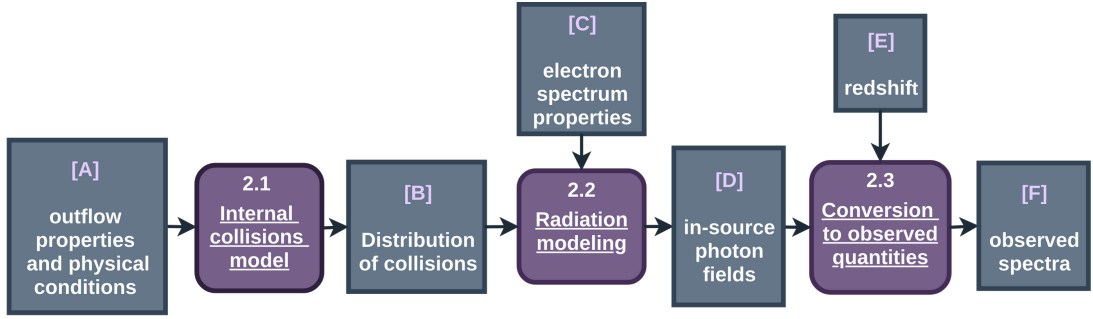}}
\caption{Flowchart of the full modelling process. Grey boxes represent physical quantities (input/output), purple ones modelling steps, which are described in Sections \ref{sec:internal_coll_model}-\ref{sec:convert_to_observed}. The outflow properties and physical conditions [A], which are wind luminosity, $\Gamma$ and mass distribution, number of plasma layers, engine activity time and microphysics parameters $\epsilon_\mathrm{B}, \epsilon_\mathrm{e}$ are input for the fireball modelling. This in turn yields a distribution of collisions of plasma layers [B] which describes the energy dissipated at a certain time and distance from the source and the corresponding bulk Lorentz factor of the region. Those properties, as well as assumptions on the injected particle spectra [C] are used for the radiation modelling. This returns the in-source photon densities [D]. With a given redshift [E] and assumptions on EBL absorption we can calculate the observed spectra and light curves [F].}
\label{fig:flowchart}
\end{figure*}

In this section we describe the modelling process, which is divided in several steps illustrated in \figu{flowchart}:
We model the evolution of the jet following the internal shock scenario (\cite{Daigne:1998xc, Kobayashi:1997jk}). In a similar way as in \cite{Daigne:2007qz}, we adopted the scenario in which the outflows of LL GRBs are mildly relativistic (having lower bulk Lorentz factors) and have lower wind luminosities. 
The simulation of the shock dynamics is used to derive the energy dissipated at a certain time and distance from  the source, as well as the  bulk Lorentz factor of the region. We describe the physical conditions in the shocked medium by three microphysics parameters: the fraction of energy received by non-thermal electrons ($\epsilon_\mathrm{e}$) and the magnetic field ($\epsilon_\mathrm{B}$), and the fraction of accelerated electrons ($\zeta$). With these assumptions we calculate the corresponding spectra in the comoving frame, and convert them into observed quantities. These different steps are described in \Secs~\ref{sec:internal_coll_model}--\ref{sec:convert_to_observed}.

In \Sec~\ref{sec:parameters} we list the input parameters for the benchmarks introduced in Section~\ref{sec:reference_grbs}.

\subsection{Internal shock model}
\label{sec:internal_coll_model}
Here we limit ourselves to a short description of the most relevant formulas while referring to \cite{Daigne:1998xc, Bosnjak:2008bd} for a more detailed view. 

A relativistic outflow of a given mass density and velocity profile is approximated by a series of discrete layers with Lorentz factors $\Gamma_i$ and masses $m_i$, separated by a (constant) ejection time $\Delta t_\mathrm{ej}$, corresponding to our discretization width. From here on we will assume an initially constant wind luminosity of the outflow, given by $L_\mathrm{wind}$. This simplified model was validated by the hydrodynamical calculation in \cite{Daigne:2000xg, Rudolph:2019ccl}. As a fast part of the outflow catches up with a slower one, energy is released in collisions between the layers. 
The collision between a fast (subscript $f$) and a slow (subscript $s$) layer 
creates a new, merged shell of mass $m_m = m_f + m_s$ of Lorentz factor
\begin{align}
 \Gamma_\mathrm{m} \simeq \sqrt{\frac{m_\mathrm{f} \Gamma_\mathrm{f} + m_\mathrm{s} \Gamma_\mathrm{s}}{m_\mathrm{f}/ \Gamma_\mathrm{f} + m_\mathrm{s}/ \Gamma_\mathrm{s}}} \, , 
 \label{equ:gamma_m}
\end{align}
which continues propagating in the outflow and may undergo subsequent collisions.

We assume that during the collision process most of the energy is released as the lighter of the two shells sweeps up a mass comparable to its own; in this case of equal masses of the colliding shells \equ{gamma_m} simplifies to 
\begin{align}
    \Gamma_\mathrm{r} = \sqrt{\Gamma_\mathrm{f} \Gamma_\mathrm{s}} \, .
\end{align}
From energy conservation, the internal energy released during the collision can be calculated as 
\begin{align}
     E_\mathrm{diss} = m c^2 (\Gamma_\mathrm{f} + \Gamma_\mathrm{s} - 2 \Gamma_\mathrm{r}) \, ,
\label{equ:ediss}
\end{align}
where $m$ is the smaller of the two masses ($m = \min{(m_\mathrm{f}, m_\mathrm{s})}$). Within this model, $\Gamma_\mathrm{r}$ is the Lorentz factor of the radiation-emitting plasma. 

If we define the contrast of Lorentz factors as $\kappa=\Gamma_\mathrm{f}/\Gamma_\mathrm{s}$, we may calculate the comoving mass density $\rho^\prime$ and dissipated energy per unit mass $\epsilon_\mathrm{diss}^\prime$ during the collision as
\begin{align}
    \rho^\prime &\simeq \frac{L_\mathrm{wind}}{4 \pi R_\mathrm{coll}^2 \Gamma_r^2 c^3} \\
    \epsilon^\prime_\mathrm{diss} &\simeq \frac{\left( \sqrt{\kappa} -1 \right)^2}{2 \sqrt{\kappa}}c^2 \, ,
\end{align}
where $R_\mathrm{coll}$ is the distance from the source at which the two shells collide. Primed quantities refer to the comoving frame of the plasma.
During the dynamical evolution of the outflow, the distribution of collisions results in a distribution of $\rho^\prime$, $\epsilon^\prime_\mathrm{diss}$ and $\Gamma_\mathrm{r}$ as a function of radius $R_\mathrm{coll}$ which is fully determined by the initial Lorentz factor distribution and the wind luminosity.

We identify the expansion time (due to adiabatic cooling) as dynamical time-scale of the system for each collision: 
\begin{align}
    t^\prime_{\rm ex} = \frac{R_\mathrm{coll}}{c \Gamma_\mathrm{r}} \, .
\end{align}
Finally, the collision in a GRB of redshift $z$ and ocurring at a time $t$ (in the source frame) starts to be observed at
\begin{align}
     T_\mathrm{obs} = (1+ z) (t - \frac{R_\mathrm{coll}}{\mathrm{c}}) \, 
\label{equ:obs_time}
 \end{align}

\subsubsection{Particle acceleration in the shocked medium}
\label{sec:shocked_medium_conditions}
In each layer, non-thermal electrons receive the energy $E_e = \epsilon_\mathrm{e} E_\mathrm{diss}$, whereas the magnetic field is supplied by a fraction $\epsilon_\mathrm{B}$ of the internal energy.
For the radiation modelling we assume that no baryons are accelerated.

As a pre-set assumption we impose $\epsilon_\mathrm{e} = 1/3 $ in all collisions and GRBs (as suggested for relativistic shocks). 

The relativistic electrons are assumed to follow a power-law of index $-p$ between a minimum and maximum Lorentz factor ($\gamma_\mathrm{e, min}$, $\gamma_\mathrm{e, max}$):
\begin{align}
    n(\gamma_e) \simeq (p-1) \frac{n_\mathrm{e}^\mathrm{acc}}{\gamma_\mathrm{e, min}}\left(\frac{\gamma_e}{\gamma_\mathrm{e, min}}\right)^{-p} \, , 
\label{equ:initial_electrons}
\end{align}
where $\gamma_e$ is the electron Lorentz factor. In the following, we assume only a fraction of the electrons in the outflow to be accelerated to  relativistic velocities and calculate the number of accelerated electrons as $n_\mathrm{e}^\mathrm{acc} = \zeta \frac{\rho^\prime}{m_p}$. As in \cite{Bosnjak:2014hya}, we set the fraction of accelerated electrons $\zeta$ to be proportional to the dissipated energy per unit mass $\epsilon^\prime_\mathrm{diss}$ (via the parameter $\zeta_0$, $\zeta = \zeta_0 \cdot \frac{\epsilon^\prime_\mathrm{diss}}{100 \mathrm{MeV/ proton}} \simeq \zeta_0 \frac{\epsilon^\prime_\mathrm{diss}}{9.58 \cdot 10^{19}c^2} $).

Under this assumption, the minimum Lorentz factor of electrons remains constant throughout the evolution and is given by:  
\begin{equation}
    \gamma_\mathrm{e, min} = \frac{p-2}{p-1}\frac{\epsilon_\mathrm{e}}{\zeta}\frac{m_p}{m_e} \frac{\epsilon^\prime_\mathrm{diss}}{c^2}
    \propto \frac{p-2}{p-1}\frac{\epsilon_\mathrm{e}}{\zeta_0}\frac{m_p}{m_e} \ .
    \label{equ:gamma_min}
\end{equation}
This formula demonstrates that the electron spectrum can equivalently described by either $\gamma_{\mathrm{e, min}}$ or $\zeta_0$; we use the latter in this work.

The maximum Lorentz factor is determined balancing radiative losses and acceleration. For this we assume the acceleration time-scale to be defined as 
\begin{equation}
t^\prime_\mathrm{acc} = \frac{1}{\eta} \frac{E^\prime}{c B^\prime e} \, ,
\label{equ:acctime}
\end{equation}
(where $E^\prime$ is the energy of the particle in the shell frame and $B^\prime$ the comoving magnetic field strength)
as suggested by diffusive shock acceleration and set $\eta = 1 $ for electrons.

\subsubsection*{Synchrotron peak energy and microphysics parameters}
\label{sec:microphysics_parameters}

The observed synchrotron peak in a single layer is (in the fast-cooling approximation) given by
\begin{equation}
    E_\mathrm{peak} \simeq 17 \ \mathrm{eV} \left(\frac{\Gamma_r}{10} \right) \left(\frac{B^\prime}{100 \,  \mathrm{G}} \right) \left(\frac{\gamma_\mathrm{e, min}}{1000} \right)^2 \, .
\end{equation}
Given that the magnetic field in the comoving frame can be calculated as $B^\prime = \sqrt{8 \pi \epsilon_\mathrm{B} \rho^\prime \epsilon^\prime_\mathrm{diss}}$, it is straightforward to see that the observed peak depends on both $\zeta_0$ and $\epsilon_\mathrm{B}$ (under condition that it is not affected by inverse Compton scatterings). 

For the reproduction of a given GRB we choose those microphysics parameters such that the synchrotron peak at the maximum of the pulse (the region where most of the energy is dissipated) matches the observed one. For a fixed dynamical jet evolution (setting $\rho^\prime$, $\epsilon^\prime_\mathrm{diss}$, $\Gamma_r$, etc.) this creates a relation between $\zeta_0$ and $\epsilon_\mathrm{B}$, leaving only one of them as a free parameter. 

\subsection{Radiation Modeling}

With these assumptions on the accelerated electrons, we calculate the spectrum produced in each layer as a result of the collision occurring between layers of different velocities. These layers correspond to our radiation zone. The overall final spectrum and light curve are given by the superposition of all spectra produced in individual layers, taking into account the time at which collisions occur. Note that the layers are determined by our discretization scheme, and the result does not depend on that as long as enough layers are used and the properties between adjacent layers change slowly enough (compared to the expansion time-scale). 

The evolution of the particle spectra and photon fields in each layer are modeled with the time-dependent radiative Code \software{AM$^3$} (\cite{Gao:2016uld}) which includes synchrotron radiation, synchrotron self-absorption, inverse Compton scattering in both Thomson and Klein-Nishina regimes, and $\gamma \gamma$-absorption. The (low-energy) thermal component is not modeled or taken into account in the current work. As detailed above, the thermal emission (if observed) represents only a small fraction of the energy budget; also its signature might be in any way distinguished from the jetted emission.
While some studies (\cite{Bosnjak:2008bd, Bosnjak:2014hya, Globus:2014fka}) neglect the effect of secondary lepton pairs created by $\gamma \gamma$-absorption, they are explicitly included in our approach. Being treated the same way as primary leptons, those pairs contribute by synchrotron and inverse Compton radiation (see Appendix~\ref{appendix:radiative_processes} for a detailed discussion).
In addition to the effects described above we account for adiabatic cooling due to the expansion of the shell with the cooling time given by $t^\prime_{\rm ex}$. We note that adiabatic cooling is expected to impact the electron spectra mostly in the low-energy ranges (where the contribution to the observed photon spectra is small), representing the longest cooling time-scale of electrons (see \figu{time-scaleplot_ele}).

The initial distribution at $t^\prime = 0 $ of accelerated electrons in the comoving volume is in our model given by \equ{initial_electrons} and no further injection of electrons beyond this point is assumed. This means that we do not make any assumptions on how or on what time-scale the radiation zone is filled with electrons, which depends on the details of the acceleration process.

We follow the particle spectra and radiative processes over the dynamical time-scale $t^\prime_{\rm ex}$ and take the spectrum at $t^\prime = t^\prime_\mathrm{ex}$ as final, emitted spectrum. Since this represents the slowest cooling time-scale of electrons, at $t^\prime = t^\prime_\mathrm{ex}$ all electrons are expected to have cooled. As an result, they will not contribute anymore to the photon spectrum.
This approach also (roughly) corresponds to the assumption of photons escaping over the dynamical/expansion time-scale. However, this treatment is only approximate - with a full treatment of the escape in the radiative calculations (incorporating an escape term of $t^\prime_\mathrm{escape} =t^\prime_\mathrm{ex}$ in the partial differential equations and integrating over the escaped spectra), the temporal evolution of the spectra within single layers would also leave an imprint on the observed spectra.

\subsection{Conversion to observed spectra and light curves}
\label{sec:convert_to_observed}
Given an total energy density of photons ($u_{\nu^\prime}^\prime = \frac{\nu^\prime dN}{d\nu^\prime dV^\prime}$, where $V^\prime = 4 \pi R^2 \Delta t_\mathrm{ej} c \Gamma_r$ is the comoving volume of the emitting layer and $\nu^\prime = \nu^\prime \cdot \Gamma_r /(1+z)$ is the comoving photon energy), one may compute the differential observed fluence for a single collision as
\begin{align}
    \nu F_{\nu} =u_{\nu^\prime}^\prime V^\prime \frac{\Gamma_r \nu^\prime}{4\pi (1+z)D^2} \, ,
\end{align}
with the comoving distance $D$.

For the calculation of time-dependent quantities such as the light curves and the time evolution of spectra, we follow \cite{Granot:1998ep} and take into account the curvature of the emitting surface for each collision. 

Given our focus on the (expected) high-energy component, we explicitly incorporate absorption due to interactions with the EBL in our simulations which are expected to impact the spectra in the TeV-range for redshifts $z>0.1$ \cite{H.E.S.S.:2017odt}. EBL absorption is calculated using the open-source \software{gammapy} package (\cite{Deil:2017yey, Nigro:2019hqf}), using the model of \cite{Dominguez:2010bv}. 

\subsection{Parameter space for reference GRBs}
\label{sec:parameters}

\begin{table*}
\fontsize{8}{10}\selectfont
\caption{Observed properties for the reference GRBs (isotropic equivalent emitted $\gamma$-ray Energy $E_{\rm \gamma, iso}$, duration $T_{90}$, observed peak energy  $E_{\rm peak}$ and redshift $z$), as well as input parameters to our model used to reproduce alike events sp-GRB, ul-GRB and hl-GRB (maximum and minimum of the initial Lorentz factor distribution ($\Gamma_{\rm initial, max}, \Gamma_{\rm initial, min}$), the source luminosity $L_{\rm wind} $ , engine activity time $t_{\rm eng} $ and the number of initial layers $N_{\rm shells}$).}	
	\begin{tabular}[b]{c | r | c  | c | c}
		& & GRB 980425 & GRB 100316D  & GRB 120714B\\ \hline
		\parbox[t]{1.7mm}{\multirow{4}{*}{\rotatebox[origin=c]{90}{Observed}}}
		& $E_{\rm \gamma, iso}$ [erg] & $1.6 \cdot 10^{48}$ &  $3.9 \cdot 10^{49}$ &  $5.9 \cdot 10^{50}$\\
		& $T_{90}$ [s] & 35 & 1300 & 159\\
		& $E_{\rm peak}$ [keV] & 122 & 30 & 101\\
		& $z$ & 0.0085 & 0.059  & 0.3984 \\ \hline
		& & sp-GRB & ul-GRB  & hl-GRB \\ \hline
		\parbox[t]{1.7mm}{\multirow{5}{*}{\rotatebox[origin=c]{90}{Input}}}
		& $\Gamma_{\rm initial, max}, \Gamma_{\rm initial, min}$ & 40, 10 & 40, 10 & 80, 20 \\
		& $L_{\rm wind} $ [erg/s] & $2.5 \cdot 10^{48}$ & $5.8 \cdot 10^{48}$ & $ 3  \cdot 10^{50}$\\
		& $N_{\rm shells}$ & 1000 & 1000 & 1000 \\
		& $t_{\rm eng} $ [s] & 40 & 1000 & 130 \\ 
	\end{tabular}
    \label{tab:params_modelinput}
\end{table*}

\begin{figure*}
\centering
\subfloat[][sp-GRB]{\includegraphics[width=.3 \textwidth, valign = b ]{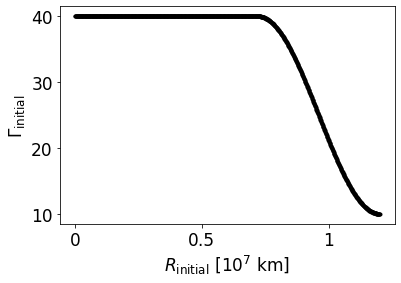}}
\subfloat[][ul-GRB]{\includegraphics[width=.3 \textwidth, valign = b]{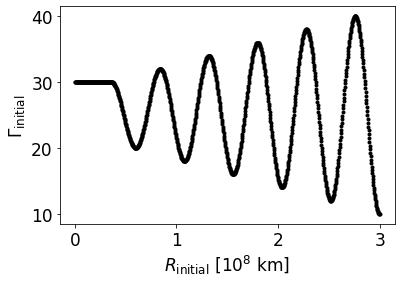}}
\subfloat[][hl-GRB]
{\includegraphics[width=.3 \textwidth, valign = b]{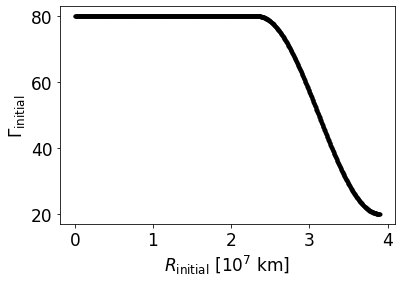}}
\caption{Initial jet Lorentz factor distributions for the three models.}
\label{fig:shell_dist}
\end{figure*}

\begin{table*}
	\caption{Microphysics parameters and maximum values retrieved from the fireball evolution: $\zeta_0$, he minimimum electron Lorentz factor $\gamma_\mathrm{e, min}$, the fraction of accelerated electrons  at the maximum of the pulse (where most of the energy is dissipated) $\zeta_\mathrm{max}$ 
	and the maximum magnetic field $B^\prime_\mathrm{max}$ for different choices of $\epsilon_\mathrm{B}$ for our models. We define $\zeta =  \min \left\{ \zeta_0 \cdot \frac{\epsilon^\prime_\mathrm{diss}}{100 \mathrm{MeV/ proton}}, 1 \right\}$}
\fontsize{7}{9}\selectfont
\setlength{\tabcolsep}{3pt}
    \begin{tabular}[b]{c | c c c c | c c c c | c c c c }
		& \multicolumn{4}{c | }{sp-GRB} & \multicolumn{4}{c | }{ul-GRB} & \multicolumn{4}{c}{hl-GRB}\\ \hline

		$\epsilon_\mathrm{B}$ & 
		$\zeta_0 / 10^{-4}$ & $\gamma_\mathrm{e, min}/ 10^{4}$ & $\zeta_\mathrm{max}/ 10^{-4}$ & $B^\prime_\mathrm{max}$ [G] &
		$\zeta_0 / 10^{-4}$& $\gamma_\mathrm{e, min}/ 10^{4}$& $\zeta_\mathrm{max}/ 10^{-4}$& $B^\prime_\mathrm{max}$ [G] &
		$\zeta_0 / 10^{-4}$& $\gamma_\mathrm{e, min}/ 10^{4}$& $\zeta_\mathrm{max}/ 10^{-4}$& $B^\prime_\mathrm{max}$ [G] \\ \hline
		$10^{-1}$     & 7.3 & 3.0 & 5.8 & 820 & 9.2 &2.8& 7.6  & 234 &  5.7 & 3.6 & 6.0 & 346 \\ 
		$10^{-2}$    & 4.1  &5.3 & 3.2 & 257 & 5.2 & 5.0 &4.3 & 74 & 3.2 & 6.4 & 3.4 & 110 \\ 
		$10^{-3}$   & 2.3 &9.4&1.8 & 82 & 2.9 & 8.9 & 2.4 & 23 & 1.8 & 11.3 & 1.9 & 35 \\ 
		$10^{-4}$  & 1.3 & 16.8&1.0 & 26 & 1.6 &15.9 & 1.4 & 7 & 1.0 &20.1 & 1.0 & 11 \\
    \end{tabular}
    \label{tab:params_epsb}
\end{table*}  

We summarize our assumptions on the initial jet configuration (ejection time and luminosity, Lorentz factor profile) in Table~\ref{tab:params_modelinput}, the (initial) Lorentz factor distributions are displayed in \figu{shell_dist}. The Lorentz factor distribution, engine active time and wind luminosity are closely related to observations such as the light curve and emitted isotropic energy and thus relatively well constrained. 
For example, to reproduce a multi-peaked light curve (as the one of GRB~100316D/ul-GRB), also a multi-peaked Lorentz factor distribution is necessary. 
In accordance with Lorentz factor measurements of ll-GRBs, we choose relatively low Lorentz factors of the outflow.
Note that due to their relatively low luminosities, we expect sp-GRB and ul-GRB not to have Thomson optical depths above unity, despite their low Lorentz factors.
As hl-GRB has a higher luminosity, we impose higher initial Lorentz factors in this case (see \cite{Ghirlanda:2017opl} for a correlation of GRB Lorentz factors and luminosities). The higher Lorentz factors for this prototype will also ensure the outflow to be optically thin, despite the higher luminosity.

We adjust the fraction of accelerated electrons with $\zeta_0$ such that we can fit the observed peak of the reference event as a synchrotron peak (see  \Sec~\ref{sec:microphysics_parameters} for a more detailed description) and leave $\epsilon_\mathrm{B}$ as a free parameter to study the impact of the magnetic field strength. 
The values of $\zeta_0$ and $\zeta_\mathrm{max}$ (the maximum value of $\zeta$ during the fireball evolution) and the maximum magnetic field $B^{\prime}_\mathrm{max}$ for all models and our choices of $\epsilon_\mathrm{B}$ are summarized in Table~\ref{tab:params_epsb}.

\begin{figure}[htb]
\centering
    \includegraphics[width=.43 \textwidth]
    {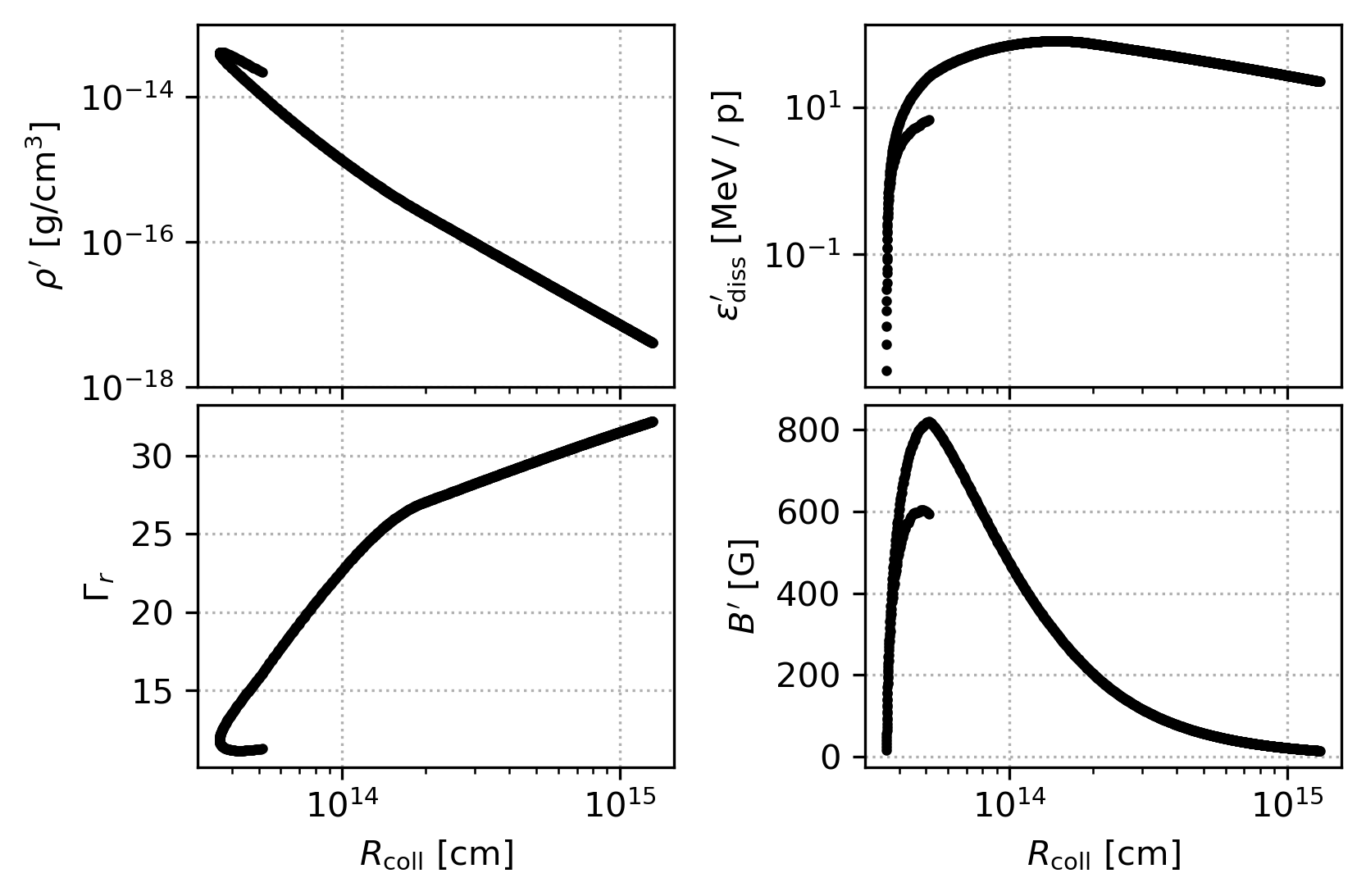}
    \caption{Dynamical evolution of the outflow for sp-GRB: $\rho^\prime$, $\epsilon^\prime_\mathrm{diss}$ and $\Gamma_\mathrm{r}$ which are fully determined by the initial Lorentz factor distribution of sp-GRB (left plot of \figu{shell_dist}) and the wind luminosity ($L_\mathrm{wind} = 2.5 \cdot 10^{48} \, \mathrm{erg/s}$), and $B^\prime$ for $\epsilon_B = 10^{-1}$.}
    \label{fig:parameter_evolution}
\end{figure}

To illustrate the dynamical evolution of the outflow, we show $\rho^\prime$, $\epsilon^\prime_\mathrm{diss}$ and $\Gamma_\mathrm{r}$ as a function of collision radius for sp-GRB in \figu{parameter_evolution}. The evolution of those parameters depends only on the initial Lorentz factor distribution of the outflow and the jet luminosity $L_\mathrm{wind}$; for $B^\prime$ it is assumed that $\epsilon_B = 10^{-1}$.

\section{Theoretical predictions for the high-energy emission}
\label{sec:theory_he_emission}

A significant emission at very high energies from low-luminosity events would make them interesting targets for Imaging Air Cherenkov Telescopes (IACTs). 
So far, the observational strategy of ground-based gamma-ray experiments has
nominally been to follow up detection by X-ray instruments, which have not been
very sensitive to ll-GRBs. 
The upcoming Cherenkov Telescope Array (CTA) will improve on the sensitivity
of existing IACTs on short time scales (\cite{2019ICRC...36..673F}).
Correspondingly, it will significantly
improve the prospects for serendipitous detection of ll-GRBs.
As these events are found at lower redshift, the effect of EBL absorption is less significant.
The prospects of detecting GeV--TeV emission from ll-GRBs are  therefore relatively high, compared to HL-GRBs.

While we describe the general, theoretical predictions on the shape of the photon spectrum in Appendix~\ref{sec:shape_theo_predic}, we here provide estimates for the expected luminosity attributed to the inverse Compton component which will give rise to a HE component by calculating the Compton Y-parameter for characteristic spectral properties of ll-GRBs.
For simplicity,  we consider the scatterings of electrons having Lorentz factors $\gamma_\mathrm{e, min}$ interacting with  photons around the peak of the synchrotron emission. 
We define the dimensionless quantity $\eta_{\mathrm{m}} = \gamma_\mathrm{e, min} h \nu_{\mathrm{peak, obs}} (1+z) /(m_e c^2 \Gamma)$ to measure if the scatterings occur in Klein-Nishina regime ($\eta_m \gtrsim 1$) or in Thomson regime ( $\eta_m < 1$) ($\nu_{\mathrm{peak, obs}}$ denotes the observed peak of the spectrum). The peak energy of synchrotron emission is dominated by electrons  at the minimum of the  distribution $\gamma_\mathrm{e, min}$ in the cases we presented, and we assume that the average value Y($\gamma_e$) can be approximated as $\bar{\mathrm{Y}} \approx \mathrm{Y}(\gamma_\mathrm{e, min})$ (\cite{Nakar:2009er}). 
\begin{enumerate}

\item[(a)] {\it Klein-Nishina regime } 
\newline
For the case where  electrons with Lorentz factors below $\gamma_\mathrm{e, min}$ cool via inverse Compton process in the Klein-Nishina regime we adopt the following approximate expression for $\mathrm{Y} (\gamma_\mathrm{e, min})$ (\cite{Duran:2012ww}): 
\begin{equation}
  \mathrm{Y}(\gamma_\mathrm{e, min}) =  t_{\mathrm{syn}}^\prime/t_{\mathrm{IC}}^\prime. 
\end{equation}
We use the following expressions for synchrotron cooling time and for the cooling time due to IC scattering in the comoving frame: 
\begin{equation}
t_{\mathrm{syn}}^\prime = \frac{6 \pi m_e c}{\sigma_T B'^2 \gamma_\mathrm{e, min}}, 
\end{equation}

\begin{equation}
t_{\mathrm{IC,KN}}^\prime \approx 4 \pi R^2 \Gamma h \nu_{\mathrm{peak, obs}} (1+z) / ( \Delta L \sigma_T/\eta_{\mathrm{m}}).   
\end{equation}
Here the cross-section for inverse Compton scatterings of photons around peak energy is smaller than the Thomson cross section by a factor $\approx \eta_{\mathrm{m}}$ (\cite{Duran:2012ww}). 
Note that the luminosity $L$ (which is derived from the observations) in the  expression for $t_{\mathrm{IC,KN}}^\prime$ is considered as the luminosity of the synchrotron component only. $\Delta$ is a numerical factor accounting for the scattering angles ($\Delta \approx$ 0.5, \citealt{Kumar:2014upa}).
This leads to the expression:  

\begin{equation}
\mathrm{Y}(\gamma_\mathrm{e, min})_{\mathrm{KN}} \approx \frac{3}{2} \left[\frac{m_e c^2}{h \nu_{\mathrm{peak, obs}} (1+z) }\right]^2  \frac{L \Delta}{c \gamma_{\mathrm{e, min}}^2 B'^2 R^2} \, .
\end{equation}
\noindent   
For the values typical of low-luminosity events (L $\approx$ 10$^{47}$ erg s$^{-1}$, E$_{\mathrm{peak, obs}}\approx$ 50 keV and $z\ll$1), we find 
\begin{equation}
 \mathrm{Y}(\gamma_\mathrm{e, min})_{\mathrm{KN}} \approx 0.03  B^{\prime -2}_{2} \gamma_{\mathrm{e,min},4}^{-2} R_{14}^{-2}.
 \label{equ:Y_KN}
\end{equation}
\noindent
We used the notation $Q=10^n Q_n$. 

\item[(b)] {\it Thomson regime } 
\newline
For Compton scatterings occurring in the Thomson regime, we follow again the simplified procedure as in \cite{Duran:2012ww}: 

\noindent 

\begin{equation}
\mathrm{Y} (\gamma_\mathrm{e, min})_\mathrm{TH} \approx \frac{3}{2} \frac{L}{cB{^\prime}^2R^2\Gamma^2} \simeq 50  B^{\prime -2}_{2} \Gamma_{1}^{-2} R_{14}^{-2}.
\label{equ:Y_TH}
\end{equation}

\end{enumerate}

\noindent
 The ratio of luminosities is given by $L_{\mathrm{IC}}/L_{\mathrm{syn}} \approx Y$. Note that $\gamma\gamma$-pair annihilation may produce a high-energy cutoff in  the spectrum, suppressing the inverse Compton component, which is not accounted for in the analytical estimate. When Klein-Nishina effects become important, the cross section for scattering photons gets smaller than the Thomson cross-section, and the scattered photon energy is reduced, which results in  $\mathrm{Y}(\gamma_\mathrm{e, min})_{\mathrm{KN}}\ll\mathrm{Y} (\gamma_\mathrm{e, min})_\mathrm{TH} $.
 
 Motivated by Equations~\ref{equ:Y_KN} and~\ref{equ:Y_TH}, in the following, for the same assumptions on the properties of the outflow, we keep the peak energy of the spectrum approximately constant and study the effect of the magnetic field on the spectral shape. In our model, the magnetic field changes throughout the dynamical evolution of the jet and is not set directly, but through the magnetic energy density via $\epsilon_B$. For Thomson scattering in fast cooling regime, we can easily relate Y to $\epsilon_B$, Y=$(\epsilon_e/\epsilon_B)^{1/2}$ , and a more general expression for Y can be found in 
e.g. \cite{kumarpanaitescu}.

\section{Results: simulated spectra and light curves}
\label{sec:results}
In this section we present the results of our simulations: the predicted observed time integrated spectra, light curves and time resolved fluxes/fluences for different energy bands. The parameter sets used for sp-GRB, ul-GRB and hl-GRB are motivated by the three real GRBs which guide our simulations. For all GRBs, we explore the impact of the magnetic field strength by varying $\epsilon_\mathrm{B}$. 

Our models reproduce various observed features of the reference GRBs, such as the light curve structure and the spectral shape. The predictions on the multiwavelength features  of  the  presented models will be possible to test by future facilities. 
This includes the fluence and flux in different energy bands (including optical and very high energy emission) and a possible delayed onset of the HE component.

\begin{figure*}
\centering
\makebox[\textwidth][c]{
\subfloat[sp-GRB]{\includegraphics[width=.33 \textwidth]{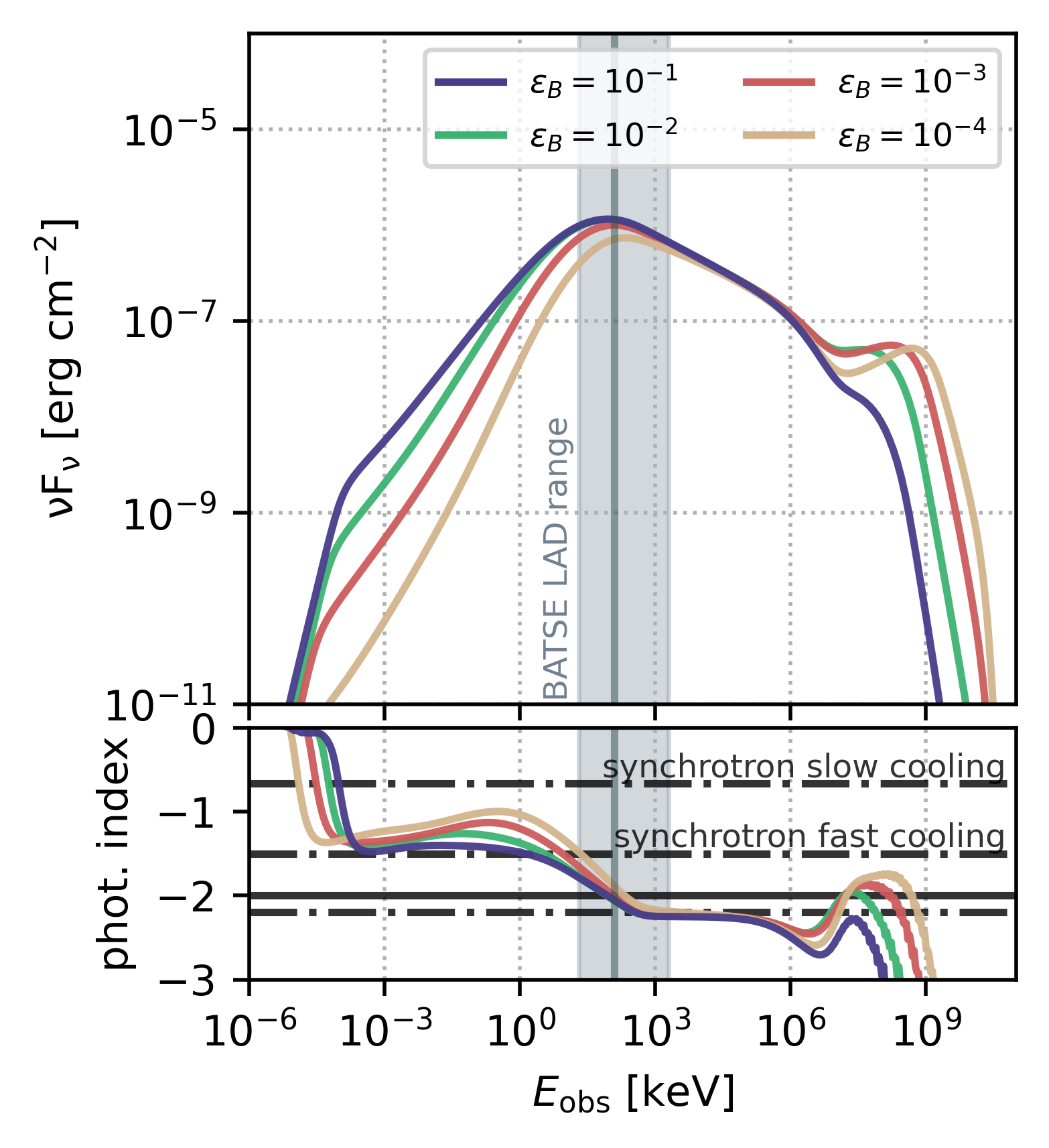}}
\subfloat[ul-GRB]{\includegraphics[width=.33 \textwidth]{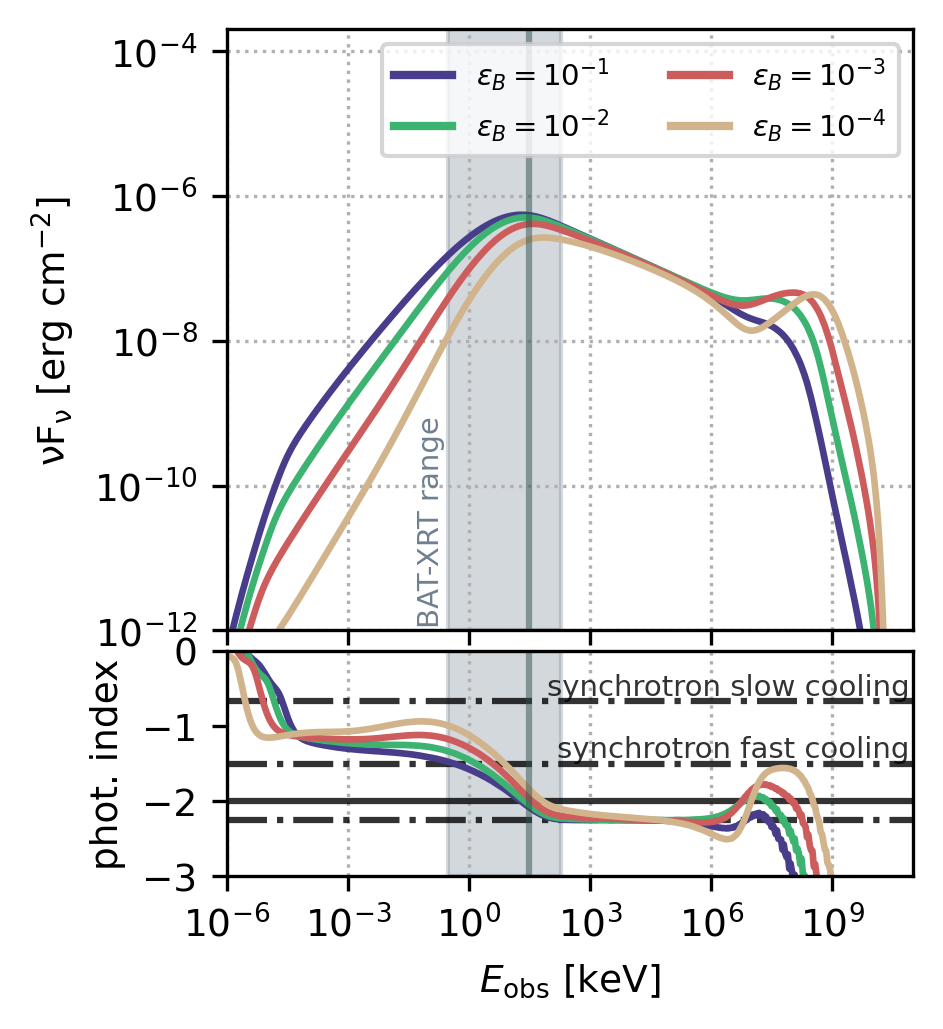}}
\subfloat[hl-GRB]{\includegraphics[width=.33 \textwidth]{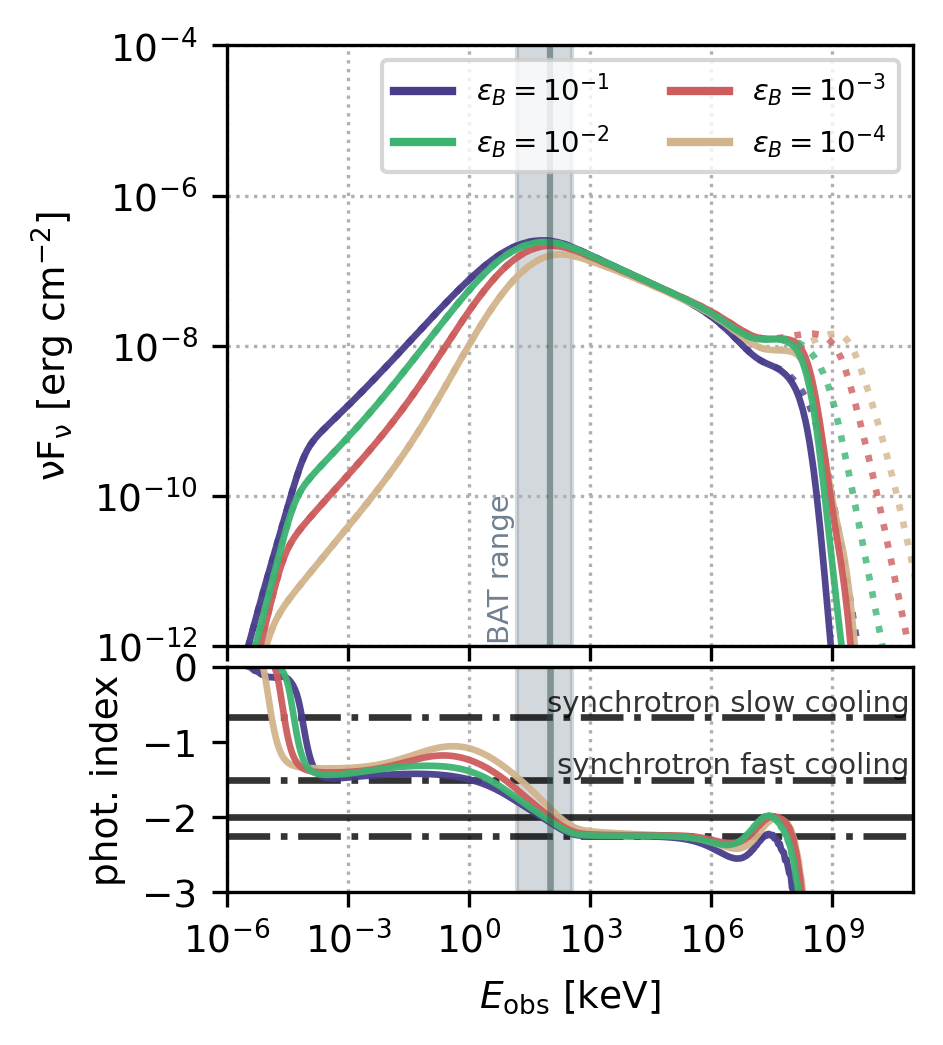}}}
\caption{Time integrated spectra $\nu F_\nu \propto E^2 dN/dE$ for sp-GRB, ul-GRB and hl-GRB for $\epsilon_\mathrm{B} \in \{ 10^{-4}, 10^{-3}, 10^{-2}, 10^{-1}\}$. 
The grey bands mark the energy range of the X-ray instruments used to detect the reference events, with the observed peak energy indicated by the vertical line. For hl-GRB we show the spectra without EBL absorption as dotted curves. \newline
 The lower panel shows the spectral index of the photon flux $dN/dE$ (for a power-law shape), where the dashed lines correspond to the synchrotron predictions ($- 2/3$, $ - 1.5$ and $ - 2.25$ ) and the solid line indicates the position of maxima/minima of $\nu F_\nu$.}
\label{fig:time_integrated_spectra}
\end{figure*}

\subsection{Time integrated spectra}

In \figu{time_integrated_spectra} we show the time integrated spectra $\nu F_\nu$ for the benchmark scenarios. The grey bands indicate the energy range of the X-ray instruments used for the initial detection of the reference events. For all GRBs we explore realisations for $\epsilon_\mathrm{B} \in \{ 10^{-4}, 10^{-3}, 10^{-2}, 10^{-1}\}$.
\noindent The observed sub-MeV peak is generated by synchrotron radiation and fluxes in the regime of the observing instrument are (by construction) comparable. The exception are the simulations obtained for $\epsilon_\mathrm{B}= 10^{-4}$, where the fluxes around the synchrotron peak are much lower than for the other choices of $\epsilon_\mathrm{B}$ and may be in disagreement with the observed fluxes of the reference events. 
In principle the fluxes in this case could be enhanced by imposing a higher jet luminosity, however at the cost of a reduced comparability to the other results due to changed dynamical properties of the outflow. We show in Appendix~\ref{appendix:analytical_estimates} that for $\epsilon_\mathrm{B}= 10^{-4}$ for large parts of the dynamical evolution of the jet, the physical conditions in the shocked plasma correspond to the slow-cooling regime. 
As this further aggravates the efficiency problem of the internal shock model, we suggest that $\epsilon_\mathrm{B}= 10^{-4}$ is not a realistic parameter assumption,
and will show the corresponding results in the following only for completeness. 

The shape of the spectrum is dominated by three features: At the lowest energies, synchrotron self-absorption results in a depletion of the flux and a spectral break. In the intermediate keV-regime, the observed peak energy is produced by synchrotron radiation.
At the highest energies above $\sim~10^8$~keV, inverse Compton radiation results in a second spectral peak. The change of magnetic field, determined by $\epsilon_\mathrm{B}$, shows up mainly in two points: low values of $\epsilon_\mathrm{B}$ lead to higher fluxes in the regime above $\sim~10^8$~keV and lower fluxes below the synchrotron peak, especially in the optical regime. 

The systematic dependence of the high-energy component on $\epsilon_\mathrm{B}$ matches expectations, as the Compton-Y parameter is inversely proportional to $B^\prime$ (and thus $\epsilon_B$) in both Klein-Nishina and Thomson regime (see the approximate formulas \equ{Y_KN} and \equ{Y_TH}). 
We quantify this behaviour by calculating the ratio of the luminosity in inverse Compton to synchrotron component for sp-GRB: numerical values range between L$_{\mathrm{IC}}$/L$_{\mathrm{syn}}$ = 0.02 for $\epsilon_\mathrm{B}$=10$^{-1}$ to L$_{\mathrm{IC}}$/L$_{\mathrm{syn}}$ = 0.08 for $\epsilon_\mathrm{B}$=10$^{-4}$. 

The time integration was performed on the duration of energy dissipation process. Thus, the depletion of the flux below the  spectral peak in the observed spectrum is attributed to two effects: one is the contribution of the low peak energy spectra  generated in the late shocks, and the other one is the effect of IC scatterings occuring in Klein-Nishina regime as shown in  \figu{microphysics_parameters} (see  also \citealt{Daigne:2010fb}). 
As VHE emission might be not be observed due to EBL absorption, the systematic dependence of the optical flux on $\epsilon_\mathrm{B}$ could play a significant role in constraining the magnetic field and can potentially help with the rejection of models and parameter sets (as recently shown in \cite{Samuelsson:2020upt, Oganesyan:2019fpa}). This will however require ll-GRBs to be within the sensitivity range of optical instruments. 
If we compare the optical flux of ul-GRB to the most stringent UVOT $u$-band limits of $1.9 \cdot 10^{-13}$~erg/$\mathrm{cm^2}$/s for a duration of 1300~s (see Section~\ref{sec:reference_grbs}), we find tension with this limit for all results except $\epsilon_B = 10^{-4}$. For the least stringent UVOT limit both $\epsilon_B = 10^{-3}$ and $\epsilon_B = 10^{-4}$ are not ruled out.  We point out that the UVOT limits were obtained for exposure times much shorter than the burst duration and should be taken cautiously for the scenario of a temporally variable source (as ul-GRB). Additionally, these upper limits  depend on the level of absorption in the host galaxy, which is often not well determined. Nevertheless, the results illustrate how optical measurements could potentially constrain parameters of the model. 

sp-GRB and ul-GRB are not significantly affected by EBL absorption, due to their low redshifts. This is different for hl-GRB, where we additionally show the un-absorbed spectra as dotted lines (right plot of \figu{time_integrated_spectra}). In this case, emission above ${\sim0.1~\mathrm{TeV}}$ is strongly suppressed. We conclude that events at these redshifts are likely not to be observed in the HE regime.

The lower panels of \figu{time_integrated_spectra} show the spectral index of the photon flux. 
The dashed lines show the synchrotron predictions for the fast-cooling ($ - 2/3 $) and slow-cooling low-energy slope ($ - 3/2$) below the spectral peak, in addition to the high-energy spectral index ($ - 2.25 $) corresponding to $p = 2.5$.
We find that inverse Compton scatterings in the Klein-Nishina regime affect the low-energy slope $\alpha$: in that case values of $\alpha$ up to --1 can be achieved (\citealt{Daigne:2010fb}); for a detailed discussion of the theoretical predictions of the photon spectral shape see Appendix~\ref{sec:shape_theo_predic}.
This systematic effect on $\alpha$ is common to all benchmark scenarios.
As a consequence, $\alpha$ may 
be used to draw (more robust) conclusions on the magnetic field strength/ the equipartition parameter $\epsilon_B$ in this framework. 
As the spectral slope changes as a function of energy, the fit energy range will have a large impact on the fit result -- an effect which should be taken into account when comparing these predictions to observed data.

\subsection{Time-dependent observational signatures}

Multiwavelength observations of GRBs are critical, both for their
detection and their subsequent interpretation. In particular,
observation of temporal correlations  of the emission in different bands
mitigates the challenge
of detecting the short prompt stage of these events.
In order to illustrate the potential for discovery, we present the predictions for the fluxes  and fluences in different energy  regimes -corresponding to existing and upcoming  instruments-  as a function  of  observation time. 

\begin{figure*}
\centering
\makebox[\textwidth][c]{
\subfloat[(a) Energy ranges]{\includegraphics[width=.33 \textwidth]{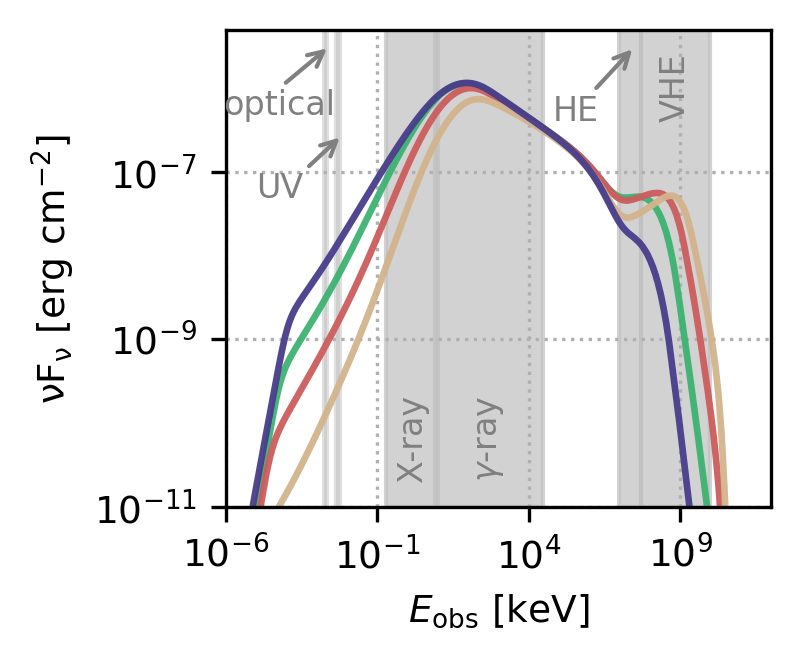}}
\subfloat[(b) Flux ]{\includegraphics[width=.33 \textwidth]{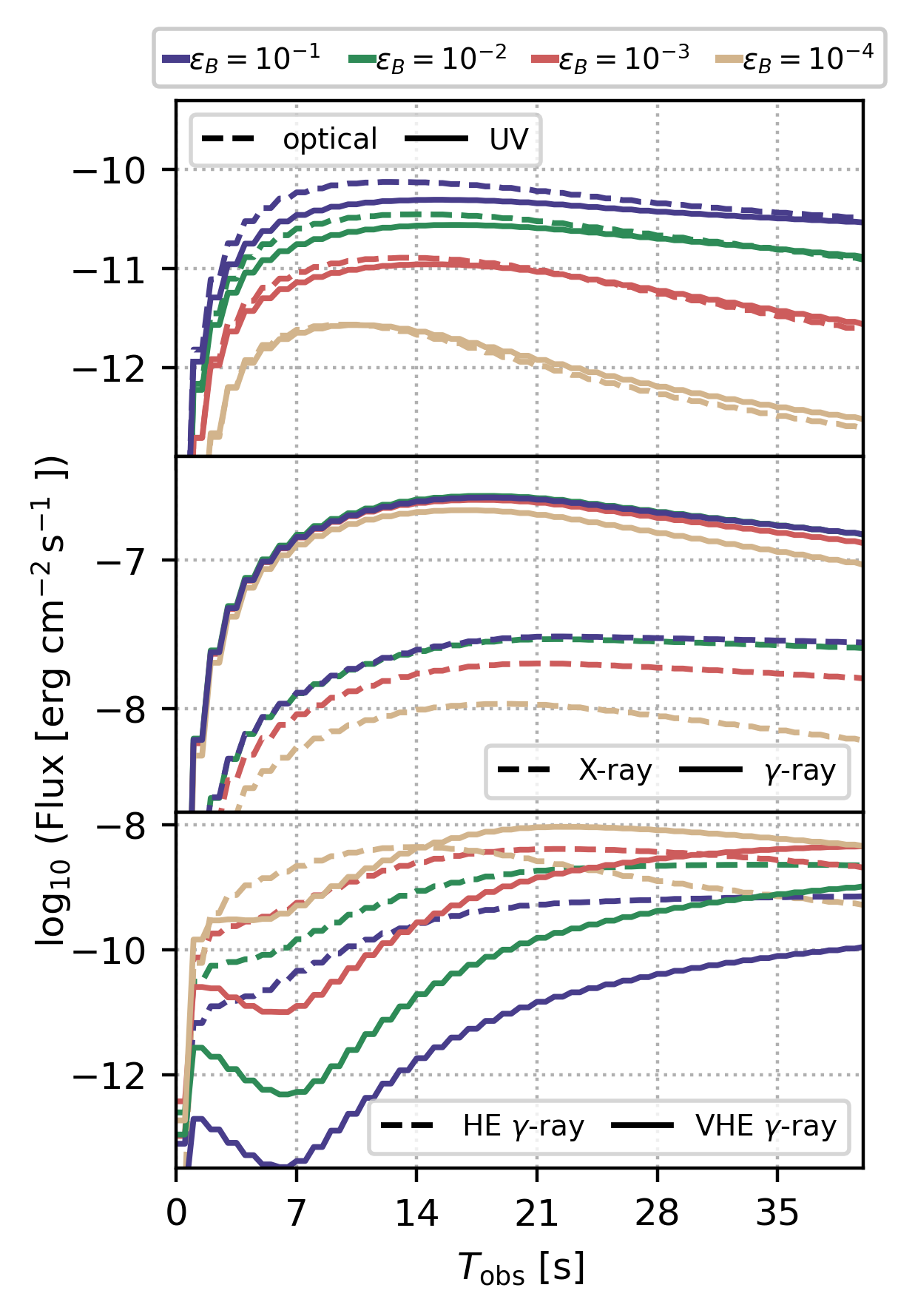}}
\subfloat[(c) Fluence ]{\includegraphics[width=.33 \textwidth]{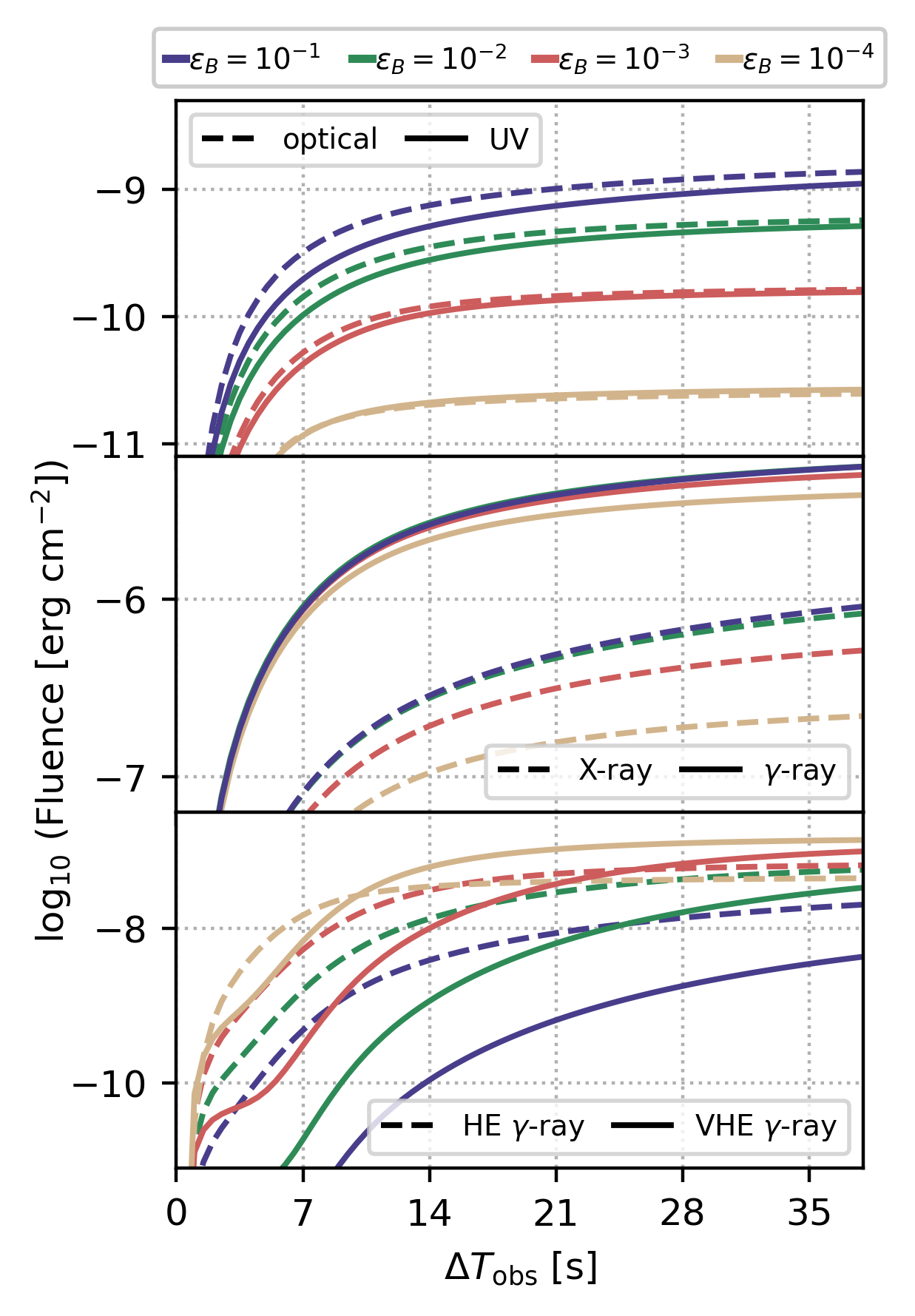}}}
\caption{(a) Observed spectrum of sp-GRB, showing different energy regimes; 
(b) flux and (c) fluence as a function of observation duration, for different choices of $\epsilon_\mathrm{B}$ (see \figu{time_integrated_spectra}). In (c) $\Delta T_{\rm obs}$ denotes the time which has passed since the start of the observation $T_0$ and the fluence is obtained by integrating the flux between the $T_0$ and $\Delta T_{\rm obs}$. The energy ranges/ wavelength bands are: Optical (560 - 730 nm, corresponding to the ZTF $r$-band), UV (220 - 280 nm, corresponding to \textsc{ULTRASAT}), X-ray (0.1 - 10~keV, corresponding to \textit{Swift XRT}), $\gamma$-ray (8~keV - 30~MeV, corresponding to \textit{Fermi GBM}), HE $\gamma$-ray (50 - 100~GeV), VHE $\gamma$-ray (100~GeV - 10~TeV).}
\label{fig:fluence_energybands}
\end{figure*}

As the observed fluences are roughly of the same order of magnitude for the three prototypes (see Section \ref{sec:reference_grbs}), we investigate the discovery potential in different energy bands by calculating the corresponding predicted signals for sp-GRB (\figu{fluence_energybands}).  

The top panels indicate
optical and UV emission, corresponding to the energy bands observable by survey telescopes, such as ZTF (\cite{Bellm_2018}) and LSST (\cite{Ivezic:2008fe})  (energy band \eg \, of the ZTF $r$-band 560 - 730 nm) ,
or a satellite such as the planned ULTRASAT mission (220 - 280 nm) (\cite{Sagiv:2013rma}).
The middle panels correspond to energy bands of X-ray and gamma-ray instruments, such as \textit{Swift XRT}  (0.1 - 10~keV)  (\cite{Burrows:2005gfa}), and \textit{Fermi GBM}(8~keV - 30~MeV)  (\cite{Meegan:2009qu}) .
The ranges presented correspond also roughly to the energy ranges of the upcoming SVOM-mission, which consists of the $\gamma$-ray monitor GRM (30~keV to 5000~keV, \cite{Dong2010GRM}), the X-ray to $\gamma$-ray telescope ECLAIR (4~keV to 250~keV, \cite{Godet:2014ava}), the soft X-ray telescope MXT (0.2~keV to 10 keV, \cite{Perinati2012MXT}) and the optical telescope VT (\cite{Wu2012VT}). 
The bottom panels represent the sensitivity range of ground-based Cherenkov telescopes.

We show predictions for the observable
flux and fluence. 
The realistic sensitivity of the different experiments is not trivial
to estimate, as it depends on multiple factors,
such as the period of day over which the event occurs; 
the pointing of a particular telescope
upon receiving an alert, etc. However, it is possible to relate our
predictions to the nominal capabilities of relevant instruments.
While we account for EBL absorption in the HE regime, the effect of extinction in the host galaxy and our own Galaxy (which is relevant for the optical and UV regime) is not taken into account.

Without extinction we predict for sp-GRB a range of
flux in the optical band of ${\sim10^{-9}\text{--}10^{-11}~\mathrm{erg/cm^2/s}}$ within 30~s , which roughly 
corresponds to an AB magnitude 
within~${\sim16\text{--}21}$~mag for the $i$-band of ZTF. 
These prompt optical flashes are
therefore within the detection sensitivity of the telescope, which
is characterized by a corresponding 30~s limiting magnitude of ${\sim20.5}$. 
Similarly, our predicted UV fluence without extinction corresponds 
to AB magnitudes,~${\sim11\text{--}17}$. Such a signal
would be detectable with the upcoming ULTRASAT,
which has an expected limiting magnitude of ${\sim22}$
(over~5~minute exposures) in the corresponding band.
We also point out that the ratio of fluences in the optical and UV band ($ F_\mathrm{optical}/F_\mathrm{UV}$) changes for the different magnetic fields and could be used to discriminate between scenarios: While for $\epsilon_B = 10^{-1}$ we find $F_\mathrm{optical}/F_\mathrm{UV} = 1.24 $ it drops to $F_\mathrm{optical}/F_\mathrm{UV}= 1.04$  for $\epsilon_B = 10^{-3}$ and $F_\mathrm{optical}/F_\mathrm{UV}= 0.93$  for $\epsilon_B = 10^{-4}$. 

We predict a fluence of ${\sim10^{-7}\text{--}10^{-6}~\mathrm{erg/cm^2}}$ over 40~s in the X-ray regime (\textit{Swift XRT}; 0.1--10~keV) for the different choices of $\epsilon_\mathrm{B}$. For context, \cite{Burrows:2005gfa} derive an XRT sensitivity limit of ${\sim3 \cdot 10^{-11}~\mathrm{erg/cm^2/s}}$ given
an exposure of 10~s, well below the required 
detection threshold of an event such as sp-GRB.

Considering the HE and VHE emission, 
the upcoming CTA has an
expected sensitivity of $\sim10^{-8}\text{--}10^{-10}~\mathrm{erg/cm^2/s}$
over $\sim10$~s intervals (\cite{2019ICRC...36..673F}) in the
energy ranges we consider. Our predicted fluence
for most choices of $\epsilon_\mathrm{B}$ will therefore be within the
detection capabilities of this observatory. For this particular example, resolving the light curve with a ground-based
IACT, while challenging, might be possible. However, the nature of the emission
at these energies depends strongly on the redshift of the source, 
given the potential high impact of EBL absorption.

\begin{figure}
\centering
    \includegraphics[width=.43 \textwidth]
    {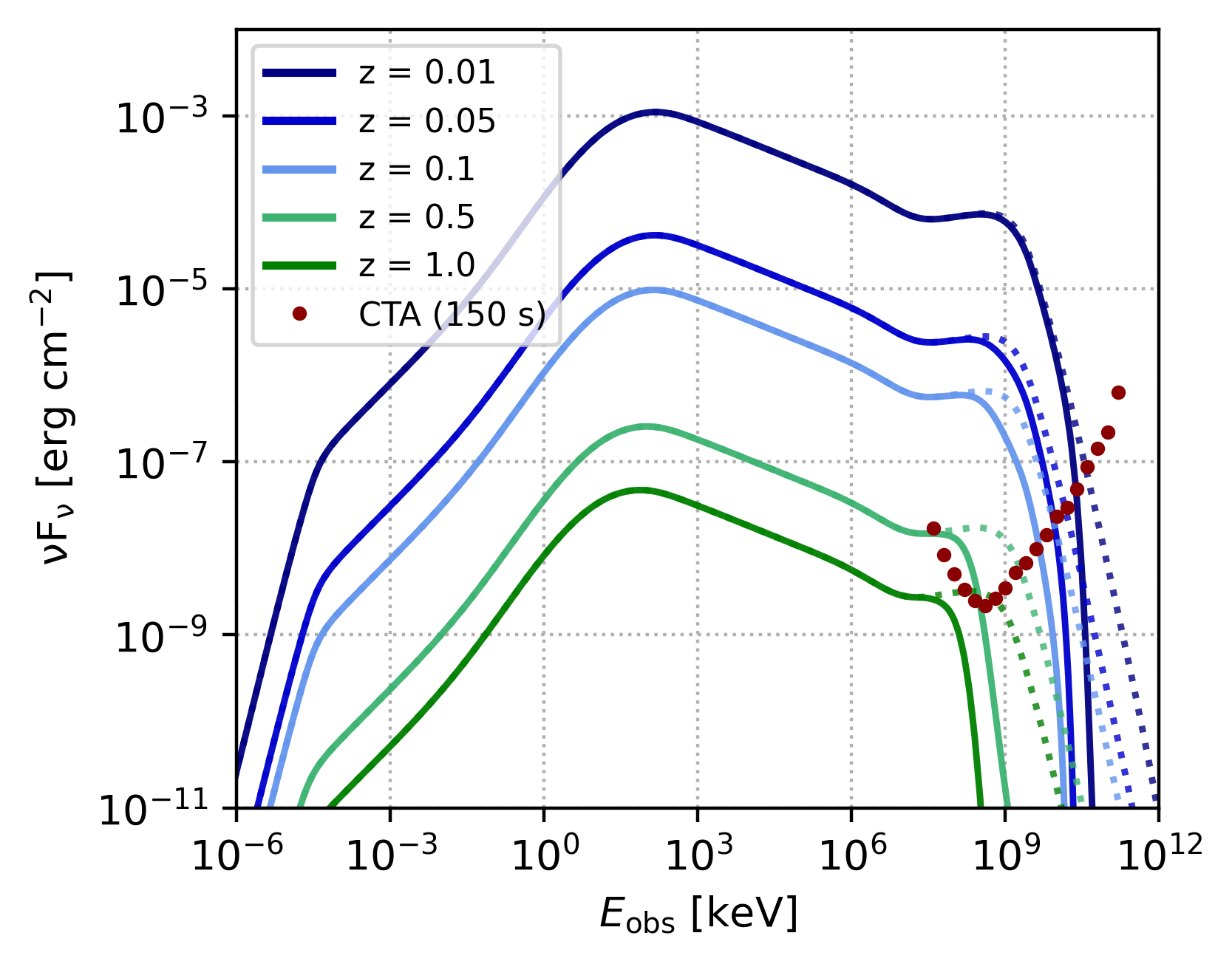}
    \caption{ Predicted observed spectra for the same source prototype (hl-GRB with $\epsilon_\mathrm{B} = 10^{-3}$) placed at different redshifts. Dotted (solid) lines reperesent the spectra without (with) EBL absorption. The red markers represent the minimal fluence nominally detectable by CTA for an observation duration of 150s. }
    \label{fig:impact_of_ebl}
\end{figure}

To illustrate this point, we show the predicted spectra as a function of energy for hl-GRB
in \figu{impact_of_ebl}, assuming the same ll-GRB source to be placed at different redshifts.
The Figure shows the observed spectra for each
redshift, with and without accounting for EBL absorption. The observed duration scales with $(1+z)$ (see \equ{obs_time}), where for $z=0.01$ we calculate a duration of $\sim $~150~s.
The figure also presents the corresponding differential sensitivity of CTA for 150s intervals. It is defined here as the minimal fluence of a source, in order for it to be detectable with at least $5\sigma$ significance within a given energy range. The sensitivity is derived for the Northern site of the observatory using
the \textit{ctools} simulation package (\cite{2016A&A...593A...1K}).
We use instrument response functions optimized for short (30 minutes) observations at zenith angles of $20^{\circ}$. The position of the putative source is displaced by $0.5^{\circ}$ from the centre of the field of view of the instrument.

As one may infer, for redshifts of $z = 0.5$ and $z=1.0$, the
observable emission above 1~TeV is strongly attenuated.
We conclude that for a HE component to be detected,
these events should be within redshift, $z< 0.5$.
Fortunately, the expected rate of occurrence of ll-GRBs is relatively higher in the local Universe (\citealt{Liang:2006ci}). 

\begin{figure*}
\centering
\makebox[\textwidth][c]{
\subfloat[sp-GRB]{\includegraphics[width=.33 \textwidth]{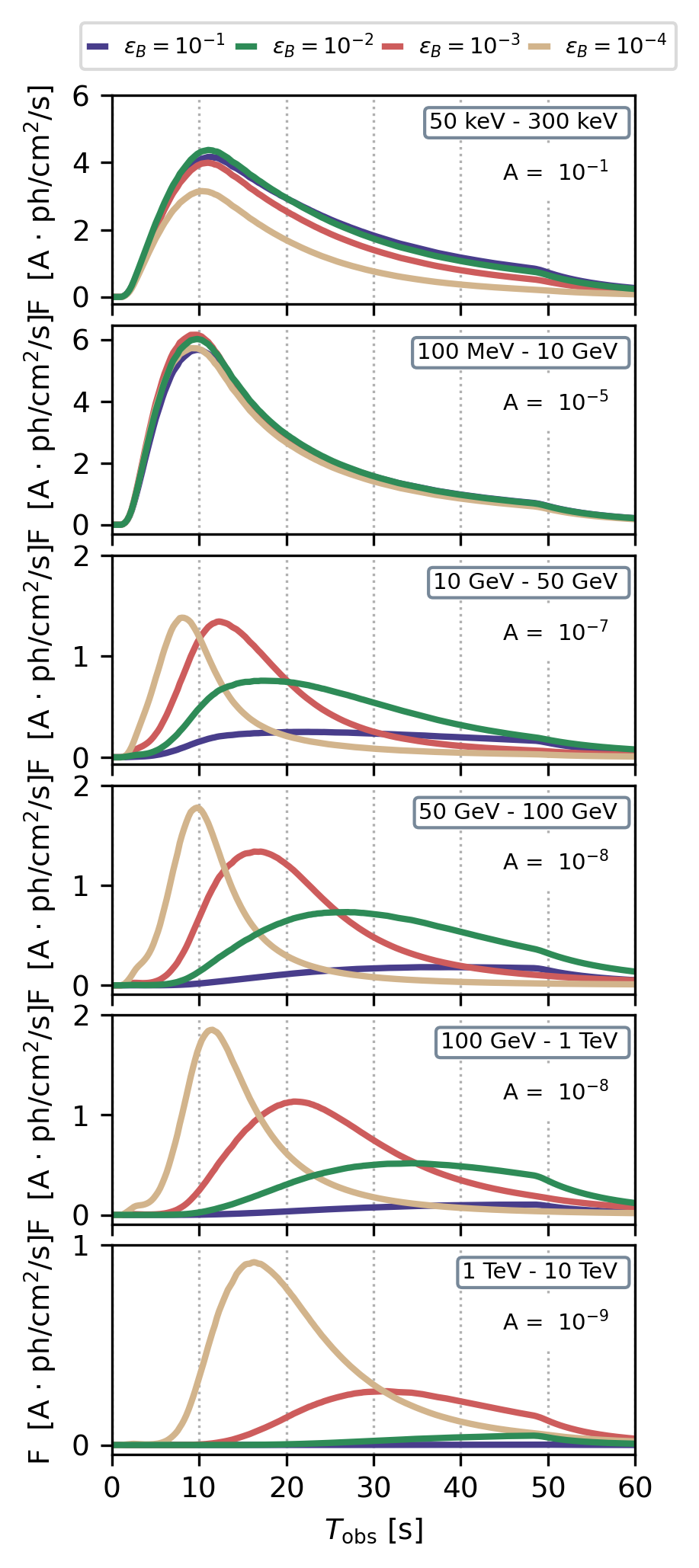}}
\subfloat[ul-GRB]{\includegraphics[width=.33 \textwidth]{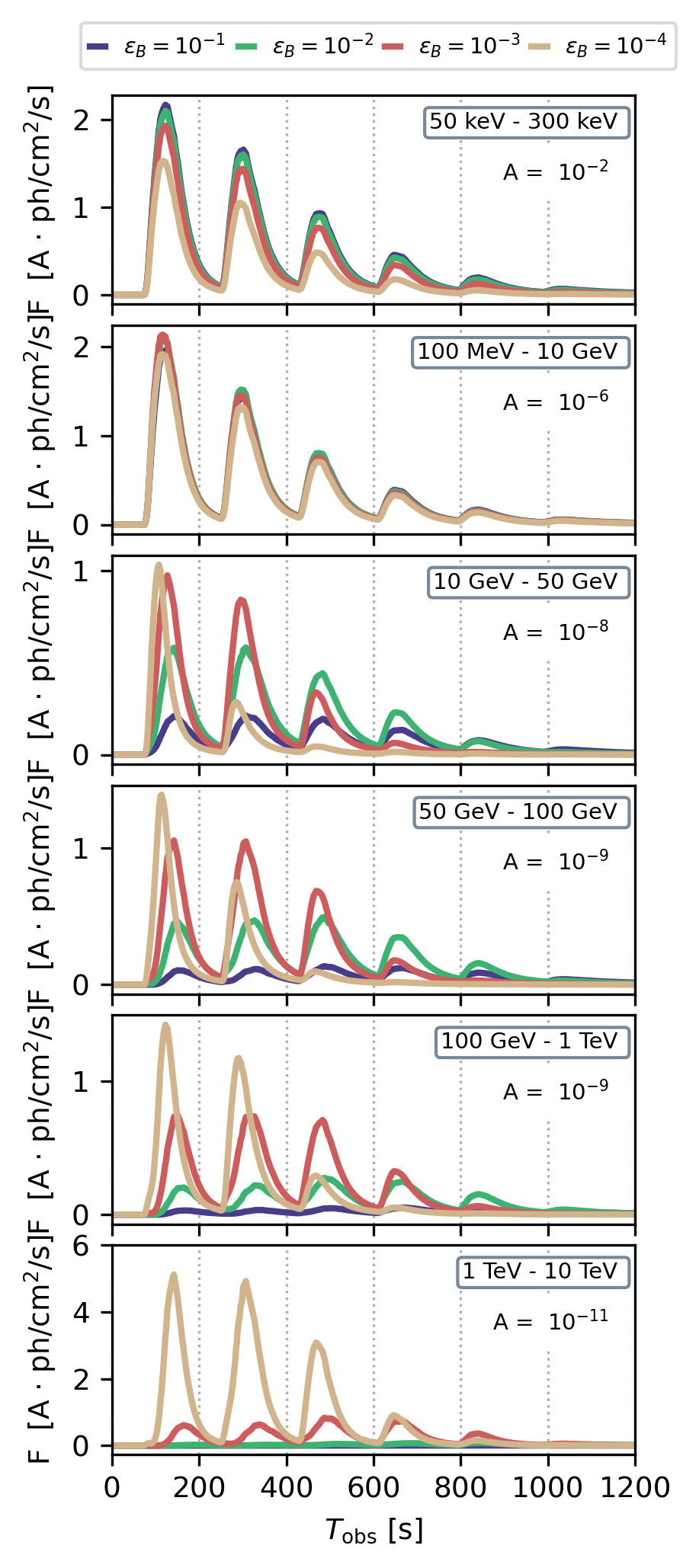}}
\subfloat[hl-GRB]{\includegraphics[width=.33 \textwidth]{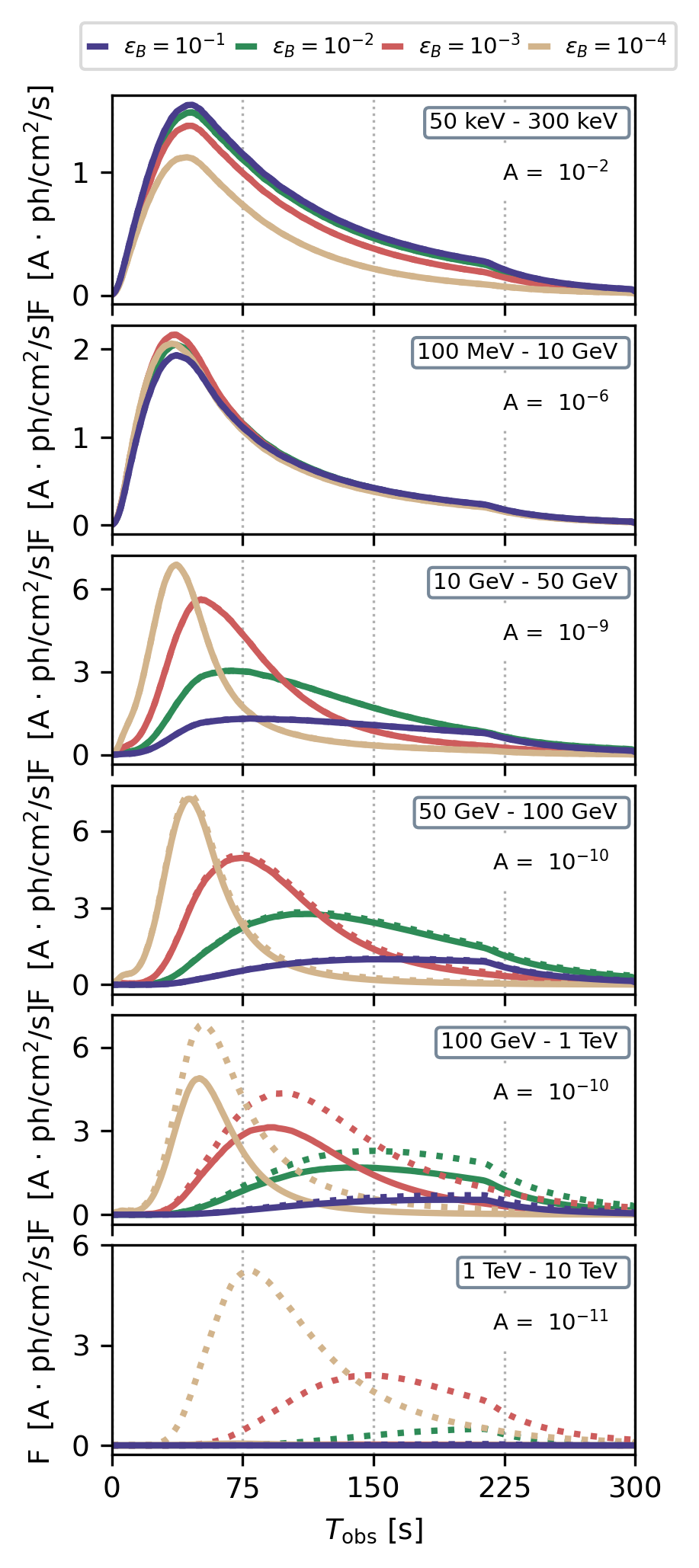}}
}

\caption{Light curves for the $\gamma$-ray and different HE/VHE $\gamma$-ray regimes for sp-GRB, ul-GRB and hl-GRB. We show the results for  different choices of $\epsilon_\mathrm{B}$ (see top plots for the different color labels). We shift the origin of the x-axis such that the observation starts at $T_\mathrm \simeq 0 \, \mathrm{s}$ and for better comparison re-normalize the fluxes by multiplying with the factors A indicated on each plot. As fluctuations on small time-scales are caused by the finite number of shells in our simulations, we smooth the light curves by applying a moving average filter. For hl-GRB the results without EBL absorption are shown as dotted curves.}
\label{fig:light_curves}
\end{figure*}

Even more intriguing is the possibility of exploring the time dependence 
of the photon rates in the HE and VHE bands. We therefore
compare the simulated light curves for 
sp-GRB, ul-GRB and hl-GRB in \figu{light_curves} for different  HE and VHE $\gamma$-ray bands. 

For all models, the general features
of the temporal structure of the reference 
GRB sub-MeV light curve is very well reproduced. A smooth 
single-peaked temporal profile is predicted 
for sp-GRB (GRB~980425) and hl-GRB (GRB~120714B), and a 
multi-peaked light curve with decreasing pulse maximum for ul-GRB (GRB~100316D). While EBL absorption plays no significant role for sp-GRB and ul-GRB, we again see that it suppresses the photon fluxes above 1~TeV by at least two orders of magnitude in the case of hl-GRB.
In accordance with \figu{time_integrated_spectra} we observe a strong dependence of the HE flux has on the 
magnetic field strength, where choices of low  $\epsilon_\mathrm{B}$ 
lead to higher fluxes for all models. This difference 
becomes especially noticeable above 100~GeV. 
For all models we notice an early, weak peak in the HE/VHE regime (see also the flux below 7~s in \figu{fluence_energybands}). We attribute this to the reverse shock (in contrast to the forward shock, which produces the main emission in single peaks) passing through the ejecta, but point out that due to its relatively low flux this early peak might not be observed.

It is noteworthy that the HE emission 
shows a delayed onset with increasing $\epsilon_\mathrm{B}$ in all scenarios.
This is an example of how the different observed light curves may be used
to constrain the physical processes at play.
The early signal in a single-peaked light curve is related to collisions close to the source. These are subject to strong $\gamma \gamma$ - 
absorption, which potentially suppress the HE component (\cite{Hascoet:2011gp,Bustamante:2016wpu}).
This suppression could potentially be slowed by continuous up-scatterings of photons which contribute to a  high energy component. 
For this, however, relativistic electrons need to be present in the region. 
As we don't consider a steady injection term but instead follow a cooling electron distribution, this may be realized if electron cooling time-scales are large. This is the case for low-$\epsilon_\mathrm{B}$, where the synchrotron 
cooling time-scale is long; it is in fact the dominating cooling
time-scale for high-energy electrons for low-$\epsilon_\mathrm{B}$  (see Appendix~\ref{app:ele_cooling}.)
Another way of preventing an early suppression of the HE flux due to $\gamma \gamma$ absorption may be a continuous injection of accelerated particles (ensuring the continuous presence of relativistic electrons in emission region). The latter could be fueled by either a slow enough acceleration process or by the injection of relativistic electrons from (neighbouring) collisions and plasma layers.
While thus for low magnetic fields an early and strong HE-peak ($>$ 10 GeV) is predicted, it will become wider and peak later in time with increasing $\epsilon_B$.
The wide peak might be connected to the high(er) efficiency in late collisions further away from the source for high $\epsilon_B$.

Overall, the signals observable in different energy bands can influence the observational strategies for future experiments. 
For instance, while it may be challenging 
to detect these events at ${>\text{TeV}}$ energies, the emission between
50~GeV and 1~TeV is within the sensitivity window of CTA.
Furthermore, one may consider the different predictions, related \eg to different
choices of $\epsilon_\mathrm{B}$ in different
energy bands; these illustrate how observations of the time-structure
of ll-GRBs may be used to constrain their physical modelling.
Considering our three reference models, it is also interesting to note that our models accommodate a rich phenomenology, which may largely be attributed to the properties of the engine (\eg the Lorentz factor distribution, 
engine active time and wind luminosity).
Time-resolved observations may therefore serve as a direct probe for properties of the central engine.

\section{Implications for the connection to UHECRs and neutrinos}
\label{sec:uhecr}
ll-GRBs have been recently studied as the sources of the UHECR nuclei in \cite{Zhang:2017moz}, and may even describe the PeV neutrino flux and UHECRs across the ankle simultaneously~(\cite{Boncioli:2018lrv}). 

We point out that current IceCube neutrino limits suggest that standard high-luminosity GRBs cannot be sources of UHECR and high-energy neutrinos (\cite{Abbasi:2012zw, Aartsen:2014aqy, IceCube:2016ipa, Aartsen:2017wea}). However, note that an actual fit of UHECRs~\cite{Biehl:2017zlw} (in combination with these limits) within a more extensive parameter space study points towards regions indicative for ll-GRBs (at lower luminosities) or magnetic reconnection models (at larger production radii). Furthermore, the limits apply to one zone collision models (where all properties of the emission regions are alike), whereas multi-collision models point towards different production regions for the different messengers~\cite{Bustamante:2014oka}. As a consequence, \cite{Heinze:2020zqb} have demonstrated that a fit to the UHECR spectrum is still viable in multizone internal shock models with neutrino predictions in reach of upcoming instruments such as IceCube-Gen2 (\cite{IceCube:2014gqr}), where typical baryonic loadings ($\xi=\epsilon_\mathrm{CR}/\epsilon_{e}$) between about 50 and 100 have been found. Choosing instead the bightests GRBs~130427A and 160625B, (\cite{Fraija:2017mlx, Gao:2013fra}) have limited the baryonic loading to $\xi \lesssim 0.5$; it is therefore likely that not all GRBs carry the same baryonic loading. Finally note that for exposures substantially longer than ~100sec (i.e., for long GRBs), atmospheric neutrino backgrounds can no longer be neglected (\cite{IceCube:2018omy}). Dedicated analyses for ll-GRBs lasting longer than 100s are therefore required, which are beyond the scope of the current limits.

Further, doubts about the maximal UHECR energy being reachable in  models with the photon peak energy produced from synchrotron emission of accelerated electrons have been raized in \cite{Samuelsson:2018fan, Samuelsson:2020upt}. The model in our study is such a model, however we applied a multizone model which includes the dynamical evolution of the outflow as well 
: In contrast to one-zone models, which allow for one energy dissipation zone representative for the complete burst, we account for different emission regions along the jet that may be production regions for different particle species.
We therefore demonstrate that high UHECR energies can be, in principle, obtained in our approach, and we qualitatively discuss the requirements for the UHECR description in comparison to~\cite{Boncioli:2018lrv}. Note that the approach in this section is not strictly consistent in the sense that we do not include the radiation feedback from the nuclei on the Spectral Energy Distribution (SED); it can be considered a perturbative/test-particle approach, see second subsection for a more detailed discussion on its applicability. 

\subsection{Maximal energies of UHECR nuclei}

\begin{figure*}
\centering
\makebox[\textwidth][c]{
\subfloat{\includegraphics[width=.33 \textwidth] {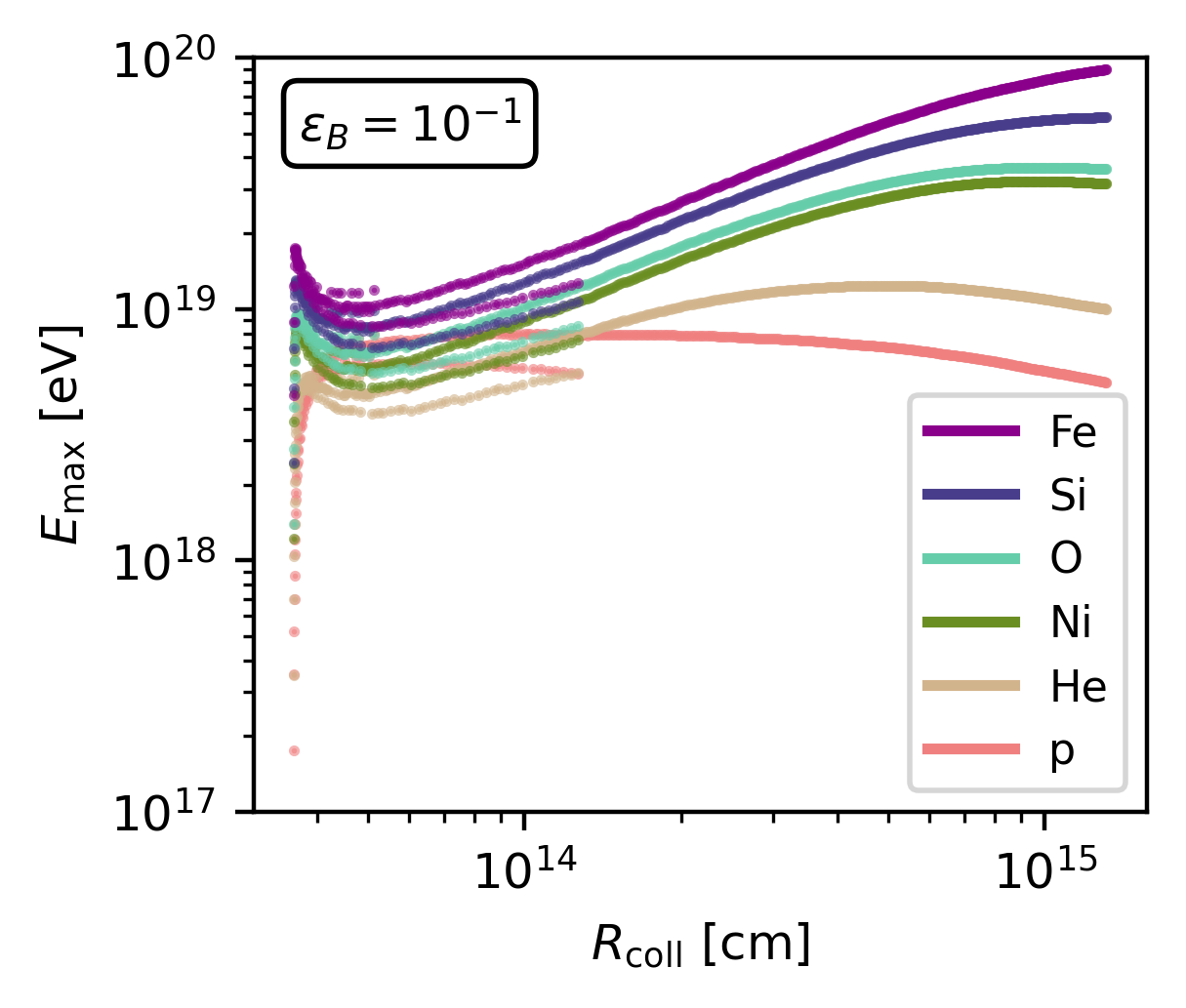}}
\subfloat{\includegraphics[width=.33 \textwidth] {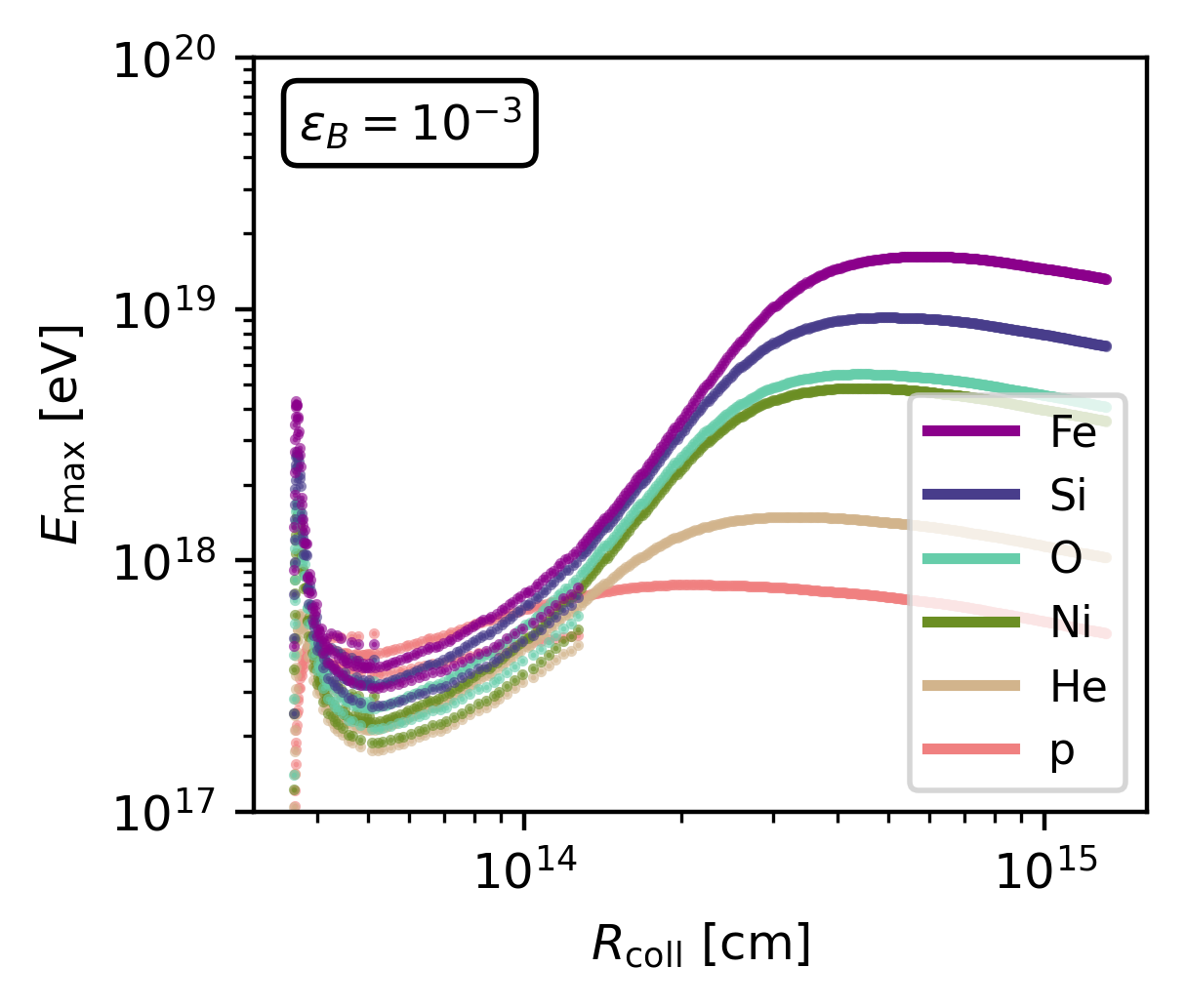}}
\subfloat{\includegraphics[width=.33 \textwidth] {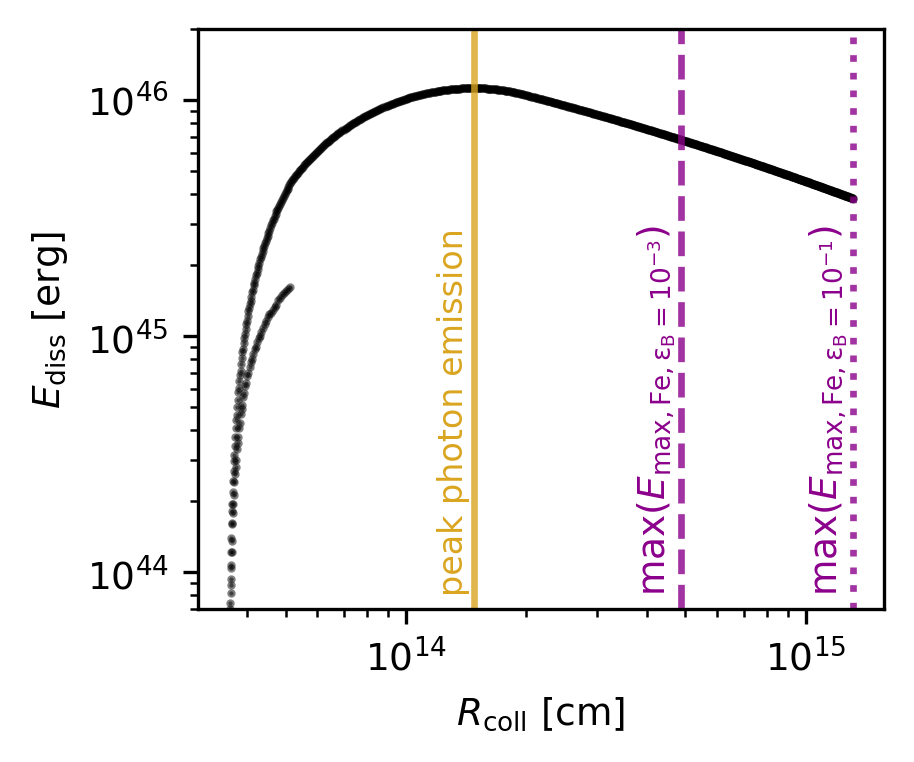}}}
\caption{Left and middle: Maximal UHECR energies in the source frame as a function of collision radius for sp-GRB for $\epsilon_\mathrm{B} = 10^{-1}$ and $\epsilon_\mathrm{B} = 10^{-3}$.  Each dot corresponds to one collision, for which the maximal UHECR energy for the indicated isotope is computed.
Right: Dissipated energy $E_\mathrm{diss}$ (\equ{ediss}) as a function of collision radius.  Vertical lines mark  the radius where the synchrotron peak is predominantly produced and the radii corresponding to the maximum of $E_\mathrm{max, Fe}$ for both choices of $\epsilon_\mathrm{B}$.}
\label{fig:max_energies_980425}
\end{figure*}

\begin{figure*}
\centering
\makebox[\textwidth][c]{
\subfloat{\includegraphics[width=.33 \textwidth] {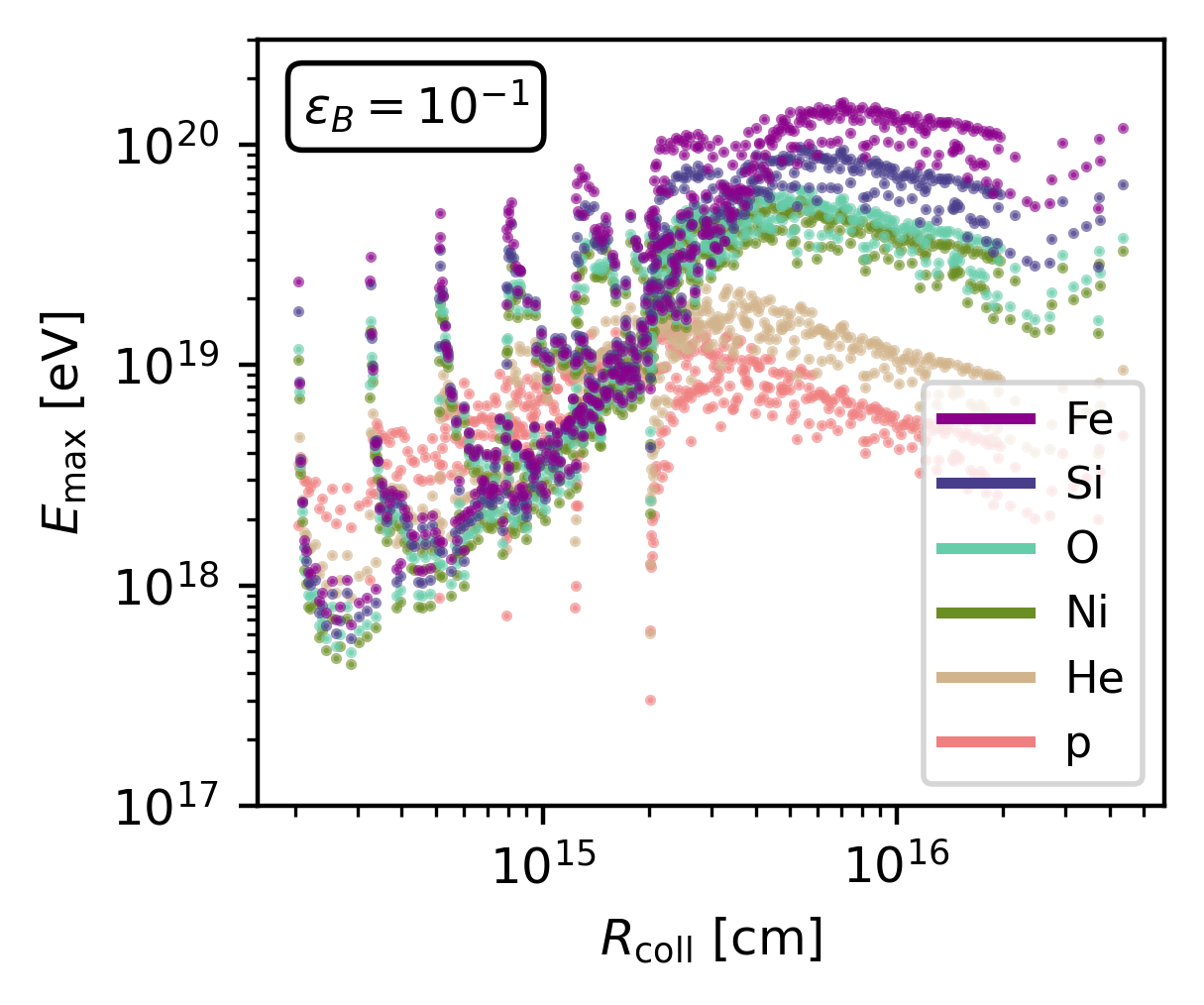}}
\subfloat{\includegraphics[width=.33 \textwidth] {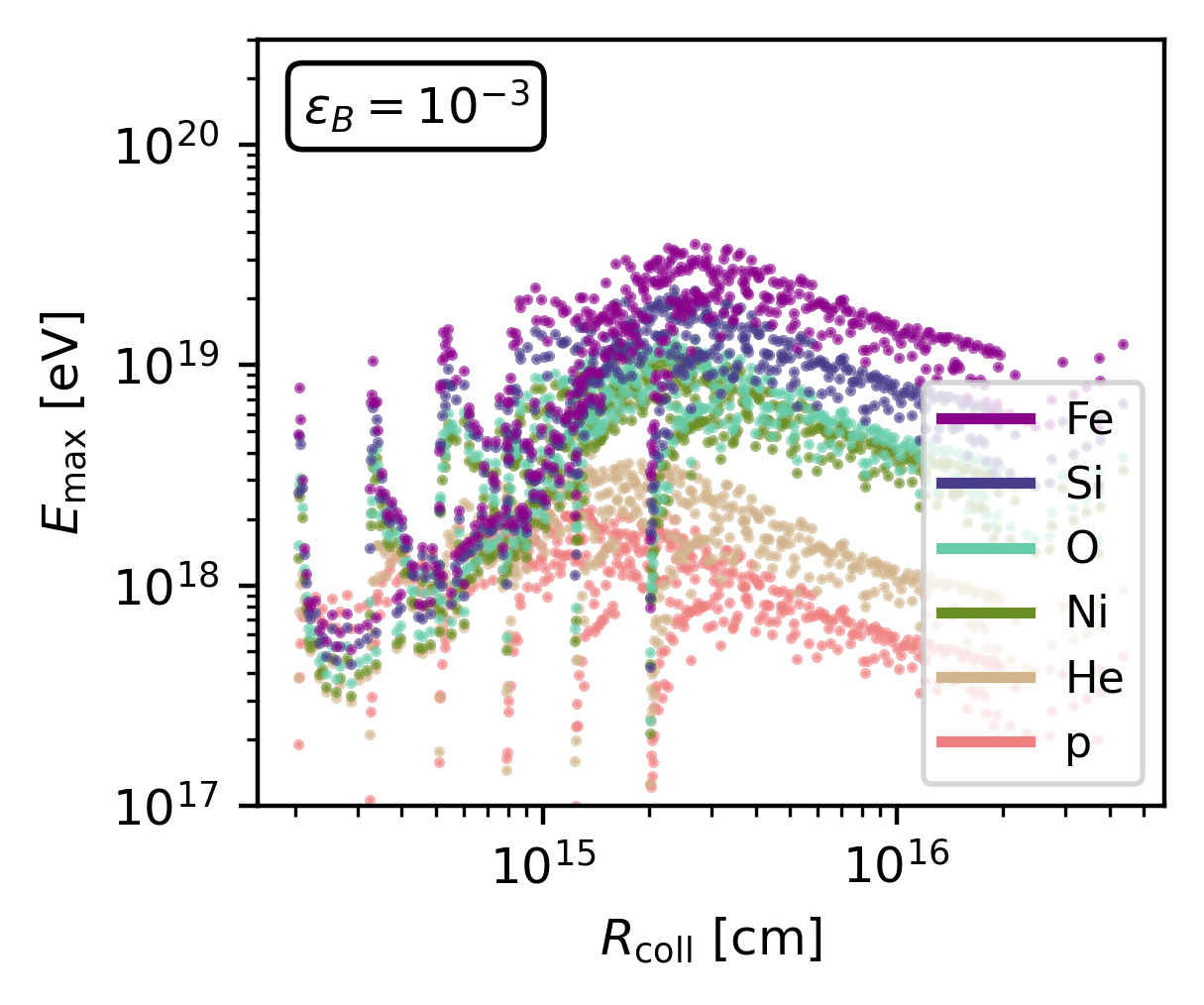}}
\subfloat{\includegraphics[width=.33 \textwidth] {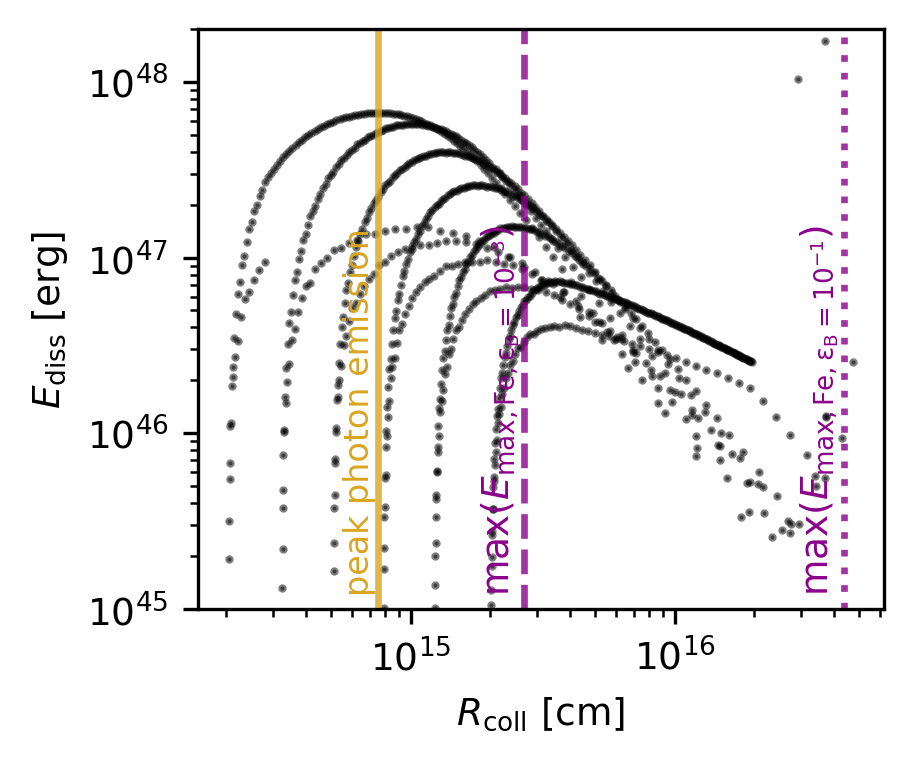}}}
\caption{Left and middle: Maximal UHECR energies in the source frame as a function of collision radius for ul-GRB for $\epsilon_\mathrm{B}= 10^{-1}$ and $\epsilon_\mathrm{B}= 10^{-3}$. Each dot corresponds to one collision, for which the maximal UHECR energy for the indicated isotope is computed.
Right: Dissipated energy $E_\mathrm{diss}$ (\equ{ediss}) as a function of collision radius.  Vertical lines mark  the radius where the synchrotron peak is predominantly produced and the radii corresponding to the maximum of $E_\mathrm{max, Fe}$ for both choices of $\epsilon_\mathrm{B}$.}
\label{fig:max_energies}
\end{figure*}

In order to estimate the maximal energy of UHECRs, we apply the radiative \software{NeuCosmA}-Code (\cite{Biehl:2017zlw}) and follow a procedure similar to \cite{Zhang:2017moz,Samuelsson:2018fan,Boncioli:2018lrv,Samuelsson:2020upt}: We balance the acceleration rate ${t^\prime}_\mathrm{acc}^{-1}=\eta \, c/R^\prime_\mathrm{L}$ (see \equ{acctime}, with the Larmor radius $R^\prime_\mathrm{L}$) with the energy losses (photo-hadronic cooling, photo-disintegration cooling, synchrotron cooling and adiabatic cooling), assuming efficient acceleration ($\eta=1$); see last paragraph of this section for a critical discussion of the acceleration efficiency. 
We point out that, since $pp$ reactions are negligible in this case,
these maximal energies are independent of energy density of accelerated baryons
as long as the photon fields are not perturbed by hadronic contributions. 

As the acceleration efficiency (and accordingly the maximal cosmic-ray energy) depends on the magnetic field, we proceed similarly to Section~\ref{sec:results} and impose different magnetic field strengths (set by $\epsilon_\mathrm{B}$). We choose $\epsilon_\mathrm{B} = 10^{-3}$ and $\epsilon_\mathrm{B} = 10^{-1}$ as feasible examples sufficiently distinct to clearly show the impact of $\epsilon_\mathrm{B}$. For the sake of simplicity, we limit ourselves to sp-GRB (as an example for a single-peaked, regularly long ll-GRB) and ul-GRB (as an example for an ultra-long ll-GRB, recently discussed as UHECR sources). 
Note that compared to earlier studies, we  a) simulate the different collisions explicitly (instead of using an effective one zone model) and b) compute the spectral energy distribution from first principles. As discussed earlier, deviations from the pure synchrotron assumption are expected from additional radiation processes, such as inverse Compton scattering. 

Our results are shown in \figu{max_energies_980425} and \figu{max_energies}, where the maximally achievable UHECR energy is shown for each collision as ``dot''.

We find that iron nuclei (protons) may be accelerated to energies of up to $\simeq$ $ 10^{11}$~GeV ($10^{10}$~GeV), where acceleration for both GRBs is significantly stronger in the case of strong magnetic fields ($\epsilon_\mathrm{B} = 10^{-1}$). All nuclear species except for protons show a strong dependence of the maximal energy on the collision radius. This can be understood by comparing the time-scales for different energy losses which show that nuclei heavier than protons experience strong disintegration close to the source (see Appendix~\ref{appendix:hadronic_rates}). Therefore their maximal energies are significantly lower at small radii -- for protons, which only experience pion-geneneration losses, the radial dependence is not as strong. Further away from the source, small radiation densities make adiabatic cooling the dominating energy loss for all nuclei.

The results are roughly consistent with the mentioned earlier studies, given that slightly different assumptions and parameters are used. 
For example, \cite{Samuelsson:2018fan} conclude that for ll-GRBs the ``highest obtainable energies are $<10^{19}$~eV and $<10^{20}$~eV for protons an iron respectively, regardless of the model'' and find a maximum energy of iron of approximately $10^{21}$~eV for ll-GRBs with comparable parameters (see their Figure~7, lower right plot)-- in consistency with our \figu{max_energies}. 

UHECR data (spectrum and composition measured by the Pierre Auger observatory) from ll-GRBs are explicitly fitted in \cite{Zhang:2017moz,Boncioli:2018lrv}. For instance,
the maximal silicon energy of our $\epsilon_\mathrm{B} = 10^{-1}$ ($\epsilon_\mathrm{B} = 10^{-3}$)  examples is 
(at large enough distances to the source) about a factor of two higher (lower) than the maximal energy obtained at the best-fit point of \cite{Boncioli:2018lrv} (see their Fig. 2,  left panel), and well within the values obtained in the fit region . It is therefore expected that the maximal energies are sufficient to describe UHECRs if the acceleration is efficient ($\eta \rightarrow 1$ in \equ{acctime}). This can  be seen more explicitly in \cite{Heinze:2019jou} who describe spectrum and composition of the UHECRs 
using a model where the maximal energy scales with rigidity. It has been demonstrated that in a 3D fit including the relevant parameters (spectrum, source evolution, rigidity), a rigidity $r = E/Ze \simeq 2-3 \cdot 10^9 \, \mathrm{GV}$ is needed to describe UHECRs -- almost uncorrelated with source evolution and spectral index. This translates into a silicon energy of $\simeq 3-4 \cdot 10^{10} \, \mathrm{GeV}$, in  consistency with \figu{max_energies}.  Note that the maximally measured UHECR energy  only represents the very end of the UHECR spectrum known with extremely limited statistics, whereas the fit of UHECR data  (shape and composition) is driven by lower energies (around $10^{10} - 10^{11} \, \mathrm{GeV}$) where the statistical uncertainties are small. 

We point out that past studies, such as \cite{Boncioli:2018lrv,  Samuelsson:2018fan, Samuelsson:2020upt} apply simplified one-zone models, where the observed gamma-ray emission and the maximal cosmic-ray energy are computed for the same radius. The right plots of \figu{max_energies_980425}  and \figu{max_energies} show the dissipated energy per collision as a function of collision radius. The observed prompt emission is dominated by the collisions that dissipate large amounts of energy (in the vicinity of the yellow vertical line, the `max photon emission'). While this corresponds (in both cases) to rather low radii of $\approx 10^{14} - 10^{15}$~cm, cosmic-ray nuclei attain the highest energies only at larger distances to the source. In the right plots of \figu{max_energies_980425}  and \figu{max_energies} purple vertical lines indicate the radius of the highest maximal energies for iron nuclei ($\max{(E_\mathrm{Fe, max}})$, for both choices of $\epsilon_\mathrm{B}$). For $\epsilon_\mathrm{B} = 10^{-1}$ (which yields the higher maximal energies), the maximal energy is reached at the largest radii possible, roughly an order of magnitude further out than the production region of the $\gamma$-ray emission. 
Although these outer collisions contribute only little to the overall photon emission and more energy is dissipated at lower radii, this study shows they may be the production regions of UHECR of the highest energies. 
The emission radii especially of sp-GRB are  comparable with past multi-collision studies of UHECR production in GRB-HL of (roughly) similar duration (\cite{Bustamante:2016wpu, Rudolph:2019ccl}). The maximal energies for protons are slightly lower, which might be explained by the lower internal energies and corresponding magnetic fields. 
We conclude that a simple one-zone model, which does not account for the distribution of dissipated energy as a function of radius when computing the maximal cosmic-ray energy, potentially underestimates the achievable maximal cosmic-ray energy in the GRB.

It remains to be discussed if the assumed acceleration time-scale $t_\mathrm{acc} \simeq R_L / c$, necessary to describe the connection with UHECRs, can be supported by theoretical arguments concerning particle acceleration.
It was suggested that cosmic rays with energies $\sim$ 10$^{20}$~eV cannot originate in the (ultra-)relativistic GRB shocks, where due to the slow acceleration rate protons cannot exceed PeV-energies (\cite{Lemoine:2009vr, Plotnikov:2012ew, Reville:2014mta}; for a recent review see e.g. \cite{Marcowith:2020vho}). 
The mildly relativistic shocks operating in GRB jets can instead be the viable candidates, but remain to explored further (e.g. \cite{Crumley:2018kvf, Ligorini:2021lbj, Marcowith:2016vzl}).
In case of magnetized outflows (\cite{Sironi:2009jw, sironispitkovskyarons2013}) find that Fermi acceleration is suppressed. In this case, alternative acceleration mechanisms such as \cite{Giannios:2010cv} may be required. Here we point out that $\epsilon_B = 10^{-1}$ (for which the largest magnetic fields are achieved in our framework) does not necessarily correspond to a high magnetisation of the outflow. 

\subsection{Requirements for the multimessenger description from a GRB population}

\begin{figure*}
\centering
\includegraphics[width=0.45\textwidth]{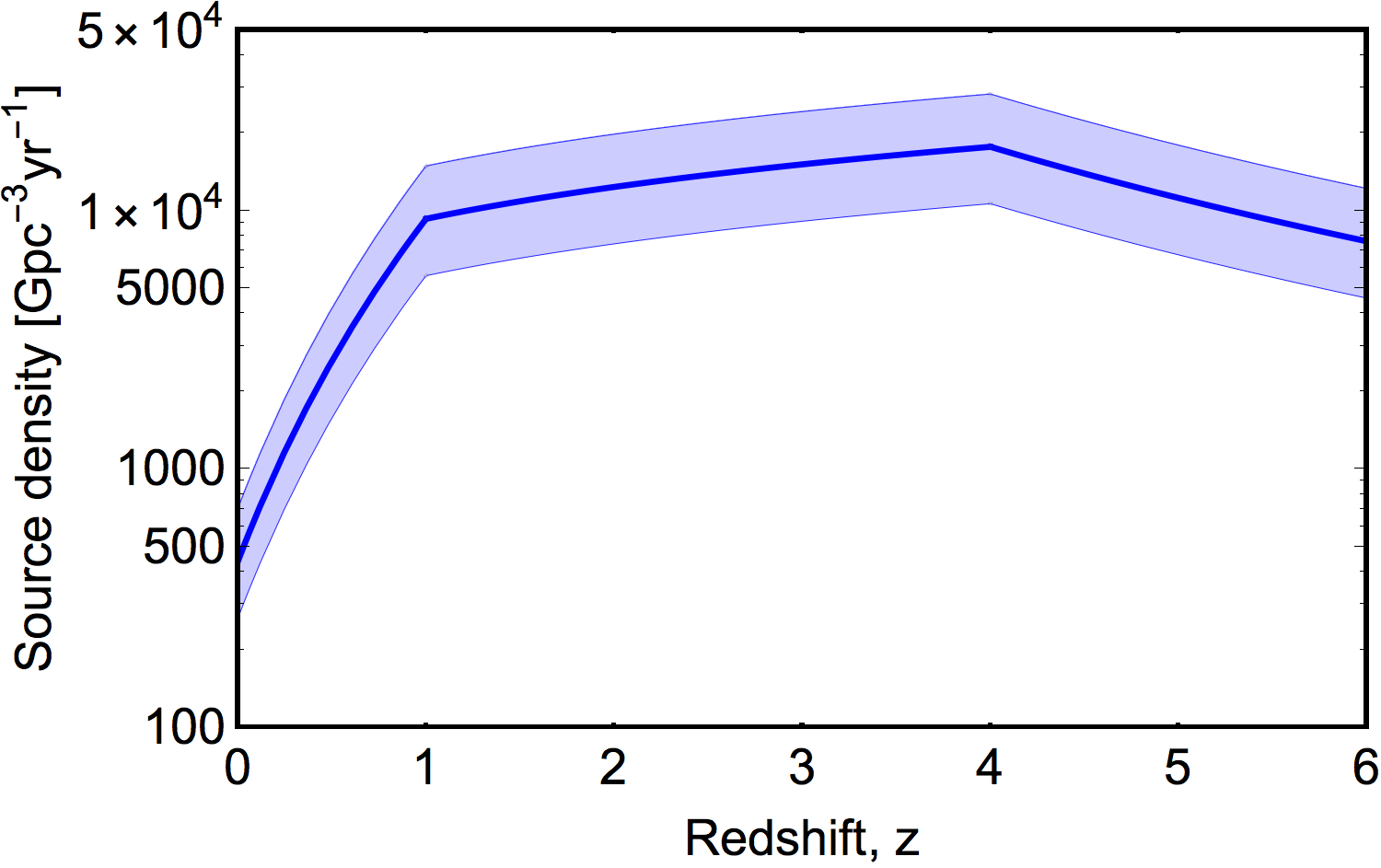}
\includegraphics[width=0.45\textwidth]{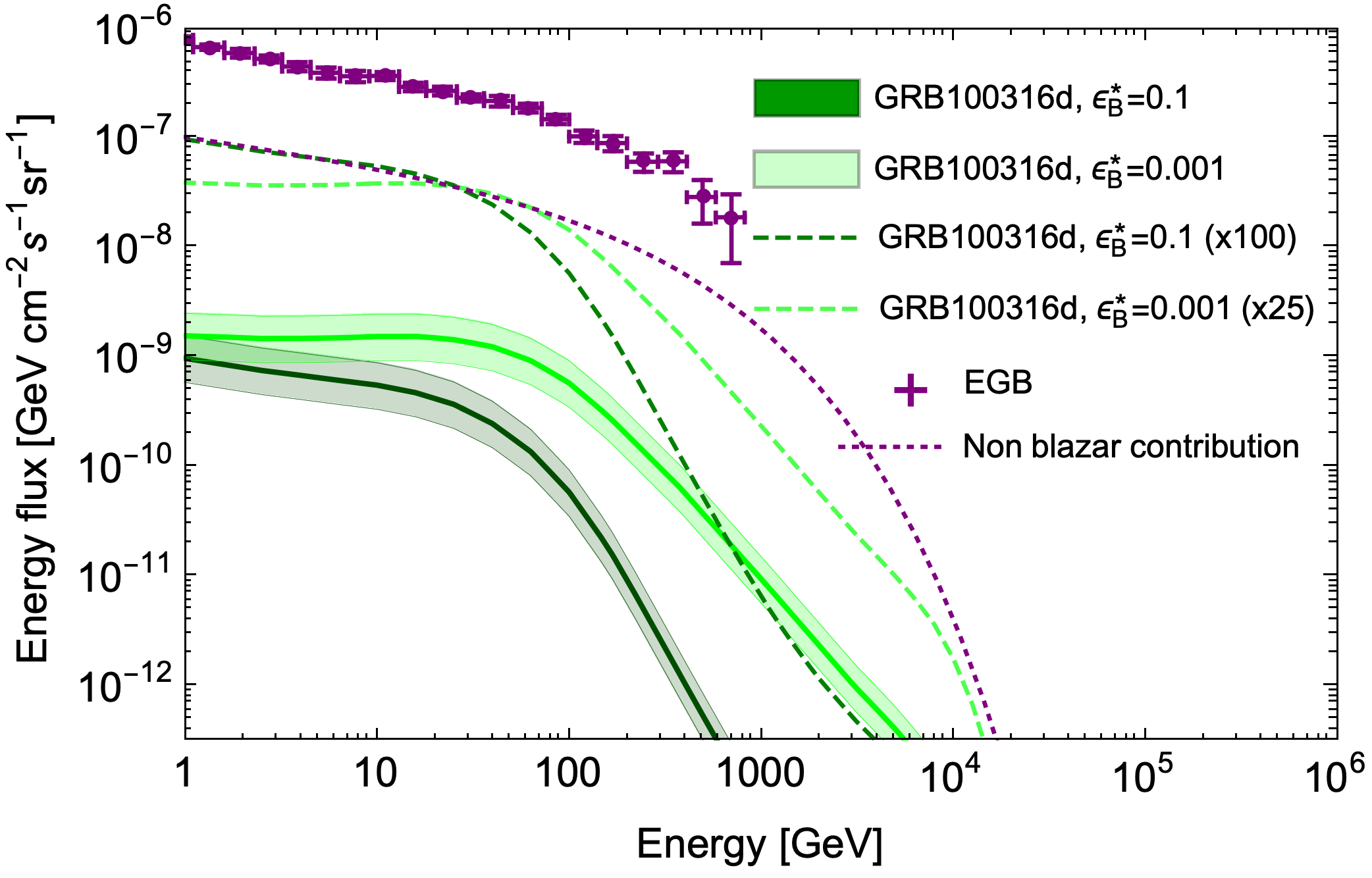}
\caption{Left panel: source evolution of ll-GRBs. The local density is taken from \citep{Sun:2015bda}, while the evolution is equal to the Star Formation Rate (given in \citep{Kistler:2009mv}) multiplied by an extra source evolution factor $(1+z)$, as it is proposed in \citep{Boncioli:2018lrv} for a better description of UHECR data. Right panel: diffuse gamma-ray emission from a population of ll-GRBs assumed to be similar to ul-GRB and comparison with the EGB (as  measured by Fermi). We show two choices of magnetic field ($\epsilon_B^{*}$ is defined such that it yields the same magnetic fields as the $\epsilon_B$ in Section~\ref{sec:results}, independent of the wind luminosity/ baryonic loading). Shaded areas illustrate the uncertainties on the local density of ll-GRBs, different color brightnesses refer to our different assumptions on the magnetic field. The dashed green curves show the corresponding EGB contributions multiplied with an enhancement factor (see legend) to saturate the non-blazar EGB contribution. \label{fig:gamma}}
\end{figure*}

Let us now consider a population of ll-GRBs and assume that they describe UHECR and neutrino data as discussed in \cite{Boncioli:2018lrv}. 
We also compute the contribution to the extragalactic diffuse gamma-ray background following \cite{Berezinsky:2016feh}, and derive constraints  for prompt emission duration and baryonic loading.

We here define the baryonic loading as $\xi = \epsilon_\mathrm{CR}/\epsilon_{e}$, where $\epsilon_\mathrm{CR}$ quantifies the energy going into non-thermal baryons ($E_\mathrm{CR} = \epsilon_\mathrm{CR} E_\mathrm{diss}$). 
Using the results of our (leptonic) radiation modelling  in Section~\ref{sec:results} naturally implies that the energy dissipated in electrons is held constant. As energy conservation additionally dictates $\epsilon_e + \epsilon_\mathrm{CR} + \epsilon_{B} \lesssim 1$, a baryonic loading $\xi \gtrsim 1 $ thus implies a increased wind luminosity $L_\mathrm{wind}^{*}$ (translating into an increased dissipated energy per collision $E_\mathrm{diss}^{*})$; In this case we need to redefine $\epsilon_e^{*} = \epsilon_e E_\mathrm{diss}/E_\mathrm{diss}^{*} $ and $\epsilon_B^{*} = \epsilon_B E_\mathrm{diss}/E_\mathrm{diss}^{*} $ to ensure that the leptonic radiation modelling yields the same results. As can be inferred from \equ{gamma_min}, the fraction of accelerated electrons would have to be adjusted accordingly.
For the sake of simplicity we assume that all ll-GRBs are similar to one of our prototypes, although the typical duration may differ. Note, however, that there are already constraints on minute-scale long transients by IceCube~(\cite{Aartsen:2018fpd}), which means that in this case the typical ll-GRB duration has to be probably longer than a few hundred seconds. The ll-GRBs with the prototype luminosities are assumed to be distributed over the Universe with a local rate  $\rho_0= 440^{+264}_{-175} \ \rm Gpc^{-3} yr^{-1}$~(\cite{Sun:2015bda}), following the star formation rate of \cite{Kistler:2009mv} with an additional evolution factor $(1+z)$ which has been proposed in \cite{Boncioli:2018lrv} to better fit UHECR data, see left panel of \figu{gamma}. 

In order to describe UHECR and neutrino data, we can rescale\footnote{The result in \cite{Boncioli:2018lrv} is degenerate in the product of these factors in \equ{uhecr}; in that paper (e. g. Fig. 2) $\rho_0= 300 \ \rm Gpc^{-3} yr$ and $T_{90}=2 \cdot 10^5 \ \rm s$ were fixed, and a baryonic loading $\xi \simeq 10$ was found at the best-fit, consequently $\frac{\xi}{10} \, \frac{\rho_0}{300 \, \mathrm{ Gpc^{-3} \, yr^{-1} }}  \, \frac{T_{90}}{2 \, 10^5 \, \mathrm{s}} \simeq 1$, which we have just rescaled here to the values used in this paper. } from \cite{Boncioli:2018lrv} (almost independent of the collision radius) 
\begin{equation}
 \frac{\xi}{10^3} \, \frac{\rho_0}{440 \, \mathrm{ Gpc^{-3} \, yr^{-1} }}  \, \frac{T_{90}}{1300 \, \mathrm{s}} \simeq 1 \, .\label{equ:uhecr}
\end{equation}
The UHECR and neutrino description is degenerate in these parameters, which means that lower baryonic loadings could be compensated by longer durations.

An independent constraint can be obtained from the contribution to the extragalactic diffuse gamma-ray background. 
Due to the source evolution and the obtained HE gamma-ray emission, this contribution may be rather significant. 
We therefore show in \figu{gamma}, right panel, the expected contribution from a population of ll-GRBs given by prototype ul-GRB, and compare it with the EGB (\cite{Ackermann:2014usa}) measured by Fermi. Note that the EGB is expected to be driven by Active Galactic Nuclei, and the expected contribution from other source classes is small. We therefore show (dotted purple curve) the 14\% of the EGB flux, since blazars already provide $86\% \pm 14\%$ of contribution to the EGB, according to \cite{TheFermi-LAT:2015ykq}. We thus identify this as the
possible maximal contribution from ll-GRBs to the EGB. Comparing the curve GRB~100316D /ul-GRB  (using $\epsilon_\mathrm{B}^{*}=10^{-1
}$ which exhibits higher  maximal energies, see left panel of \figu{max_energies}) with the dotted curve, one easily finds the constraint\footnote{We show in \figu{gamma} the GRB contribution multiplied with the corresponding factor 100 as dark-green dashed cure, to illustrate that this enhancement factor saturates the possible non-blazar contribution to the EGB (it touches the dotted curve). }
\begin{equation}
 \ \frac{\rho_0}{440 \, \mathrm{ Gpc^{-3} \, yr^{-1} }}  \, \frac{T_{90}}{1300 \, \mathrm{s}} \lesssim 100 \, \label{equ:gammaconstr}
\end{equation}
as the gamma-ray flux (for fixed luminosity) scales with these two quantities.
We note that \equ{gammaconstr} may be used to break the degeneracy in \equ{uhecr}, which yields the constraints
\begin{equation}
 \xi \gtrsim 10 \, ,   \qquad T_{90} \lesssim 10^5 \, \mathrm{s}
 \label{equ:constr}
\end{equation}
for the typical ll-GRB (and $\rho_0$ fixed), which is in fact consistent with the UHECR fit in  \cite{Boncioli:2018lrv}.

As a duration of $10^5$~s is a strong
assumption for the typical ll-GRB duration, a related question is how large the baryonic loading can be in order to not to impact the observed SED (a high value of $\xi$ would allow reducing the required $T_{90}$ in \equ{constr}). We point out that the SED may be also produced by hadronic processes (including the second peak, such as from $\pi^0$ photons or large contributions by cascade emission), therefore this part can only give some rough estimates for the model discussed in this work. We focus on the pion production efficiency (the average fraction of energy lost into pion production), whereas protons in the source may produce other effects as well (such as proton synchrotron radiation). 
Additional effects would be expected from muon and pion synchrotron and radiation from secondary leptons produced \eg via Bethe-Heitler. 

Similar to \cite{Guetta:2003wi}, it can be  analytically estimated as 

\begin{align}
    f_{\pi^0} \simeq 5 \left( \frac{L_{\gamma}}{10^{47} \, \mathrm{erg/s}} \right) \left( \frac{t_\mathrm{dyn}}{100 \, \mathrm{s}} \right)\left( \frac{R}{10^{10} \, \mathrm{km}} \right)^{-2}\left( \frac{\varepsilon_{\gamma, \mathrm{br}}}{100 \, \mathrm{keV}} \right)^{-1} \, ,
    \label{equ:fpi}
\end{align}
(for small redshifts), taking the pitch angle-averaged inclusive cross section for $\pi^0$ production from \cite{Hummer:2010vx}. Here $t_{\mathrm{dyn}}$ is a measure for the dynamical time-scale, i.e., the time the protons have to interact; it may be associated with the shell expansion time $t_{\mathrm{ex}} = t^{\prime}_{\mathrm{ex}} / \Gamma \simeq R/(c \Gamma^2)$ in our model.  For ul-GRB, we have  $t_{\mathrm{ex}} \simeq 83 \, \mathrm{s}$ (for $R \simeq 10^{10} \, \mathrm{km}$, $\Gamma \simeq 20$, $\varepsilon_{\gamma, \mathrm{br}}=30 \, \mathrm{keV}$), which leads to $f_{\pi^0} \simeq 0.04$. From  \cite{Boncioli:2018lrv} it is clear that (relatively high) pion production efficiencies are required if ll-GRBs are to power the diffuse neutrino flux. However, \cite{Hummer:2011ms} have demonstrated that the actual value can be about an order of magnitude lower because this approximation neglects the energy dependence of the mean free path of the protons and the shape of the target photon spectrum (which are included in numerical calculations).

If the photons from the $\pi^0$ decays are fed into the electromagnetic cascade in the source, they will affect the SED at the highest energies. Since the injection luminosity into HE gamma-rays from these processes is $L_\mathrm{HE \, \gamma} = L_p \, f_{\pi^0} \simeq L_\gamma \cdot \xi \cdot f_{\pi^0}$ and $L_\mathrm{HE \, \gamma}\lesssim L_\gamma/10$ from
\figu{time_integrated_spectra} (the peak for HE gamma-rays is about a factor of ten lower than the one for gamma-rays), we have
\begin{equation}
     f_{\pi^0} \, \xi \lesssim 0.1 \label{equ:fpixi}
\end{equation}
in order not to affect the SED at all from these processes. For ul-GRB, we therefore find $\xi \lesssim 3-25$ for this process to be on the safe side, where the lower number uses \equ{fpi} at face value, and the upper value includes the estimate from \cite{Hummer:2011ms}. 

We have verified with numerical simulations that for ul-GRB ($\epsilon_\mathrm{B}^{*} = 10^{-1}$) the co-acceleration of protons (with the same injection index as electrons) with a baryonic loading of $ \xi = 25$ does not lead to additional, observable signatures or significantly change the observed spectrum. A more complete examination of this (including also \eg \, the impact of a modified injection index of protons and/or the presence of heavier nuclei and an investigation of the maximal baryonic loading compatible with observed spectra) is clearly beyond the scope of this study.

As pointed out earlier, $\xi \sim 25$ would require to lower the fraction of energy transferred to non-thermal electrons ($\epsilon_e^* \sim 10^{-2}$) and the fraction of accelerated electrons ($\zeta_0^* \sim 10^{-5}$ to $ 10^{-6}$), which is closer to the non-exlcuded parameter space in \cite{Samuelsson:2020upt}. In this sense, the baryonic loading necessary to power the UHCER flux may require re-thinking some of the standard paradigms of the partition parameters within the GRB internal shock model.

\section{Summary and conclusions}
\label{sec:summary}
We have studied ll-GRBs as potential targets for multiwavelength astronomy and sources of UHECR nuclei. 
For this purpose we have selected three representative reference events out of the sample of detected ll-GRBs. The prototypes based on these reference events represent different types of ll-GRBs: {\it sp-GRB} is a single-peaked GRB of medium peak energy, low luminosity and low redshift, {\it ul-GRB} is an ultralong, multi-peaked GRB with low peak energy and {\it hl-GRB} is a single-peaked GRB with comparatively high luminosity and high redshift. 
Choosing the parameters of the outflow such that we reproduce the main features of the reference events, we self-consistently calculated the full spectral and temporal properties within the internal shock scenario and a leptonically-dominated radiation model.

We have found that ll-GRBs are indeed potential targets for multimessenger observations and could be detected by current/ future Imaging Air Cherenkov Telescopes (IACTs). This is mainly due to their low redshifts (and high local rate), which reduce the effect of EBL absorption at the highest energies.
The intensity of the HE component is (for comparable intensity of the sub-MeV synchrotron peak) directly linked to the magnetic field strength $B^\prime$. In our dynamical modelling of the outflow, $B^\prime$ varies throughout the evolution and is not set directly; Instead we control the fraction of energy supplying the magnetic field with the constant microphysics parameter $\epsilon_B$. If we require a relatively high radiative efficiency (which implies lower bounds on $\epsilon_B$), we have found a significant VHE inverse Compton component if the magnetic field is powered by a fraction of order $\epsilon_B= 10^{-3}- 10^{-2}$ of the available internal energy. The intensity of the VHE emission increases systematically with decreasing $\epsilon_B$. While this is expected in Thomson regime (recall that $\mathrm{Y_{Th}} \propto \frac{\epsilon_e}{\epsilon_B}$), we observe the same tendency if the scatterings occur in Klein-Nishina regime. Therefore, a detailed numerical modelling over a broad energy range (including a VHE regime) may be used to constrain the physical conditions in the outflow and microphysics parameters, such as $\epsilon_B$. 

We have demonstrated that alternative probes of $\epsilon_B$ could be the (a) the optical flux (which increases with increasing $\epsilon_B$), 
(b) the low-energy photon spectral index $\alpha$ (the low energy portion of the spectrum becomes steeper with increasing $\epsilon_B$) and (c) a delayed onset of the HE component (the early suppression of the VHE flux due to $\gamma \gamma$-absorption becomes larger with increasing $\epsilon_B$). We point out that those observables might not be reliable: For example, (a) the optical flux is subject to extinction in our own Galaxy and the host galaxy, which is in most cases not well constrained. On the other hand, (b) the photon index $\alpha$ is largely impacted by the fit range of the observing instrument and the location of the spectral peak within the instrument range. Especially for ll-GRBs, which have generally low peak frequencies, this introduces uncertainties on $\alpha$ measurement. Overall, it might be the joint analysis of all these observables that will make it possible to constrain the physical parameters of the sources and the processes at play, if a full numerical modelling is performed (including the dynamical evolution of parameters such as the magnetic field strength, volume and Lorentz factor).

Although using the same internal shock model commonly applied to GRB-HL, our simulations reside in a specific phase space of parameters such as dissipation radius, Lorentz factor and dissipated energy, due to the distinct properties of ll-GRBs (\cite{Daigne:2007qz}). It is the combination of these parameters that result in specific properties of the observed spectra (like the intensity of the inverse Compton component). Thus, while some of the findings outlined above may also apply to high-luminosity events (see for example the discussion of $\alpha$ in \cite{Daigne:2010fb}), our results can not easily be generalized to all GRBs.

We have also shown that ll-GRBs are able to accelerate nuclei to the UHE:
Overall, the maximal energies of iron nuclei (protons) could be as high as $\simeq 10^{11}$~GeV ($10^{10}$~GeV). The highest maximal energies were achieved for large magnetic fields (set by large $\epsilon_B$), for which the inverse Compton efficiency was found to be low. This means that in our model, a ll-GRB can {\bf either} accelerate UHECR to the highest energies {\bf or} have high fluxes in the VHE regime.
The high maximal energies in our model were possible by decoupling the production regions of the prompt $\gamma$-ray emission and the most energetic UHECR nuclei. The former is produced at intermediate radii, the latter in outer regions. The difference in radius is roughly an order of magnitude, questioning the validity of simplified one-zone models. However, outer collisions contribute less to the overall gamma-ray emission and more energy is dissipated at lower radii, for which the maximal cosmic-ray energies are lower. We point out that in order to calculate the true emitted spectra, additional assumptions on the injection spectrum and escape functions are required. 

If ll-GRBs are to power the UHECR flux, a given local rate results in requirements for the typical energy emitted in UHECRs per GRB. This energy output is controlled by the product of burst duration and baryonic loading, which are thus degenerate. We have shown that the contribution to the diffuse gamma-ray background breaks this degeneracy and leads to an upper limit on the typical duration and a lower limit on the baryonic loading. An additional upper limit may come from hadronic signatures in the spectral energy distribution due to photo-pion production. Overall, we have estimated that  basic consistency arises for a baryonic loading around 10-25. However, the energy partition parameters such as the fraction energy transferred to non-thermal electrons that correspond to this baryonic loading in combination with the results of our leptonic radiation modelling differ from the values discussed in the literature - a problem which is not specific to ll-GRBs, but applies to long GRBs as well if they are the dominant sources of UHECRs (see \eg \, \cite{Heinze:2020zqb}).
Additionally, these results should be verified by including the hadronic and nuclear processes fully self-consistent in the spectral modelling. 

We conclude that ll-GRBs are potential targets for multiwavelength and multimessenger astronomy. They are potentially within the sensitivity range of IACTs, and future instruments such as CTA and a multiwavelength coverage by optical/ UV surveys may help to constrain the physical processes at play in the source. 
ll-GRBs also fulfill the energetic requirements to accelerate cosmic-rays to UHE; To constrain the maximal baryonic loading in agreement with gamma-ray observations further studies with detailed radation modelling are required.

\subsubsection*{Acknowledgments}

This work has been supported by the European Research Council (ERC) under the European Union’s Horizon 2020 research and innovation programme (Grant No. 646623).
The work was supported by the International Helmholtz-Weizmann Research School for Multimessenger Astronomy, largely funded through the Initiative and Networking Fund of the Helmholtz Association. \v ZB acknowledges the support of  the Deutscher Akademischer Austauschdienst (DAAD) scholarship programme.
This work was conducted in the context of the CTA Consortium. 
We use CTA instrument response functions, provided by the CTA Consortium and Observatory (version \textit{prod3b-v2}). (See~\url{http://www.cta-observatory.org/science/cta-performance/}.)
We would also like to thank the CTA Consortium, for providing an initial review of the manuscript, as well as Anatoli Fedynitch and Filip Samuelsson for useful comments and discussion. 
\subsubsection*{Data availability}
Data available on request.

\bibliography{references}

\begin{thebibliography}{}
\makeatletter
\relax
\def\mn@urlcharsother{\let\do\@makeother \do\$\do\&\do\#\do\^\do\_\do\%\do\~}
\def\mn@doi{\begingroup\mn@urlcharsother \@ifnextchar [ {\mn@doi@}
  {\mn@doi@[]}}
\def\mn@doi@[#1]#2{\def\@tempa{#1}\ifx\@tempa\@empty \href
  {http://dx.doi.org/#2} {doi:#2}\else \href {http://dx.doi.org/#2} {#1}\fi
  \endgroup}
\def\mn@eprint#1#2{\mn@eprint@#1:#2::\@nil}
\def\mn@eprint@arXiv#1{\href {http://arxiv.org/abs/#1} {{\tt arXiv:#1}}}
\def\mn@eprint@dblp#1{\href {http://dblp.uni-trier.de/rec/bibtex/#1.xml}
  {dblp:#1}}
\def\mn@eprint@#1:#2:#3:#4\@nil{\def\@tempa {#1}\def\@tempb {#2}\def\@tempc
  {#3}\ifx \@tempc \@empty \let \@tempc \@tempb \let \@tempb \@tempa \fi \ifx
  \@tempb \@empty \def\@tempb {arXiv}\fi \@ifundefined
  {mn@eprint@\@tempb}{\@tempb:\@tempc}{\expandafter \expandafter \csname
  mn@eprint@\@tempb\endcsname \expandafter{\@tempc}}}

\bibitem[\protect\citeauthoryear{Aartsen et~al.}{Aartsen
  et~al.}{2014}]{IceCube:2014gqr}
Aartsen M.~G.,  et~al., 2014, arXiv e-prints

\bibitem[\protect\citeauthoryear{Aartsen et~al.}{Aartsen
  et~al.}{2015}]{Aartsen:2014aqy}
Aartsen M.~G.,  et~al., 2015, \mn@doi [Astrophys. J.]
  {10.1088/2041-8205/805/1/L5}, 805, L5

\bibitem[\protect\citeauthoryear{Aartsen et~al.}{Aartsen
  et~al.}{2016}]{IceCube:2016ipa}
Aartsen M.~G.,  et~al., 2016, \mn@doi [Astrophys. J.]
  {10.3847/0004-637X/824/2/115}, 824, 115

\bibitem[\protect\citeauthoryear{Aartsen et~al.}{Aartsen
  et~al.}{2017}]{Aartsen:2017wea}
Aartsen M.~G.,  et~al., 2017, \mn@doi [Astrophys. J.]
  {10.3847/1538-4357/aa7569}, 843, 112

\bibitem[\protect\citeauthoryear{Aartsen et~al.}{Aartsen
  et~al.}{2019a}]{IceCube:2018omy}
Aartsen M.~G.,  et~al., 2019a, \mn@doi [Phys. Rev. Lett.]
  {10.1103/PhysRevLett.122.051102}, 122, 051102

\bibitem[\protect\citeauthoryear{Aartsen et~al.}{Aartsen
  et~al.}{2019b}]{Aartsen:2018fpd}
Aartsen M.~G.,  et~al., 2019b, \mn@doi [Phys. Rev. Lett.]
  {10.1103/PhysRevLett.122.051102}, 122, 051102

\bibitem[\protect\citeauthoryear{Abbasi et~al.}{Abbasi
  et~al.}{2012}]{Abbasi:2012zw}
Abbasi R.,  et~al., 2012, \mn@doi [Nature] {10.1038/nature11068}, 484, 351

\bibitem[\protect\citeauthoryear{Abdalla et~al.}{Abdalla
  et~al.}{2017}]{H.E.S.S.:2017odt}
Abdalla H.,  et~al., 2017, \mn@doi [Astron. Astrophys.]
  {10.1051/0004-6361/201731200}, 606, A59

\bibitem[\protect\citeauthoryear{Abdalla et~al.}{Abdalla
  et~al.}{2021}]{Collaboration:2021fro}
Abdalla H.,  et~al., 2021, \mn@doi [Science] {10.1126/science.abe8560}, 372,
  1081

\bibitem[\protect\citeauthoryear{Ackermann et~al.}{Ackermann
  et~al.}{2015}]{Ackermann:2014usa}
Ackermann M.,  et~al., 2015, \mn@doi [Astrophys. J.]
  {10.1088/0004-637X/799/1/86}, 799, 86

\bibitem[\protect\citeauthoryear{Ackermann et~al.}{Ackermann
  et~al.}{2016}]{TheFermi-LAT:2015ykq}
Ackermann M.,  et~al., 2016, \mn@doi [Phys. Rev. Lett.]
  {10.1103/PhysRevLett.116.151105}, 116, 151105

\bibitem[\protect\citeauthoryear{Aloy, Cuesta-Martínez  \& Obergaulinger}{Aloy
  et~al.}{2018}]{Aloy:2018czj}
Aloy M.-A.,  Cuesta-Martínez C.~F.,   Obergaulinger M.,  2018, \mn@doi [Mon.
  Not. Roy. Astron. Soc.] {10.1093/mnras/sty1212}, 478, 3576

\bibitem[\protect\citeauthoryear{Amati}{Amati}{2006}]{Amati:2006ky}
Amati L.,  2006, \mn@doi [Mon. Not. Roy. Astron. Soc.]
  {10.1111/j.1365-2966.2006.10840.x}, 372, 233

\bibitem[\protect\citeauthoryear{Amati, Della~Valle, Frontera, Malesani,
  Guidorzi, Montanari  \& Pian}{Amati et~al.}{2007}]{Amati:2006ru}
Amati L.,  Della~Valle M.,  Frontera F.,  Malesani D.,  Guidorzi C.,  Montanari
  E.,   Pian E.,  2007, \mn@doi [Astron. Astrophys.]
  {10.1051/0004-6361:20065994}, 463, 913

\bibitem[\protect\citeauthoryear{Asano \& Inoue}{Asano \&
  Inoue}{2007}]{Asano:2007my}
Asano K.,  Inoue S.,  2007, \mn@doi [Astrophys. J.] {10.1086/522939}, 671, 645

\bibitem[\protect\citeauthoryear{{Asano} \& {M{\'e}sz{\'a}ros}}{{Asano} \&
  {M{\'e}sz{\'a}ros}}{2011}]{2011ApJ...739..103A}
{Asano} K.,  {M{\'e}sz{\'a}ros} P.,  2011, \mn@doi [\apj]
  {10.1088/0004-637X/739/2/103}, \href
  {https://ui.adsabs.harvard.edu/abs/2011ApJ...739..103A} {739, 103}

\bibitem[\protect\citeauthoryear{Band et~al.}{Band et~al.}{1993}]{Band:1993eg}
Band D.,  et~al., 1993, \mn@doi [Astrophys. J.] {10.1086/172995}, 413, 281

\bibitem[\protect\citeauthoryear{Bellm et~al.,}{Bellm
  et~al.}{2018}]{Bellm_2018}
Bellm E.~C.,  et~al., 2018, \mn@doi [Publications of the Astronomical Society
  of the Pacific] {10.1088/1538-3873/aaecbe}, 131, 018002

\bibitem[\protect\citeauthoryear{Berezinsky \& Kalashev}{Berezinsky \&
  Kalashev}{2016}]{Berezinsky:2016feh}
Berezinsky V.,  Kalashev O.,  2016, \mn@doi [Phys. Rev.]
  {10.1103/PhysRevD.94.023007}, D94, 023007

\bibitem[\protect\citeauthoryear{Biehl, Boncioli, Fedynitch  \& Winter}{Biehl
  et~al.}{2018}]{Biehl:2017zlw}
Biehl D.,  Boncioli D.,  Fedynitch A.,   Winter W.,  2018, \mn@doi [Astron.
  Astrophys.] {10.1051/0004-6361/201731337}, 611, A101

\bibitem[\protect\citeauthoryear{Boncioli, Biehl  \& Winter}{Boncioli
  et~al.}{2019}]{Boncioli:2018lrv}
Boncioli D.,  Biehl D.,   Winter W.,  2019, \mn@doi [Astrophys. J.]
  {10.3847/1538-4357/aafda7}, 872, 110

\bibitem[\protect\citeauthoryear{Bosnjak, Daigne  \& Dubus}{Bosnjak
  et~al.}{2009}]{Bosnjak:2008bd}
Bosnjak Z.,  Daigne F.,   Dubus G.,  2009, \mn@doi [Astron. Astrophys.]
  {10.1051/0004-6361/200811375}, 498, 677

\bibitem[\protect\citeauthoryear{Bo\v{s}njak \& Daigne}{Bo\v{s}njak \&
  Daigne}{2014}]{Bosnjak:2014hya}
Bo\v{s}njak v.,  Daigne F.,  2014, \mn@doi [Astron. Astrophys.]
  {10.1051/0004-6361/201322341}, 568, A45

\bibitem[\protect\citeauthoryear{Bromberg, Nakar  \& Piran}{Bromberg
  et~al.}{2011}]{Bromberg:2011}
Bromberg O.,  Nakar E.,   Piran T.,  2011, \mn@doi [Astrophys. J. Lett.]
  {10.1088/2041-8205/739/2/L55}, 739, L55

\bibitem[\protect\citeauthoryear{Burrows et~al.}{Burrows
  et~al.}{2005}]{Burrows:2005gfa}
Burrows D.~N.,  et~al., 2005, \mn@doi [Space Sci. Rev.]
  {10.1007/s11214-005-5097-2}, 120, 165

\bibitem[\protect\citeauthoryear{Bustamante, Baerwald, Murase  \&
  Winter}{Bustamante et~al.}{2015}]{Bustamante:2014oka}
Bustamante M.,  Baerwald P.,  Murase K.,   Winter W.,  2015, \mn@doi [Nature
  Commun.] {10.1038/ncomms7783}, 6, 6783

\bibitem[\protect\citeauthoryear{Bustamante, Murase, Winter  \&
  Heinze}{Bustamante et~al.}{2017}]{Bustamante:2016wpu}
Bustamante M.,  Murase K.,  Winter W.,   Heinze J.,  2017, \mn@doi [Astrophys.
  J.] {10.3847/1538-4357/837/1/33}, 837, 33

\bibitem[\protect\citeauthoryear{Campana et~al.}{Campana
  et~al.}{2006}]{Campana:2006qe}
Campana S.,  et~al., 2006, \mn@doi [Nature] {10.1038/nature04892}, 442, 1008

\bibitem[\protect\citeauthoryear{Cano et~al.}{Cano
  et~al.}{2017a}]{Cano:2017sab}
Cano Z.,  et~al., 2017a, \mn@doi [Astron. Astrophys.]
  {10.1051/0004-6361/201731005}, 605, A107

\bibitem[\protect\citeauthoryear{Cano, Wang, Dai  \& Wu}{Cano
  et~al.}{2017b}]{Cano:2016ccp}
Cano Z.,  Wang S.-Q.,  Dai Z.-G.,   Wu X.-F.,  2017b, \mn@doi [Adv. Astron.]
  {10.1155/2017/8929054}, 2017, 8929054

\bibitem[\protect\citeauthoryear{Chand et~al.}{Chand
  et~al.}{2020}]{Chand:2020wqt}
Chand V.,  et~al., 2020, ArXiv eprints

\bibitem[\protect\citeauthoryear{Crumley, Caprioli, Markoff  \&
  Spitkovsky}{Crumley et~al.}{2019}]{Crumley:2018kvf}
Crumley P.,  Caprioli D.,  Markoff S.,   Spitkovsky A.,  2019, \mn@doi [Mon.
  Not. Roy. Astron. Soc.] {10.1093/mnras/stz232}, 485, 5105

\bibitem[\protect\citeauthoryear{Cummings et~al.,}{Cummings
  et~al.}{2012}]{GCN13481}
Cummings J.~R.,  et~al., 2012, GCN Circ. 13481

\bibitem[\protect\citeauthoryear{D'Elia et~al.}{D'Elia
  et~al.}{2018}]{DElia:2018xrz}
D'Elia V.,  et~al., 2018, \mn@doi [Astron. Astrophys.]
  {10.1051/0004-6361/201833847}, 619, A66

\bibitem[\protect\citeauthoryear{Daigne \& Mochkovitch}{Daigne \&
  Mochkovitch}{1998}]{Daigne:1998xc}
Daigne F.,  Mochkovitch R.,  1998, \mn@doi [Mon. Not. Roy. Astron. Soc.]
  {10.1046/j.1365-8711.1998.01305.x}, 296, 275

\bibitem[\protect\citeauthoryear{Daigne \& Mochkovitch}{Daigne \&
  Mochkovitch}{2000}]{Daigne:2000xg}
Daigne F.,  Mochkovitch R.,  2000, Astron. Astrophys., 358, 1157

\bibitem[\protect\citeauthoryear{Daigne \& Mochkovitch}{Daigne \&
  Mochkovitch}{2007}]{Daigne:2007qz}
Daigne F.,  Mochkovitch R.,  2007, \mn@doi [Astron. Astrophys.]
  {10.1051/0004-6361:20066080}, 465, 1

\bibitem[\protect\citeauthoryear{Daigne, Bosnjak  \& Dubus}{Daigne
  et~al.}{2011}]{Daigne:2010fb}
Daigne F.,  Bosnjak Z.,   Dubus G.,  2011, \mn@doi [Astron. Astrophys.]
  {10.1051/0004-6361/201015457}, 526, A110

\bibitem[\protect\citeauthoryear{Deil et~al.}{Deil et~al.}{2018}]{Deil:2017yey}
Deil C.,  et~al., 2018, \mn@doi [PoS] {10.22323/1.301.0766}, ICRC2017, 766

\bibitem[\protect\citeauthoryear{Dominguez et~al.}{Dominguez
  et~al.}{2011}]{Dominguez:2010bv}
Dominguez A.,  et~al., 2011, \mn@doi [Mon. Not. Roy. Astron. Soc.]
  {10.1111/j.1365-2966.2010.17631.x}, 410, 2556

\bibitem[\protect\citeauthoryear{{Dong}, {Wu}, {Li}, {Zhang}  \&
  {Zhang}}{{Dong} et~al.}{2010}]{Dong2010GRM}
{Dong} Y.,  {Wu} B.,  {Li} Y.,  {Zhang} Y.,   {Zhang} S.,  2010, \mn@doi
  [Science China Physics, Mechanics, and Astronomy]
  {10.1007/s11433-010-0011-7}, \href
  {https://ui.adsabs.harvard.edu/abs/2010SCPMA..53S..40D} {53, 40}

\bibitem[\protect\citeauthoryear{Duran, Bosnjak  \& Kumar}{Duran
  et~al.}{2012}]{Duran:2012ww}
Duran R.,  Bosnjak Z.,   Kumar P.,  2012, \mn@doi [Mon. Not. Roy. Astron. Soc.]
  {10.1111/j.1365-2966.2012.21533.x}, 424, 3192

\bibitem[\protect\citeauthoryear{Fan, Zhang, Xu, Liang  \& Zhang}{Fan
  et~al.}{2011}]{Fan:2010br}
Fan Y.-Z.,  Zhang B.-B.,  Xu D.,  Liang E.-W.,   Zhang B.,  2011, \mn@doi
  [Astrophys. J.] {10.1088/0004-637X/726/1/32}, 726, 32

\bibitem[\protect\citeauthoryear{Fioretti, Ribeiro, Humensky, Bulgarelli,
  Maier, Moralejo  \& Nigro}{Fioretti et~al.}{2021}]{2019ICRC...36..673F}
Fioretti V.,  Ribeiro D.,  Humensky T.~B.,  Bulgarelli A.,  Maier G.,  Moralejo
  A.,   Nigro C.,  2021, \mn@doi [PoS] {10.22323/1.358.0673}, ICRC2019, 673

\bibitem[\protect\citeauthoryear{{Foley}, {McGlynn}, {Hanlon}, {McBreen}  \&
  {McBreen}}{{Foley} et~al.}{2008}]{foley08}
{Foley} S.,  {McGlynn} S.,  {Hanlon} L.,  {McBreen} S.,   {McBreen} B.,  2008,
  \mn@doi [\aap] {10.1051/0004-6361:20078399}, \href
  {https://ui.adsabs.harvard.edu/abs/2008A&A...484..143F} {484, 143}

\bibitem[\protect\citeauthoryear{Fraija}{Fraija}{2014}]{Fraija:2013cha}
Fraija N.,  2014, \mn@doi [Mon. Not. Roy. Astron. Soc.]
  {10.1093/mnras/stt2036}, 437, 2187

\bibitem[\protect\citeauthoryear{Fraija et~al.,}{Fraija
  et~al.}{2017}]{Fraija:2017mlx}
Fraija N.,  et~al., 2017, \mn@doi [Astrophys. J.] {10.3847/1538-4357/aa8a72},
  848, 15

\bibitem[\protect\citeauthoryear{Fraija, Veres, Beniamini, Galvan-Gamez,
  Metzger, Duran  \& Becerra}{Fraija et~al.}{2021}]{Fraija:2020vsa}
Fraija N.,  Veres P.,  Beniamini P.,  Galvan-Gamez A.,  Metzger B.~D.,  Duran
  R.~B.,   Becerra R.~L.,  2021, \mn@doi [Astrophys. J.]
  {10.3847/1538-4357/ac0aed}, 918, 12

\bibitem[\protect\citeauthoryear{Frontera et~al.}{Frontera
  et~al.}{2000}]{Frontera:1999ew}
Frontera F.,  et~al., 2000, \mn@doi [Astrophys. J. Suppl.] {10.1086/313316},
  127, 59

\bibitem[\protect\citeauthoryear{Gao, Kashiyama  \& M\'esz\'aros}{Gao
  et~al.}{2013}]{Gao:2013fra}
Gao S.,  Kashiyama K.,   M\'esz\'aros P.,  2013, \mn@doi [Astrophys. J. Lett.]
  {10.1088/2041-8205/772/1/L4}, 772, L4

\bibitem[\protect\citeauthoryear{Gao, Pohl  \& Winter}{Gao
  et~al.}{2017}]{Gao:2016uld}
Gao S.,  Pohl M.,   Winter W.,  2017, \mn@doi [Astrophys. J.]
  {10.3847/1538-4357/aa7754}, 843, 109

\bibitem[\protect\citeauthoryear{Ghirlanda et~al.,}{Ghirlanda
  et~al.}{2018}]{Ghirlanda:2017opl}
Ghirlanda G.,  et~al., 2018, \mn@doi [Astron. Astrophys.]
  {10.1051/0004-6361/201731598}, 609, A112

\bibitem[\protect\citeauthoryear{Ghisellini, Celotti  \& Lazzati}{Ghisellini
  et~al.}{2000}]{Ghisellini:1999wu}
Ghisellini G.,  Celotti A.,   Lazzati D.,  2000, \mn@doi [Mon. Not. Roy.
  Astron. Soc.] {10.1046/j.1365-8711.2000.03354.x}, 313, 1

\bibitem[\protect\citeauthoryear{Ghisellini, Ghirlanda, Mereghetti, Bosnjak,
  Tavecchio  \& Firmani}{Ghisellini et~al.}{2006}]{Ghisellini:2006zh}
Ghisellini G.,  Ghirlanda G.,  Mereghetti S.,  Bosnjak Z.,  Tavecchio F.,
  Firmani C.,  2006, \mn@doi [Mon. Not. Roy. Astron. Soc.]
  {10.1111/j.1365-2966.2006.10972.x}, 372, 1699

\bibitem[\protect\citeauthoryear{Ghisellini, Ghirlanda  \&
  Tavecchio}{Ghisellini et~al.}{2007}]{Ghisellini:2007ya}
Ghisellini G.,  Ghirlanda G.,   Tavecchio F.,  2007, \mn@doi [Mon. Not. Roy.
  Astron. Soc.] {10.1111/j.1745-3933.2007.00396.x}, 382, 77

\bibitem[\protect\citeauthoryear{Giannios}{Giannios}{2010}]{Giannios:2010cv}
Giannios D.,  2010, \mn@doi [Mon. Not. Roy. Astron. Soc.]
  {10.1111/j.1745-3933.2010.00925.x}, 408, 46

\bibitem[\protect\citeauthoryear{Globus, Allard, Mochkovitch  \&
  Parizot}{Globus et~al.}{2015}]{Globus:2014fka}
Globus N.,  Allard D.,  Mochkovitch R.,   Parizot E.,  2015, \mn@doi [Mon. Not.
  Roy. Astron. Soc.] {10.1093/mnras/stv893}, 451, 751

\bibitem[\protect\citeauthoryear{Godet et~al.}{Godet
  et~al.}{2014}]{Godet:2014ava}
Godet O.,  et~al., 2014, \mn@doi [Proc. SPIE Int. Soc. Opt. Eng.]
  {10.1117/12.2055507}, 9144, 914424

\bibitem[\protect\citeauthoryear{Granot, Piran  \& Sari}{Granot
  et~al.}{1999}]{Granot:1998ep}
Granot J.,  Piran T.,   Sari R.,  1999, \mn@doi [Astrophys. J.]
  {10.1086/306884}, 513, 679

\bibitem[\protect\citeauthoryear{Gruber et~al.}{Gruber
  et~al.}{2014}]{Gruber:2014iza}
Gruber D.,  et~al., 2014, \mn@doi [Astrophys. J. Suppl.]
  {10.1088/0067-0049/211/1/12}, 211, 12

\bibitem[\protect\citeauthoryear{Guetta, Hooper, Alvarez-Muniz, Halzen  \&
  Reuveni}{Guetta et~al.}{2004}]{Guetta:2003wi}
Guetta D.,  Hooper D.,  Alvarez-Muniz J.,  Halzen F.,   Reuveni E.,  2004,
  \mn@doi [Astropart. Phys.] {10.1016/S0927-6505(03)00211-1}, 20, 429

\bibitem[\protect\citeauthoryear{Hascoet, Daigne, Mochkovitch  \&
  Vennin}{Hascoet et~al.}{2012}]{Hascoet:2011gp}
Hascoet R.,  Daigne F.,  Mochkovitch R.,   Vennin V.,  2012, \mn@doi [Mon. Not.
  Roy. Astron. Soc.] {10.1111/j.1365-2966.2011.20332.x}, 421, 525

\bibitem[\protect\citeauthoryear{Heinze, Fedynitch, Boncioli  \& Winter}{Heinze
  et~al.}{2019}]{Heinze:2019jou}
Heinze J.,  Fedynitch A.,  Boncioli D.,   Winter W.,  2019, \mn@doi [Astrophys.
  J.] {10.3847/1538-4357/ab05ce}, 873, 88

\bibitem[\protect\citeauthoryear{Heinze, Biehl, Fedynitch, Boncioli, Rudolph
  \& Winter}{Heinze et~al.}{2020}]{Heinze:2020zqb}
Heinze J.,  Biehl D.,  Fedynitch A.,  Boncioli D.,  Rudolph A.,   Winter W.,
  2020, \mn@doi [Mon. Not. Roy. Astron. Soc.] {10.1093/mnras/staa2751}, 498,
  5990

\bibitem[\protect\citeauthoryear{Hummer, Ruger, Spanier  \& Winter}{Hummer
  et~al.}{2010}]{Hummer:2010vx}
Hummer S.,  Ruger M.,  Spanier F.,   Winter W.,  2010, \mn@doi [Astrophys. J.]
  {10.1088/0004-637X/721/1/630}, 721, 630

\bibitem[\protect\citeauthoryear{Hummer, Baerwald  \& Winter}{Hummer
  et~al.}{2012}]{Hummer:2011ms}
Hummer S.,  Baerwald P.,   Winter W.,  2012, \mn@doi [Phys. Rev. Lett.]
  {10.1103/PhysRevLett.108.231101}, 108, 231101

\bibitem[\protect\citeauthoryear{Irwin \& Chevalier}{Irwin \&
  Chevalier}{2016}]{Irwin:2015rbf}
Irwin C.~M.,  Chevalier R.~A.,  2016, \mn@doi [Mon. Not. Roy. Astron. Soc.]
  {10.1093/mnras/stw1058}, 460, 1680

\bibitem[\protect\citeauthoryear{Ivezi\'c et~al.}{Ivezi\'c
  et~al.}{2019}]{Ivezic:2008fe}
Ivezi\'c v.,  et~al., 2019, \mn@doi [Astrophys. J.] {10.3847/1538-4357/ab042c},
  873, 111

\bibitem[\protect\citeauthoryear{Kaneko et~al.,}{Kaneko
  et~al.}{2006}]{Kaneko:2006mt}
Kaneko Y.,  et~al., 2006, \mn@doi [Astrophys. J.] {10.1086/508324}, 654, 385

\bibitem[\protect\citeauthoryear{Kistler, Yuksel, Beacom, Hopkins  \&
  Wyithe}{Kistler et~al.}{2009}]{Kistler:2009mv}
Kistler M.~D.,  Yuksel H.,  Beacom J.~F.,  Hopkins A.~M.,   Wyithe J. S.~B.,
  2009, \mn@doi [Astrophys. J.] {10.1088/0004-637X/705/2/L104}, 705, L104

\bibitem[\protect\citeauthoryear{Klose et~al.}{Klose
  et~al.}{2019}]{Klose:2018ftc}
Klose S.,  et~al., 2019, \mn@doi [Astron. Astrophys.]
  {10.1051/0004-6361/201832728}, 622, A138

\bibitem[\protect\citeauthoryear{{Kn{\"o}dlseder} et~al.,}{{Kn{\"o}dlseder}
  et~al.}{2016}]{2016A&A...593A...1K}
{Kn{\"o}dlseder} J.,  et~al., 2016, \mn@doi [\aap]
  {10.1051/0004-6361/201628822}, \href
  {https://ui.adsabs.harvard.edu/abs/2016A&A...593A...1K} {593, A1}

\bibitem[\protect\citeauthoryear{Kobayashi, Piran  \& Sari}{Kobayashi
  et~al.}{1997}]{Kobayashi:1997jk}
Kobayashi S.,  Piran T.,   Sari R.,  1997, \mn@doi [Astrophys. J.]
  {10.1086/512791}, 490, 92

\bibitem[\protect\citeauthoryear{Kovacevic et~al.,}{Kovacevic
  et~al.}{2014}]{Kovacevic:2014isa}
Kovacevic M.,  et~al., 2014, \mn@doi [Astron. Astrophys.]
  {10.1051/0004-6361/201424700}, 569, A108

\bibitem[\protect\citeauthoryear{{Kumar} \& {Panaitescu}}{{Kumar} \&
  {Panaitescu}}{2003}]{kumarpanaitescu}
{Kumar} P.,  {Panaitescu} A.,  2003, \mn@doi [\mnras]
  {10.1111/j.1365-2966.2003.07138.x}, \href
  {https://ui.adsabs.harvard.edu/abs/2003MNRAS.346..905K} {346, 905}

\bibitem[\protect\citeauthoryear{Kumar \& Zhang}{Kumar \&
  Zhang}{2014}]{Kumar:2014upa}
Kumar P.,  Zhang B.,  2014, \mn@doi [Phys. Rept.]
  {10.1016/j.physrep.2014.09.008}, 561, 1

\bibitem[\protect\citeauthoryear{Lemoine \& Pelletier}{Lemoine \&
  Pelletier}{2010}]{Lemoine:2009vr}
Lemoine M.,  Pelletier G.,  2010, \mn@doi [Mon. Not. Roy. Astron. Soc.]
  {10.1111/j.1365-2966.2009.15869.x}, 402, 321

\bibitem[\protect\citeauthoryear{Liang, Zhang  \& Dai}{Liang
  et~al.}{2007}]{Liang:2006ci}
Liang E.,  Zhang B.,   Dai Z.~G.,  2007, \mn@doi [Astrophys. J.]
  {10.1086/517959}, 662, 1111

\bibitem[\protect\citeauthoryear{Lien et~al.}{Lien et~al.}{2016}]{Lien:2016zny}
Lien A.,  et~al., 2016, \mn@doi [Astrophys. J.] {10.3847/0004-637X/829/1/7},
  829, 7

\bibitem[\protect\citeauthoryear{Ligorini et~al.,}{Ligorini
  et~al.}{2021}]{Ligorini:2021lbj}
Ligorini A.,  et~al., 2021, \mn@doi [Mon. Not. Roy. Astron. Soc.]
  {10.1093/mnras/stab220}, 502, 5065

\bibitem[\protect\citeauthoryear{Liu, Wang  \& Dai}{Liu
  et~al.}{2011}]{Liu:2011cua}
Liu R.-Y.,  Wang X.-Y.,   Dai Z.-G.,  2011, \mn@doi [Mon. Not. Roy. Astron.
  Soc.] {10.1111/j.1365-2966.2011.19590.x}, 418, 1382

\bibitem[\protect\citeauthoryear{{Liu}, {Wu}  \& {Lu}}{{Liu}
  et~al.}{2012}]{liu2012}
{Liu} X.-W.,  {Wu} X.-F.,   {Lu} T.,  2012, \mn@doi [\aj]
  {10.1088/0004-6256/143/5/115}, \href
  {https://ui.adsabs.harvard.edu/abs/2012AJ....143..115L} {143, 115}

\bibitem[\protect\citeauthoryear{{Mandal} \& {Eichler}}{{Mandal} \&
  {Eichler}}{2010}]{2010ApJ...713L..55M}
{Mandal} S.,  {Eichler} D.,  2010, \mn@doi [Astrophys. J. L.]
  {10.1088/2041-8205/713/1/L55}, \href
  {https://ui.adsabs.harvard.edu/abs/2010ApJ...713L..55M} {713, L55}

\bibitem[\protect\citeauthoryear{Marcowith et~al.}{Marcowith
  et~al.}{2016}]{Marcowith:2016vzl}
Marcowith A.,  et~al., 2016, \mn@doi [Rept. Prog. Phys.]
  {10.1088/0034-4885/79/4/046901}, 79, 046901

\bibitem[\protect\citeauthoryear{Marcowith, Ferrand, Grech, Meliani, Plotnikov
  \& Walder}{Marcowith et~al.}{2020}]{Marcowith:2020vho}
Marcowith A.,  Ferrand G.,  Grech M.,  Meliani Z.,  Plotnikov I.,   Walder R.,
  2020, \mn@doi [Liv. Rev. Comput. Astrophys.] {10.1007/s41115-020-0007-6}, 6,
  1

\bibitem[\protect\citeauthoryear{Meegan et~al.}{Meegan
  et~al.}{2009}]{Meegan:2009qu}
Meegan C.,  et~al., 2009, \mn@doi [Astrophys. J.]
  {10.1088/0004-637X/702/1/791}, 702, 791

\bibitem[\protect\citeauthoryear{Murase \& Ioka}{Murase \&
  Ioka}{2013}]{Murase:2013ffa}
Murase K.,  Ioka K.,  2013, \mn@doi [Phys. Rev. Lett.]
  {10.1103/PhysRevLett.111.121102}, 111, 121102

\bibitem[\protect\citeauthoryear{Murase, Ioka, Nagataki  \& Nakamura}{Murase
  et~al.}{2006}]{Murase:2006mm}
Murase K.,  Ioka K.,  Nagataki S.,   Nakamura T.,  2006, \mn@doi [Astrophys.
  J.] {10.1086/509323}, 651, L5

\bibitem[\protect\citeauthoryear{Murase, Ioka, Nagataki  \& Nakamura}{Murase
  et~al.}{2008}]{Murase:2008mr}
Murase K.,  Ioka K.,  Nagataki S.,   Nakamura T.,  2008, \mn@doi [Phys.Rev.]
  {10.1103/PhysRevD.78.023005}, D78, 023005

\bibitem[\protect\citeauthoryear{{Nakar}}{{Nakar}}{2015}]{nakar2015}
{Nakar} E.,  2015, \mn@doi [Astrophys. J.] {10.1088/0004-637X/807/2/172}, \href
  {https://ui.adsabs.harvard.edu/abs/2015ApJ...807..172N} {807, 172}

\bibitem[\protect\citeauthoryear{{Nakar} \& {Sari}}{{Nakar} \&
  {Sari}}{2012}]{2012ApJ...747...88N}
{Nakar} E.,  {Sari} R.,  2012, \mn@doi [ApJ] {10.1088/0004-637X/747/2/88},
  \href {https://ui.adsabs.harvard.edu/abs/2012ApJ...747...88N} {747, 88}

\bibitem[\protect\citeauthoryear{Nakar, Ando  \& Sari}{Nakar
  et~al.}{2009}]{Nakar:2009er}
Nakar E.,  Ando S.,   Sari R.,  2009, \mn@doi [Astrophys. J.]
  {10.1088/0004-637X/703/1/675}, 703, 675

\bibitem[\protect\citeauthoryear{Nigro et~al.}{Nigro
  et~al.}{2019}]{Nigro:2019hqf}
Nigro C.,  et~al., 2019, \mn@doi [Astron. Astrophys.]
  {10.1051/0004-6361/201834938}, 625, A10

\bibitem[\protect\citeauthoryear{{Oganesyan}, {Nava}, {Ghirlanda}  \&
  {Celotti}}{{Oganesyan} et~al.}{2017}]{oganesyan17}
{Oganesyan} G.,  {Nava} L.,  {Ghirlanda} G.,   {Celotti} A.,  2017, \mn@doi
  [\apj] {10.3847/1538-4357/aa831e}, \href
  {https://ui.adsabs.harvard.edu/abs/2017ApJ...846..137O} {846, 137}

\bibitem[\protect\citeauthoryear{Oganesyan, Nava, Ghirlanda, Melandri  \&
  Celotti}{Oganesyan et~al.}{2019}]{Oganesyan:2019fpa}
Oganesyan G.,  Nava L.,  Ghirlanda G.,  Melandri A.,   Celotti A.,  2019,
  \mn@doi [Astron. Astrophys.] {10.1051/0004-6361/201935766}, 628, A59

\bibitem[\protect\citeauthoryear{{Perinati}, {Tenzer}, {Santangelo}, {Cordier},
  {Gotz}, {Fraser}  \& {Osborne}}{{Perinati} et~al.}{2012}]{Perinati2012MXT}
{Perinati} E.,  {Tenzer} C.,  {Santangelo} A.,  {Cordier} B.,  {Gotz} D.,
  {Fraser} G.~W.,   {Osborne} J.~P.,  2012, in {Takahashi} T.,  {Murray} S.~S.,
    {den Herder} J.-W.~A.,  eds,  Society of Photo-Optical Instrumentation
  Engineers (SPIE) Conference Series Vol. 8443, Space Telescopes and
  Instrumentation 2012: Ultraviolet to Gamma Ray. p. 84434T,
  \mn@doi{10.1117/12.925458}

\bibitem[\protect\citeauthoryear{Pescalli, Ghirlanda, Salafia, Ghisellini,
  Nappo  \& Salvaterra}{Pescalli et~al.}{2015}]{Pescalli:2014qja}
Pescalli A.,  Ghirlanda G.,  Salafia O.~S.,  Ghisellini G.,  Nappo F.,
  Salvaterra R.,  2015, \mn@doi [Mon. Not. Roy. Astron. Soc.]
  {10.1093/mnras/stu2482}, 447, 1911

\bibitem[\protect\citeauthoryear{Pian et~al.}{Pian et~al.}{2000}]{Pian:1999ec}
Pian E.,  et~al., 2000, \mn@doi [Astrophys. J.] {10.1086/308978}, 536, 778

\bibitem[\protect\citeauthoryear{Plotnikov, Pelletier  \& Lemoine}{Plotnikov
  et~al.}{2013}]{Plotnikov:2012ew}
Plotnikov I.,  Pelletier G.,   Lemoine M.,  2013, \mn@doi [Mon. Not. Roy.
  Astron. Soc.] {10.1093/mnras/sts696}, 430, 1280

\bibitem[\protect\citeauthoryear{{Poolakkil} et~al.,}{{Poolakkil}
  et~al.}{2021}]{poolakkil}
{Poolakkil} S.,  et~al., 2021, arXiv e-prints, \href
  {https://ui.adsabs.harvard.edu/abs/2021arXiv210313528P} {p. arXiv:2103.13528}

\bibitem[\protect\citeauthoryear{Preece, Briggs, Giblin, Mallozzi, Pendleton,
  Paciesas  \& Band}{Preece et~al.}{2002}]{Preece:2002}
Preece R.~D.,  Briggs M.~S.,  Giblin T.~W.,  Mallozzi R.~S.,  Pendleton G.~N.,
  Paciesas W.~S.,   Band D.~L.,  2002, The Astrophysical Journal, 581, 1248

\bibitem[\protect\citeauthoryear{{Rees} \& {Meszaros}}{{Rees} \&
  {Meszaros}}{1994}]{reesmeszaros94}
{Rees} M.~J.,  {Meszaros} P.,  1994, \mn@doi [\apjl] {10.1086/187446}, \href
  {https://ui.adsabs.harvard.edu/abs/1994ApJ...430L..93R} {430, L93}

\bibitem[\protect\citeauthoryear{Reville \& Bell}{Reville \&
  Bell}{2014}]{Reville:2014mta}
Reville B.,  Bell A.~R.,  2014, \mn@doi [Mon. Not. Roy. Astron. Soc.]
  {10.1093/mnras/stu088}, 439, 2050

\bibitem[\protect\citeauthoryear{Rudolph, Heinze, Fedynitch  \& Winter}{Rudolph
  et~al.}{2020}]{Rudolph:2019ccl}
Rudolph A.,  Heinze J.,  Fedynitch A.,   Winter W.,  2020, \mn@doi [Astrophys.
  J.] {10.3847/1538-4357/ab7ea7}, 893, 72

\bibitem[\protect\citeauthoryear{Sagiv et~al.}{Sagiv
  et~al.}{2014}]{Sagiv:2013rma}
Sagiv I.,  et~al., 2014, \mn@doi [Astron. J.] {10.1088/0004-6256/147/4/79},
  147, 79

\bibitem[\protect\citeauthoryear{Samuelsson, Bégué, Ryde  \&
  Pe'er}{Samuelsson et~al.}{2019}]{Samuelsson:2018fan}
Samuelsson F.,  Bégué D.,  Ryde F.,   Pe'er A.,  2019, \mn@doi [Astrophys.
  J.] {10.3847/1538-4357/ab153c}, 876, 93

\bibitem[\protect\citeauthoryear{{Samuelsson}, {B{\'e}gu{\'e}}, {Ryde}, {Pe'er}
   \& {Murase}}{{Samuelsson} et~al.}{2020}]{Samuelsson:2020upt}
{Samuelsson} F.,  {B{\'e}gu{\'e}} D.,  {Ryde} F.,  {Pe'er} A.,   {Murase} K.,
  2020, arXiv e-prints, \href
  {https://ui.adsabs.harvard.edu/abs/2020arXiv200502417S} {p. arXiv:2005.02417}

\bibitem[\protect\citeauthoryear{Sari, Piran  \& Narayan}{Sari
  et~al.}{1998}]{Sari:1997qe}
Sari R.,  Piran T.,   Narayan R.,  1998, \mn@doi [Astrophys. J. Lett.]
  {10.1086/311269}, 497, L17

\bibitem[\protect\citeauthoryear{Schulze et~al.}{Schulze
  et~al.}{2014}]{Schulze:2014fia}
Schulze S.,  et~al., 2014, \mn@doi [Astron. Astrophys.]
  {10.1051/0004-6361/201423387}, 566, A102

\bibitem[\protect\citeauthoryear{Senno, Murase  \& Meszaros}{Senno
  et~al.}{2016}]{Senno:2015tsn}
Senno N.,  Murase K.,   Meszaros P.,  2016, \mn@doi [Phys. Rev.]
  {10.1103/PhysRevD.93.083003}, D93, 083003

\bibitem[\protect\citeauthoryear{{Siellez} \& {LIGO Team}}{{Siellez} \& {LIGO
  Team}}{2018}]{siellez2018}
{Siellez} K.,  {LIGO Team} 2018, in American Astronomical Society Meeting
  Abstracts \#231. p. 107.06

\bibitem[\protect\citeauthoryear{Singer et~al.}{Singer
  et~al.}{2013}]{Singer:2013xha}
Singer L.~P.,  et~al., 2013, \mn@doi [Astrophys. J.]
  {10.1088/2041-8205/776/2/L34}, 776, L34

\bibitem[\protect\citeauthoryear{Sironi \& Spitkovsky}{Sironi \&
  Spitkovsky}{2009}]{Sironi:2009jw}
Sironi L.,  Spitkovsky A.,  2009, \mn@doi [Astrophys. J.]
  {10.1088/0004-637X/698/2/1523}, 698, 1523

\bibitem[\protect\citeauthoryear{{Sironi}, {Spitkovsky}  \& {Arons}}{{Sironi}
  et~al.}{2013}]{sironispitkovskyarons2013}
{Sironi} L.,  {Spitkovsky} A.,   {Arons} J.,  2013, \mn@doi [\apj]
  {10.1088/0004-637X/771/1/54}, \href
  {http://adsabs.harvard.edu/abs/2013ApJ...771...54S} {771, 54}

\bibitem[\protect\citeauthoryear{{Sparre} \& {Starling}}{{Sparre} \&
  {Starling}}{2012}]{sparre2012}
{Sparre} M.,  {Starling} R. L.~C.,  2012, \mn@doi [\mnras]
  {10.1111/j.1365-2966.2012.21858.x}, \href
  {https://ui.adsabs.harvard.edu/abs/2012MNRAS.427.2965S} {427, 2965}

\bibitem[\protect\citeauthoryear{Starling et~al.}{Starling
  et~al.}{2011}]{Starling:2010ed}
Starling R. L.~C.,  et~al., 2011, \mn@doi [Mon. Not. Roy. Astron. Soc.]
  {10.1111/j.1365-2966.2010.17879.x}, 411, 2792

\bibitem[\protect\citeauthoryear{Suda et~al.}{Suda et~al.}{2021}]{Suda:2021e7}
Suda Y.,  et~al., 2021, \mn@doi [PoS] {10.22323/1.395.0797}, ICRC2021, 797

\bibitem[\protect\citeauthoryear{Sun, Zhang  \& Li}{Sun
  et~al.}{2015}]{Sun:2015bda}
Sun H.,  Zhang B.,   Li Z.,  2015, \mn@doi [Astrophys. J.]
  {10.1088/0004-637X/812/1/33}, 812, 33

\bibitem[\protect\citeauthoryear{Virgili, Liang  \& Zhang}{Virgili
  et~al.}{2009}]{Virgili:2008gp}
Virgili F.,  Liang E.,   Zhang B.,  2009, \mn@doi [Mon. Not. Roy. Astron. Soc.]
  {10.1111/j.1365-2966.2008.14063.x}, 392, 91

\bibitem[\protect\citeauthoryear{Volnova et~al.}{Volnova
  et~al.}{2017}]{Volnova:2016lal}
Volnova A.~A.,  et~al., 2017, \mn@doi [Mon. Not. Roy. Astron. Soc.]
  {10.1093/mnras/stw3297}, 467, 3500

\bibitem[\protect\citeauthoryear{Waxman, Meszaros  \& Campana}{Waxman
  et~al.}{2007}]{Waxman:2007rr}
Waxman E.,  Meszaros P.,   Campana S.,  2007, \mn@doi [Astrophys. J.]
  {10.1086/520715}, 667, 351

\bibitem[\protect\citeauthoryear{Wu, Qiu  \& Cai}{Wu et~al.}{2011}]{Wu2012VT}
Wu C.,  Qiu Y.~L.,   Cai H.~B.,  2011, \mn@doi [Proceedings of the
  International Astronomical Union] {10.1017/S1743921312013646}, 7, 421–422

\bibitem[\protect\citeauthoryear{Zhang, Fan, Shen, Xu, Zhang, Wei, Burrows  \&
  Zhang}{Zhang et~al.}{2012}]{Zhang:2012jc}
Zhang B.-B.,  Fan Y.-Z.,  Shen R.-F.,  Xu D.,  Zhang F.-W.,  Wei D.-M.,
  Burrows D.~N.,   Zhang B.,  2012, \mn@doi [Astrophys. J.]
  {10.1088/0004-637X/756/2/190}, 756, 190

\bibitem[\protect\citeauthoryear{Zhang, Murase, Kimura, Horiuchi  \&
  Mészáros}{Zhang et~al.}{2018}]{Zhang:2017moz}
Zhang B.~T.,  Murase K.,  Kimura S.~S.,  Horiuchi S.,   Mészáros P.,  2018,
  \mn@doi [Phys. Rev.] {10.1103/PhysRevD.97.083010}, D97, 083010

\bibitem[\protect\citeauthoryear{Zhang, Murase, Veres  \& M\'esz\'aros}{Zhang
  et~al.}{2021}]{Zhang:2020qbt}
Zhang B.~T.,  Murase K.,  Veres P.,   M\'esz\'aros P.,  2021, \mn@doi
  [Astrophys. J.] {10.3847/1538-4357/ac0cfc}, 920, 55

\makeatother
\end{thebibliography}
\bibliographystyle{mnras}

\begin{appendices}

\section{Theoretical predictions for the photon spectrum}
\label{sec:shape_theo_predic}
We describe the theoretical predictions of a spectrum produced by an electron distribution following a power law $N_e(E) \propto E^{-p}$ of index $p > 2$ above a minimum Lorentz factor $\gamma_\mathrm{e, min}$. This is used in the following, as part of the interpretation of the simulated spectra and light curves. We pay special attention to the processes that can shape the spectrum below the synchrotron peak
and to the dependence of a potential HE signal on the GRB parameters.

For the following it is helpful to define 
\begin{equation}
    \gamma_{e,c} = \frac{6 \pi m_e c}{\sigma_t B^{\prime 2} t^\prime_\mathrm{ex}}
\end{equation}
\noindent which represents the Lorentz factor of electrons whose synchrotron cooling time-scale $t^\prime_\mathrm{syn}$ is equal to the shell expansion time-scale $t^\prime_\mathrm{ex}$.

We want to summarize basic theoretical predictions on the shape of the SED. These are generally applicable to GRBs, and not specific to ll-GRBs. However, due to the specific properties of ll-GRBs compared to high-luminosity events (lower wind luminosities, lower Lorentz factors), different regimes may be realized. In the following we will discuss the differential spectrum (usually represented as $\nu F_\nu \propto E^2 dN/dE$, where $N(E)$ is the number of photons $N$ at a given energy $E$) and the spectral index of the differential photon flux ($dN/dE$), assuming a power-law spectral shape. We will denote the slope below the peak of $\nu F_\nu$ as $\alpha$, as described \eg \, by the Band function (\cite{Band:1993eg}). We follow \cite{Sari:1997qe}  and recapitulate the shape of the pure synchrotron spectrum $\nu F_\nu$, although omitting the effect of synchrotron self-absorption. 
Generally, the spectral shape can be divided in two regimes:

\begin{enumerate}
    \item[(a)] \textit{Fast-cooling regime : $\gamma_\mathrm{e, min} \gg \gamma_{e,c}$}:
    
    \begin{align}
        \frac{\nu F_\nu}{F_{\nu, \mathrm{max}}} = \begin{cases}
        (\frac{\nu}{\nu_c})^{4/3}, & \nu < \nu_c  \\
        (\frac{\nu}{\nu_c})^{1/2}, & \nu_c < \nu < \nu_m \\		
        (\frac{\nu_m}{\nu_c})^{-1/2} (\frac{\nu}{\nu_m})^{-(p-2)/2 }, & \nu_m < \nu \\		
        	\end{cases} \, .
    \end{align} 
    Here $\nu_m = \nu(\gamma_\mathrm{e,min})$ is the synchrotron frequency of an electron with Lorentz factor $\gamma_\mathrm{e,min}$, $\nu_c = \nu(\gamma_\mathrm{e,c})$ the synchrotron frequency of an electron with Lorentz factor $\gamma_\mathrm{e,c}$.
    
    All (initial) electrons cool via synchrotron cooling on the dynamical time-scale and thus convert their energy relatively efficiently into radiation. The spectral index of the photon flux below the peak of $\nu F_\nu$ (located at $\nu_m$) is given by $\alpha = -1.5$ .
    
    \item[(b)] \textit{Slow-cooling regime : $\gamma_\mathrm{e, min} < \gamma_\mathrm{e, c}$}: 
    
    \begin{align}
        \frac{\nu F_\nu}{F_{\nu, \mathrm{max}}} = \begin{cases}
        (\frac{\nu}{\nu_m})^{4/3}, & \nu < \nu_m  \\
        \frac{\nu}{\nu_m})^{-(p-3)/2}, & \nu_m < \nu < \nu_c \\		
        (\frac{\nu_c}{\nu_m})^{-(p-3)/2} (\frac{\nu}{\nu_c})^{-(p-2)/2}, & \nu_c < \nu \\		
        	\end{cases} \, .
    \end{align}
    Not all electrons cool via synchrotron cooling on the dynamical time-scale, some thus don't convert their energy efficiently into radiation. The spectral index of the photon flux below the peak of $\nu F_\nu$ is given by $\alpha = -2/3$.
\end{enumerate}

With pure synchrotron radiation, one achieves spectral indeces $\alpha$ below the peak that are either equal to $\alpha = -1.5$ or $\alpha = -2/3$, which is not in agreement with observations (\eg \, \cite{Preece:2002,Ghisellini:1999wu}, see also current GRB catalog spectral analysis as in e.g. \cite{ poolakkil,Gruber:2014iza,Lien:2016zny}). For the slow-cooling regime one additionally suffers from a poor radiative efficiency (defined as $f_\mathrm{rad} = u_\gamma / u_e$,  the integrated (final) photon energy divided by the integral injected electron energy). This further reduces the already small overall efficiency of the internal shock model, imposing an even higher engine power. 
Note that \cite{Daigne:2010fb} also identified the \textit{marginally fast cooling} regime, for which $\gamma_\mathrm{e, min} \simeq \gamma_\mathrm{e, c}$ and low-energy spectral slopes of $ -1 < \alpha < -2/3$ can be realized while still achieving a relatively high radiative efficiency. Recently \cite{oganesyan17} fitted the sample of prompt GRB spectra observed  down to $\sim$ 0.5 keV energies by the two low energy power-laws and a break energy corresponding to the cooling break frequency; they found a small ratio of the spectral peak  energy and the cooling break frequency, suggesting a regime of moderately fast cooling. This scenario has also been explored in \cite{Oganesyan:2019fpa}, with a focus on optical fluxes.

Additional effects can shape the spectra and impact $\alpha$:
\begin{enumerate}
    \item \textit{Synchrotron self-absorption:} At the lowest energies, electrons can re-absorb photons through synchrotron self-absorption. This gives rise to an additional break in the photon spectrum.
    \item \textit{Inverse Compton radiation:} introduces a second peak at high energies. The spectral shape below spectral peak may be affected by inverse Compton scatterings, resulting in harder low energy spectral slopes (\cite{Nakar:2009er, Daigne:2010fb, Duran:2012ww}). 
    \item \textit{$\gamma \gamma$-absorption:} If the photon densities are high, $\gamma \gamma $ pairs annihilate to lepton pairs at the highest energies. Those pairs are expected to contribute via synchrotron radiation at low energies (and therefore may reshape the spectrum below the synchrotron peak) and inverse Compton radiation (at intermediate energies) (\cite{Asano:2007my, 2011ApJ...739..103A}). As densities scale inversely with the comoving volume, this is especially relevant for collisions at low radii and/or with low Lorentz factors.
\end{enumerate}

 If those theoretical predictions are compared to data/ results of spectral fits, one has to be aware that the reported slopes are usually determined by fitting either a Band empirical function or a cut-off power law, and that resulting slopes are very sensitive to the functional fitting range which is informed by the sensitivity of the experiments.
 
\section{Parameter space exploration for different values of $\epsilon_\mathrm{B}$}

\label{appendix:analytical_estimates}
\begin{figure*}
\centering
\makebox[\textwidth][c]{
\subfloat[sp-GRB]{\includegraphics[width=.33 \textwidth]{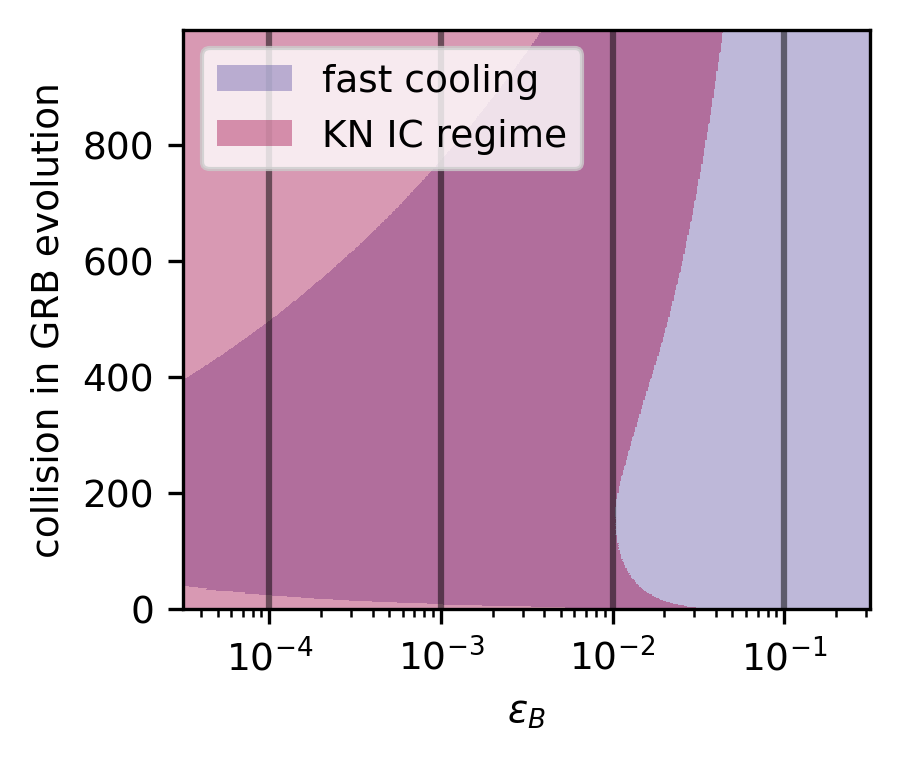}}
\subfloat[ul-GRB]{\includegraphics[width=.33 \textwidth]{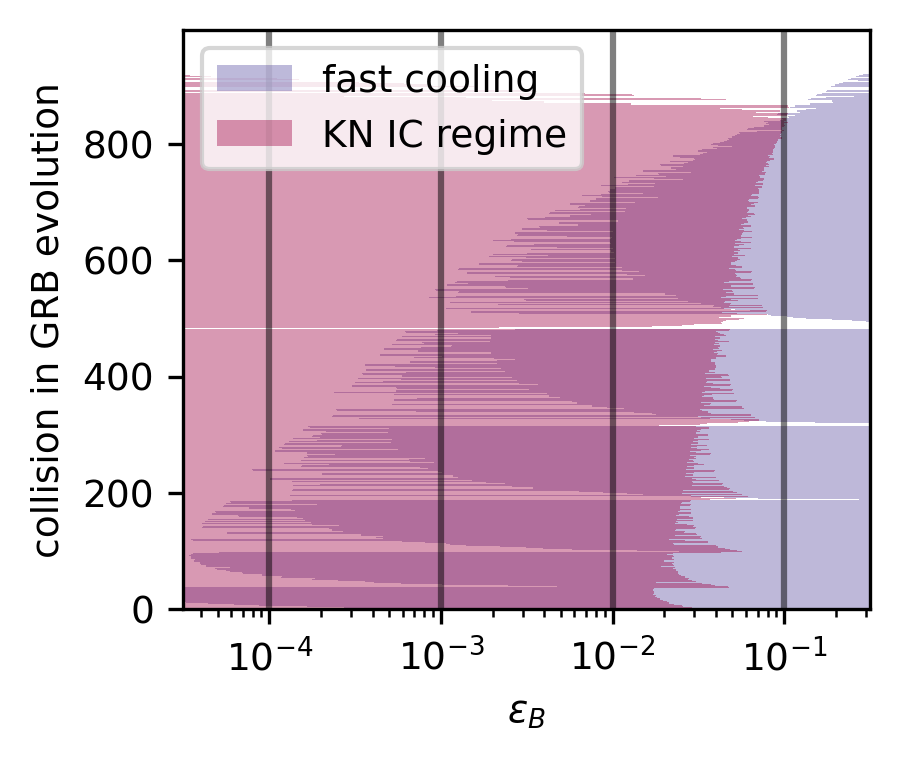}}
\subfloat[hl-GRB]{\includegraphics[width=.33 \textwidth]{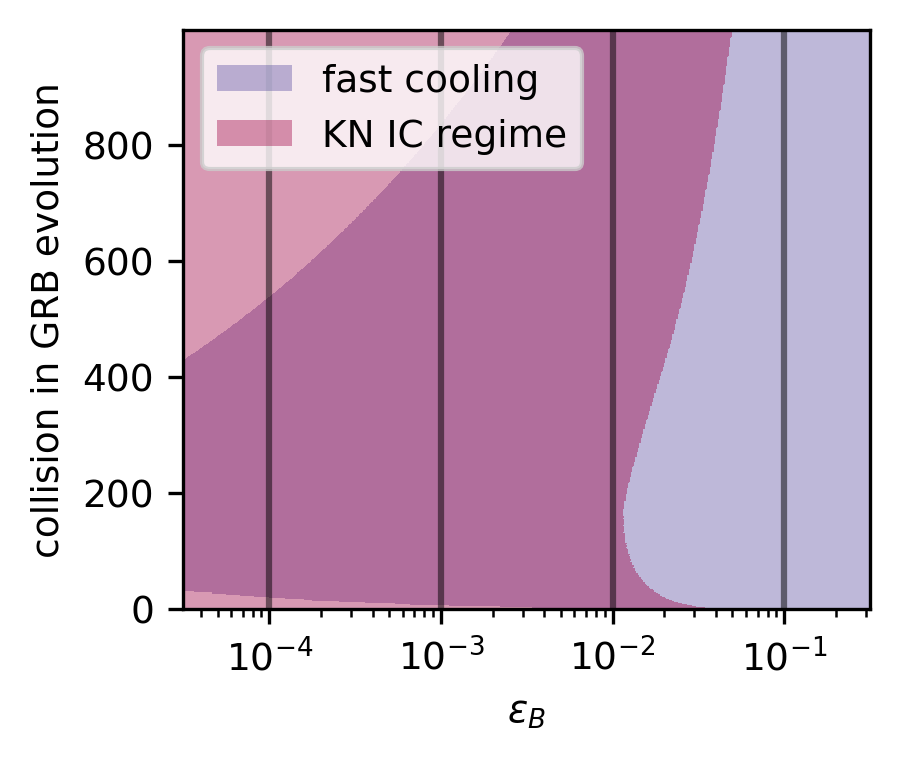}}
}
\caption{Parameter space (see text) for the fast-cooling (blue) and Klein-Nishina inverse Compton (purple) regime as a function of $\epsilon_\mathrm{B}$ for sp-GRB, GRBl-long and hl-GRB for a fixed fireball evolution (Lorentz factor distribution and injection luminosity). The dark purple region marks the overlap of both regimes. Vertical lines correspond to the choices for $\epsilon_\mathrm{B}$ considered for the SED modelling.}
\label{fig:microphysics_parameters}
\end{figure*}

In this section we show how during  the jet evolution the different emission regimes  are achieved (with different respective properties of the observed radiation, defined by analytical estimates), depending largely on $\epsilon_B$ for the three model GRBs. For a detailed discussion on the theoretical predictions of the corresponding photon spectra see Section~\ref{sec:shape_theo_predic}. For this purpose, we examine different choices of $\epsilon_\mathrm{B}$ ($x$-axis of \figu{microphysics_parameters}) and different collisions between single layers in the GRB evolution, assuming the same fireball evolution which is set by $L_\mathrm{wind}$ and Lorentz factor distribution for each GRB. The individual collisions are numbered chronologically by the time they occur in the source frame ($y$-axis of \figu{microphysics_parameters}).
For each collision and $\epsilon_B$ we evaluate the following two conditions:

\begin{enumerate}
    \item[1.] \textit{Fast cooling regime} \newline
    This regime is realized if $\gamma_\mathrm{e, c} < \gamma_\mathrm{e, min}$ is satisfied. In this case, all electrons cool on the dynamical time-scale, which translates into a relatively high radiative efficiency.
    
    The region in the parameter space fulfilling this criterium are marked \textbf{blue} in \figu{microphysics_parameters}.
    \item[2.]\textit{Inverse Compton scatterings occuring in the Klein-Nishina regime} \newline
    Following \citealt{Daigne:2010fb} and \citealt{Nakar:2009er}, we define this regime  
    as the region where $\eta_\mathrm{m}^{1/3} \leq Y_{\mathrm{Th}} \leq \eta_\mathrm{m}^3 $.
    Here we calculate the Compton parameter directly from the microphysics parameters as $Y_\mathrm{Th} \simeq [(p-2)/(p-1)] [\epsilon_\mathrm{e}/\epsilon_\mathrm{B}]$, and $\eta_\mathrm{m}  \simeq 100 \ (\gamma_\mathrm{e, min} / \mathrm{100})^{3} (B^\prime / \mathrm{3000 G})$. 
    The region in the parameter space 
    where this criterium is satisfied are marked {\bf purple} in \figu{microphysics_parameters}.
\end{enumerate}

If both criteria are met, the resulting overlap is a {\bf dark purple} region. Due to the low luminosity of the GRBs, relatively high values of $\epsilon_\mathrm{B}$ are required for the fast cooling regime condition-- especially for ul-GRB, where even for $\epsilon_\mathrm{B} = 10^{-1}$ this condition is not satisfied in all collisions.
The Klein-Nishina inverse Compton regime is generally realized for lower values of $\epsilon_\mathrm{B}$.

\section{Impact of the different radiative processes}
\label{appendix:radiative_processes}

Since the impact of secondary lepton pairs created by $\gamma \gamma$ pair annihilation was not accounted for in past studies, e.g. \cite{Bosnjak:2008bd}, we explicitly study this effect in the following section. Given that some recent studies (\cite{Oganesyan:2019fpa, Samuelsson:2020upt}) compare the observed spectra to synchrotron predicitons, we also show the obtained spectra for a pure synchrotron calculation.

\begin{figure*}
\centering
\makebox[\textwidth][c]{
\subfloat[sp-GRB]{\includegraphics[width=.33 \textwidth]{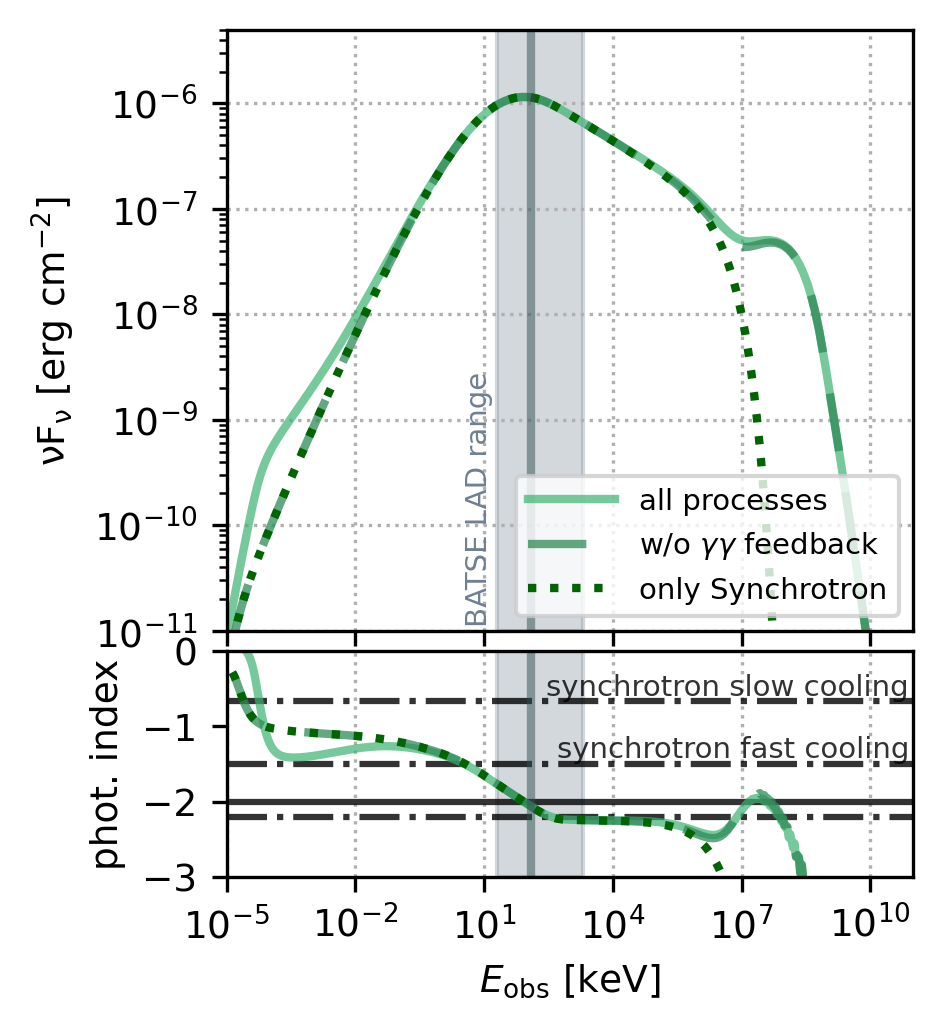}}
\subfloat[ul-GRB]{\includegraphics[width=.33 \textwidth]{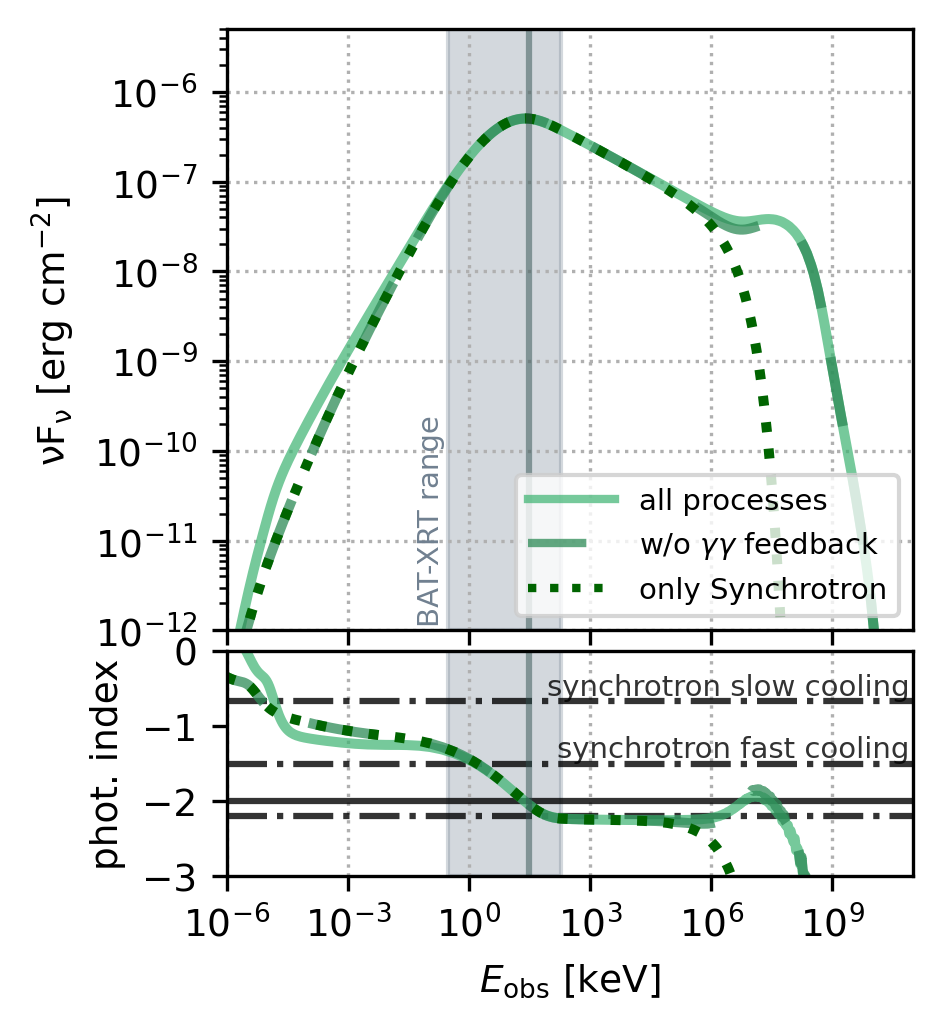}}
\subfloat[hl-GRB]{\includegraphics[width=.33 \textwidth]{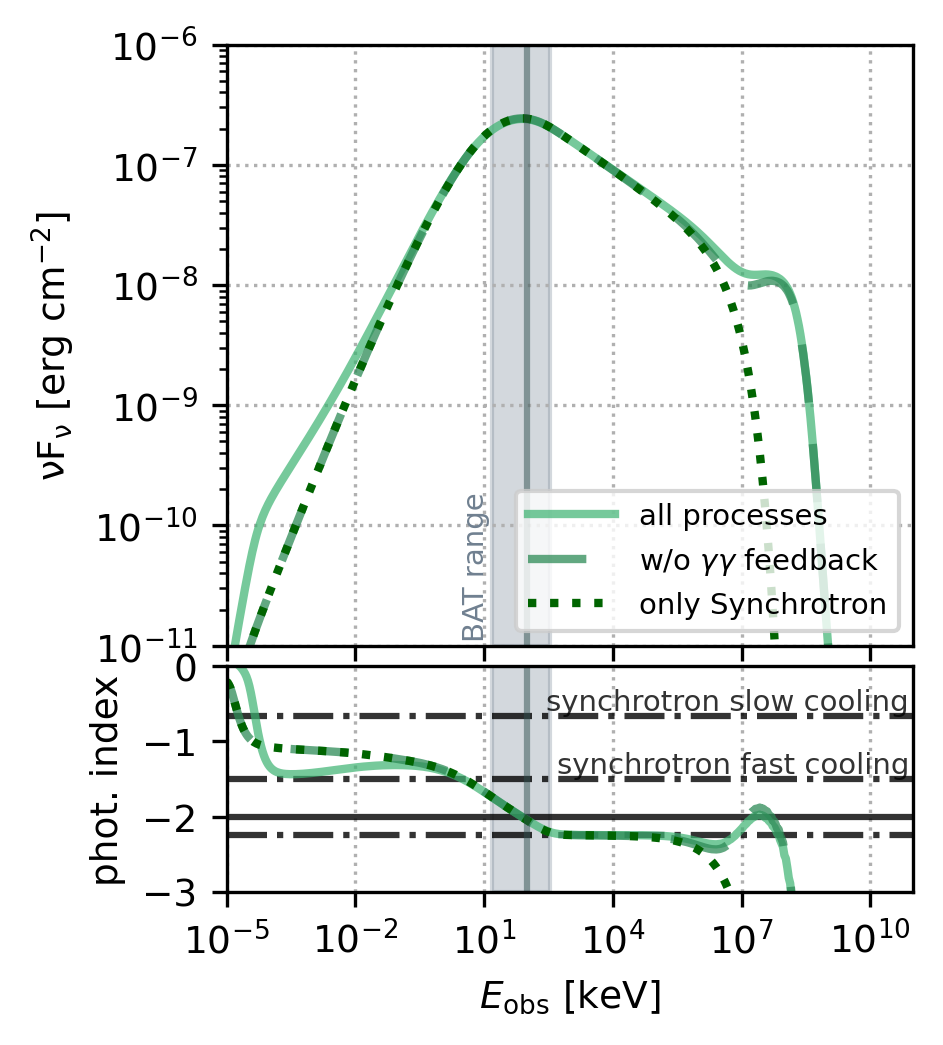}}}
\caption{Time integrated spectra $\nu F_\nu \propto E^2N(E)$ for the model GRBs for $\epsilon_B = 10^{-2}$. We show the results with/ without taking into account the effect of secondary $e^{+} / e^{-}$ produced in $\gamma \gamma$-absorption and for a pure synchrotron model. For each GRB we show the energy range of the observing instrument as a grey band and the observed peak energy as a vertical line. \newline
 The lower panel shows the photon index of $N(E)$, the dashed lines correspond to the synchrotron predictions ($ -2/3$, $ -3/2$ and $- 2.25$) , the solid line marks the position of maxima/minima of $\nu F_\nu$}
\label{fig:time_integrated_spectra_appendix}
\end{figure*}

In \figu{time_integrated_spectra_appendix} we show the time-integrated spectra for sp-GRB, ul-GRB and hl-GRB for $\epsilon_B= 10^{-2}$. Comparing to a full radiative treatment of all processes, we show the spectra obtained without taking into account secondary electrons produced by $\gamma \gamma $-annihilation, as well as the pure synchrotron prediction.
We first examine the impact of secondaries created by $\gamma \gamma $-pair production. 
Due to their synchrotron radiation, they enhance the flux in the eV-regime by roughly one order of magnitude -- stressing the importance of a full radiative treatment when comparing predictions to observations in the optical regime. 
At intermediate energies around $10^7$-$10^8$~keV, the inverse Compton scatterings of the $\gamma \gamma$-produced pairs result in a 
slight enhancement of the flux. 
As the intensity $\gamma \gamma$ absorption depends on the level of VHE emission, these additional effects will be more pronounced for a strong inverse Compton component (and thus, low $\epsilon_B$). 

The pure synchrotron case has a cutoff in the observed spectra at approximately $10^7$~keV. In the optical regime, the prediction is equal to the case without impact of secondaries from $\gamma \gamma $-pair production. Even in the case of moderate magnetic fields (as in this example), the effects of inverse Compton radiation should thus be taken into account when modelling the spectra.

\section{Time-resolved spectra}

\begin{figure*}
\centering
\includegraphics[width=.89 \textwidth]{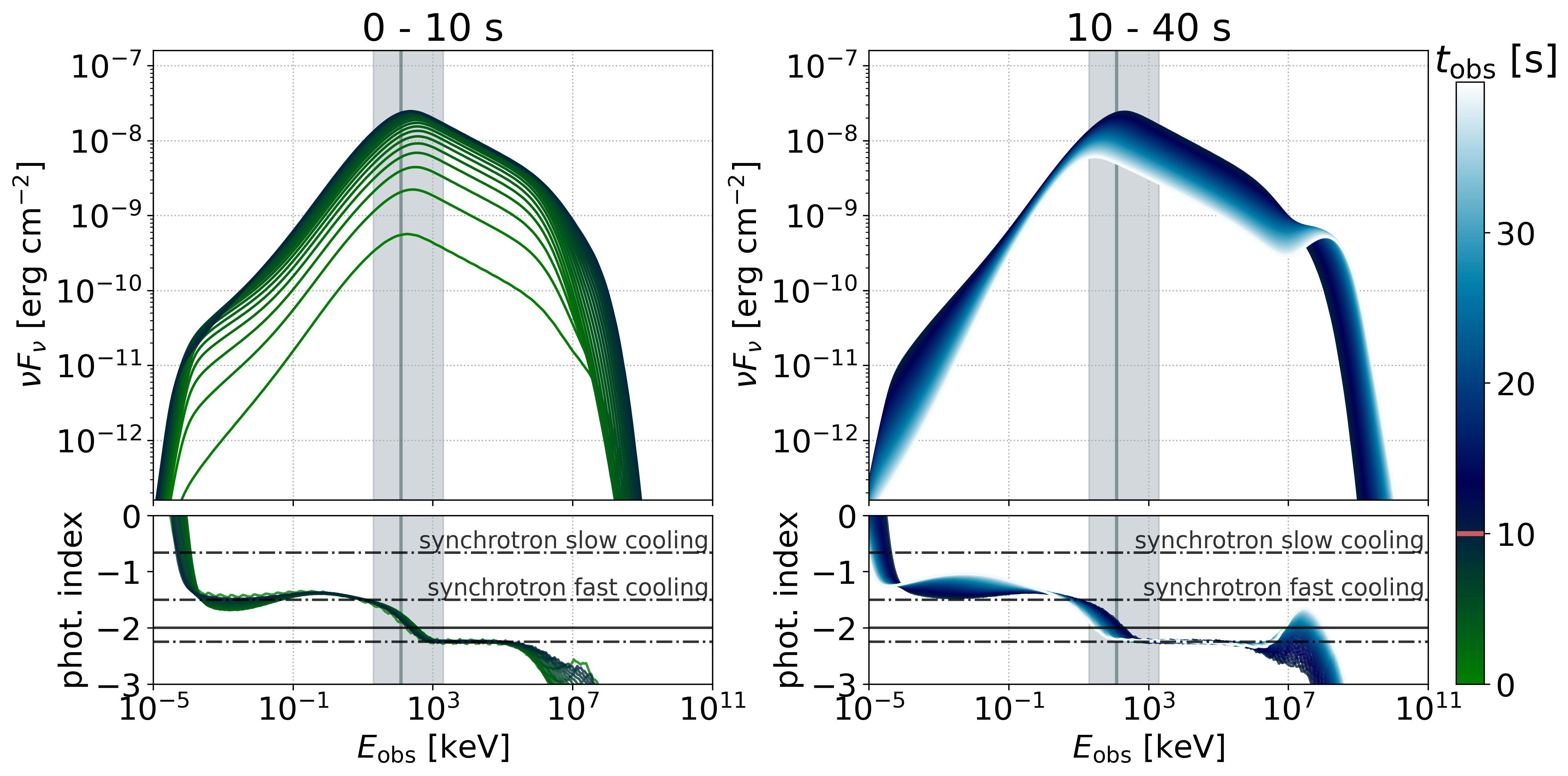}
\newline
\includegraphics[width=.89 \textwidth]{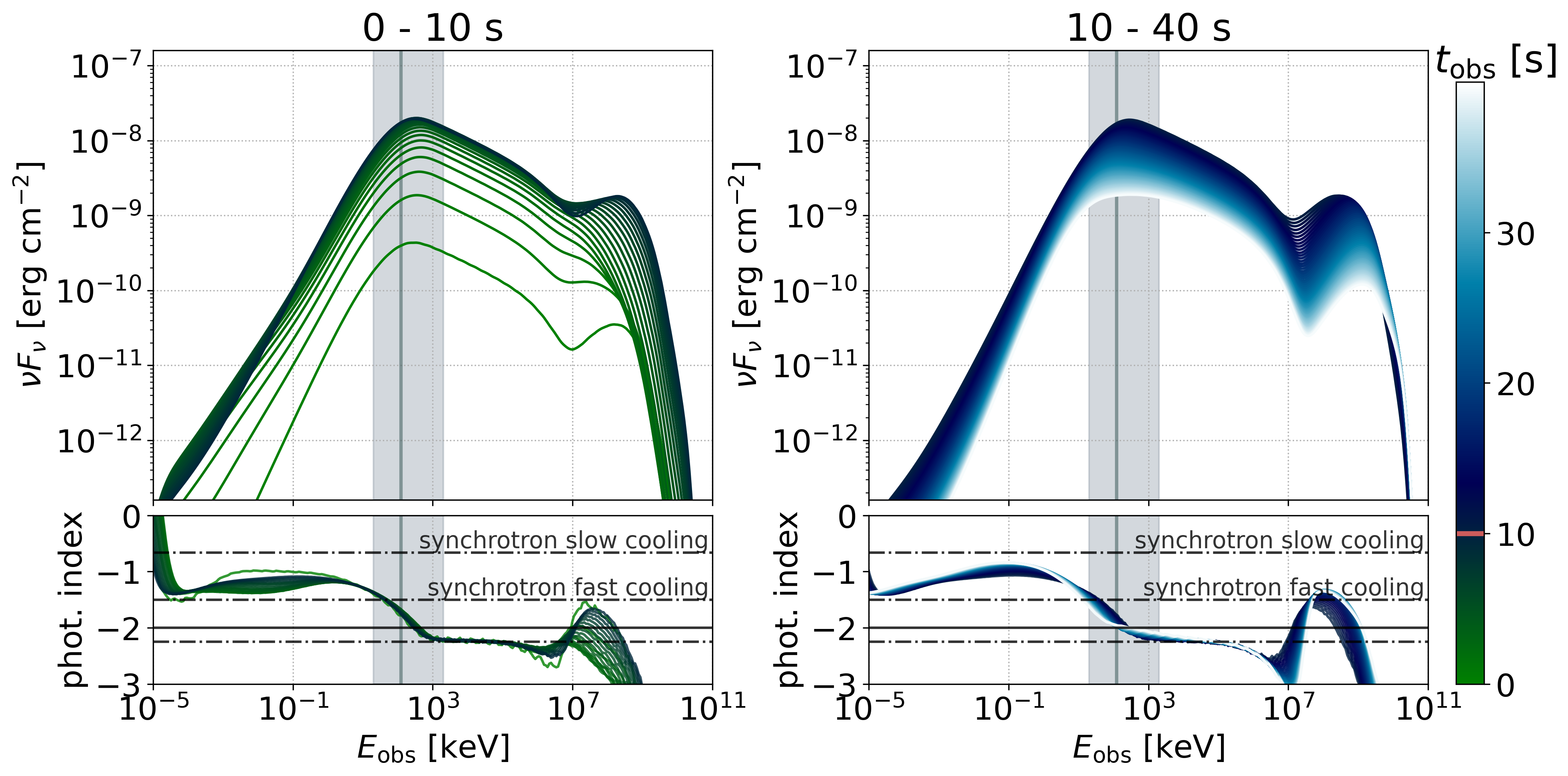}

\caption{Time-resolved spectra (integrated every 0.5 s) for sp-GRB for $\epsilon_\mathrm{B} = 10^{-2}$ (upper panel) and $\epsilon_\mathrm{B} = 10^{-4}$ (lower panel). The BATSE LAD energy range is shown as a grey band, the vertical line corresponds to the observed peak energy of the reference GRB (122~keV).
The color scale indicates the observed time of the spectrum, from early (green) to late (blue-white), see colorbar on the right hand side.  We show distinct plots for the rise of the peak ($t_{\rm obs} < 10 s$, left) and the decay ($t_{\rm obs} > 10 s$, right).
 The lower panel shows the photon index, the dashed lines correspond to the synchrotron predictions ($ -2/3$, $ -3/2$ and $- 2.25$) , the solid line marks the position of maxima/minima of $\nu F_\nu$.
 }
\label{fig:time_resolved_spectra}
\end{figure*}

We investigate the observed temporal evolution of single pulses by analyzing GRB~SP. The time-resolved spectra for this prototype are shown in \figu{time_resolved_spectra} for $\epsilon_\mathrm{B} = 10^{-2}$ (upper panel) and $\epsilon_\mathrm{B} = 10^{-4}$ (lower panel). The spectra are integrated on time scale of 0.5~s. We again point out that the overall fireball efficiency for $\epsilon_\mathrm{B} = 10^{-4}$ is very low and it thus might not be a realistic choice of parameter. However, it shows the most pronounced effects of inverse Compton scatterings, which is why we will study it here. 
Single pulses in SP-UL and the the single pulse of hl-GRB in principle show the similar behaviour.

For both cases the evolution of the peak energy $E_{\rm peak}$ shows a  trend generally observed  during the prompt GRB emission (e.g. \citealt{Kaneko:2006mt}): The  spectrum evolves from harder to softer one with time.

For the spectral index $\alpha$ we also predict an evolution with time. Observed values of $\alpha$ should be considered cautiously, keeping in mind the instrumental energy range used for spectral fit, burst brightness, and the time scale used for spectral integration, which affect the results.  

For low values of $\epsilon_\mathrm{B}$, the synchrotron cooling time is comparable to the inverse Compton cooling time even at early times. This enables efficient inverse Compton scatterings at early times, even despite high $\gamma \gamma$-absorption. As an effect, the HE component is clearly visible at early times for $\epsilon_\mathrm{B} = 10^{-4}$, but not for $\epsilon_\mathrm{B} = 10^{-2}$.

\section{Energy loss rates for different processes}

For a better understanding of the physical processes in the source we studied the energy loss rates for both leptons and hadrons.

\subsection{Leptonic loss rates}
\label{app:ele_cooling}

\begin{figure*}
\centering
\makebox[\textwidth][c]{
\subfloat[$\epsilon_\mathrm{B} = 10^{-3}$]{
\includegraphics[width=.45 \textwidth]{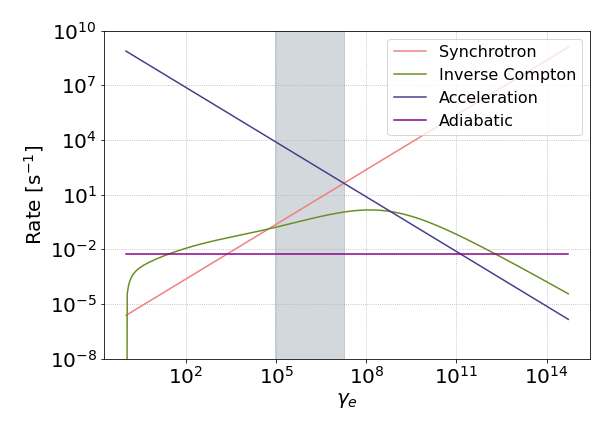}}
\subfloat[$\epsilon_\mathrm{B} = 10^{-1}$]{
\includegraphics[width=.45 \textwidth]{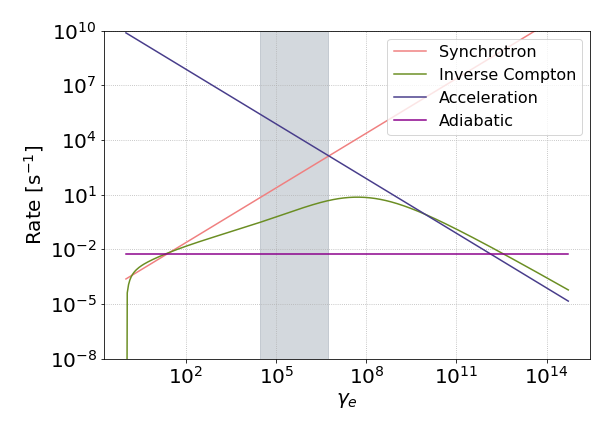}}

}
\caption{Loss rates in the plasma comoving frame for different processes for electrons at the maximum of the pulse (where most of the energy is dissipated) of sp-GRB. 
The grey shaded area marks the energy range of injected electrons. As acceleration, synchrotron  and inverse Compton losses depend on the magnetic field strength, we show the results for $\epsilon_\mathrm{B} = 10^{-3}$ and $\epsilon_\mathrm{B} = 10^{-1}$.}
\label{fig:time-scaleplot_ele}
\end{figure*}

\figu{time-scaleplot_ele} contains the energy loss rates for different leptonic processes at the maximum of the pulse (where most of the energy is disspated) for $\epsilon_\mathrm{B} = 10^{-1}$ and $\epsilon_\mathrm{B} = 10^{-3}$ for sp-GRB. 
Synchrotron losses dominate over the entire range of injection even for $\epsilon_\mathrm{B} = 10^{-3}$ (for most collisions the fast cooling regime condition is satisfied). Adiabatic cooling due to the shell expansion is sub-dominant compared to both inverse Compton and synchrotron cooling. For lower electron energies and low magnetic fields ($\epsilon_B = 10^{-4}$), inverse Compton coolings may dominate over both adiabatic and synchrotron cooling. 

\subsection{Hadronic loss rates}
\label{appendix:hadronic_rates}

As an addition to Section~6, we show the energy loss rates for iron and protons for both, an early (close to the source) and a late (far out) collision for sp-GRB ($\epsilon_\mathrm{B} = 10^{-1}$). The early collision is chosen such that $E_\mathrm{max, Fe}$ is minimal, the late collision corresponds to the maximum of $E_\mathrm{max, Fe}$.
Note that while acceleration and synchrotron losses depend on the magnetic field, we expect the hadronic processes to be almost independent of the magnetic field (given that the target photon fields are quite similar, which they are by construction around the peak).
We notice that for collisions which have small maximal cosmic-ray energies, photo-hadronic losses play an important role, while the maximal cosmic-ray energy (for iron) is achieved in a case where the energy is limited by adiabatic cooling. At this relatively late collision at lage radius the photon densities are low (as the volume scales with $R^2$). 

\begin{figure*}
\centering
\makebox[\textwidth][c]{
\subfloat[]{\includegraphics[width=.8 \textwidth]{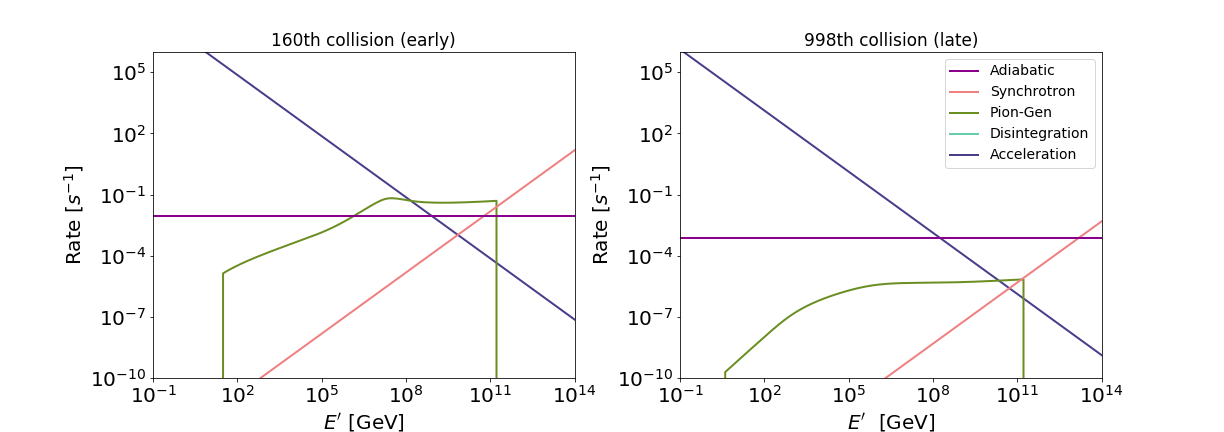}}}
\hfill
\makebox[\textwidth][c]{
\subfloat[]{\includegraphics[width=.8 \textwidth]{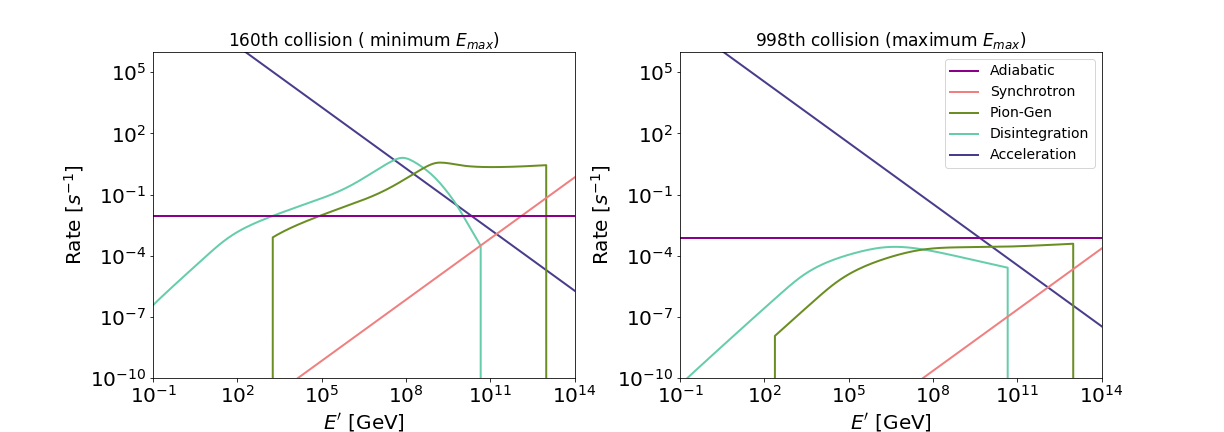}}}
\caption{Interaction rates in the shell comoving frame for GRB 980425-like burst for protons (upper panel) and iron (lower panel). We choose the 350th (530th)  collision, as for this collision the maximum energy of iron is minimal (maximal). }
\label{fig:time-scaleplots}
\end{figure*}

\end{appendices}

\end{document}